\documentclass[aps,prb,10pt,twocolumn,floatfix,showpacs,superscriptaddress]{revtex4-1}
\usepackage[colorlinks=true,citecolor=blue,linkcolor=blue,urlcolor=blue]{hyperref}
\usepackage{amsmath}
\usepackage{amssymb}
\usepackage{graphicx,amsmath,amsfonts,bm}
\usepackage{epstopdf}
\usepackage{bm}
\usepackage{bbm}
\usepackage{color}
\usepackage{relsize}
\usepackage{dsfont}
\usepackage{braket}
\usepackage[caption=false]{subfig}
\usepackage{hyperref}
\usepackage[all]{hypcap} 
\usepackage[T1]{fontenc}
\usepackage{dsfont}
\usepackage{mathbbol}

\renewcommand{\i}{\ensuremath{\mathrm{i}}}

\begin{document}
\title{Fraunhofer pattern in the presence of Majorana zero modes}

\author{F. Dominguez}
\affiliation{Institut f\"ur Mathematische Physik, Technische Universit\"at Braunschweig, 38106 Braunschweig, Germany}
\author{E. G. Novik}
\affiliation{Institute of Theoretical Physics, Technische Universit\"at Dresden, 01062 Dresden, Germany}
\author{P. Recher}
\affiliation{Institut f\"ur Mathematische Physik, Technische Universit\"at Braunschweig, 38106 Braunschweig, Germany}
\affiliation{Laboratory for Emerging Nanometrology, 38106 Braunschweig, Germany}

\date{\today}

\begin{abstract}
We propose a new platform to detect signatures of the presence of Majorana bound states (MBSs) 
in the Fraunhofer pattern of Josephson junctions featuring quantum spin Hall edge states on the normal part and Majorana bound states at the NS interfaces. 
We use a tight-binding model to demonstrate a drastic change in the periodicity of the Fraunhofer pattern when comparing trivial and non-trivial regimes. We explain these results in terms of the presence of additional parallel-spin electron-hole reflections, which due to the spin-momentum locking, occur as cross Andreev reflections, accumulating a different magnetic flux and yielding a change in the Fraunhofer periodicity. We show that this detection scheme exhibits some advantages compared to previous ones as it is robust against disorder, finite temperature and works in equilibrium. Furthermore, we introduce a scattering model that captures the main results of the microscopic calculations with MBSs and extend our discussion to the main differences found using accidental zero energy ABSs.
\end{abstract}

\maketitle

\section{Introduction}

In condensed matter, Majorana fermion quasiparticles\cite{Majorana1937a}, i.e.~$\gamma^\dagger=\gamma$ exhibit unconventional properties for the charge, which is neutral,  and the occupation number, which is not well-defined $\gamma^\dagger \gamma =\gamma\gamma^\dagger=1 $.\cite{Read2000a, Ivanov2001a,Kitaev2001a}
These exotic quasiparticles emerge as zero energy excitations in topological superconductors, among which, p-wave superconductors are the most studied platforms due to the possibility of engineering them by proximitizing semiconductor systems with strong spin-orbit interactions.\cite{Fu2009a, Lutchyn2010a, Oreg2010a}
Furthermore, Majorana zero modes exhibit a fractional nature, and therefore, they always appear in pairs as the result of delocalizing the information of a single fermion $c=(\gamma_1+i \gamma_2)/2$ onto the boundaries of the system. 
For example, in one-dimensional (1D) systems a pair of zero-dimensional bound states becomes localized at the boundaries of the topological superconductor\cite{Kitaev2001a,Lutchyn2010a,Oreg2010a}, whereas in two-dimensions, they emerge as an even number of chiral vortices.\cite{Law2009a}
Apart from their intrinsic interest, there are practical applications due to their individual charge neutrality, which protects them from the local coupling to environmental charge fluctuations, and more importantly, due to the possibility of performing computational operations by the adiabatic exchange of Majorana bound states, also known as braiding, which leads to an adiabatic state change within a degenerate ground state manifold \cite{Ivanov2001a}. For further information and references there is a collection of reviews that cover different aspects of these exotic particles, see for example Refs.~\onlinecite{Nayak2008a, Alicea2012a, Beenakker2013a, Aguado2017a, Schuray2020a, Prada2020a, Flensberg2021a}

Signatures of MBSs are found in several transport experiments. Two of the most studied experiments, the zero bias conductance in a NS junction\cite{Law2009a,Prada2012a,Li2012a} and the fractional Josephson effect \cite{Kitaev2001a} in the Josephson junction, report deviations from the theoretical predictions. In the case of the zero bias conductance, the signature consists of a quantized value $G=2e^2/h$ at zero temperature. However, most of the experimental realizations have shown a substantially suppressed value \cite{Mourik2012a,Deng2012a, Das2012a,Nichele2017a}, and only in one of them the conductance is consistently close to $2e^2/h$, exhibiting deviations below and above \cite{Hao2021a}. There are different explanations that can justify such deviations, some of them are compatible with a topological ground state, like effects of a finite temperature, or a finite coupling to the opposite MBS, while other explanations are compatible with a trivial ground state, like the scattering with quasi-Majorana bound states\cite{Fleckenstein2018a,Moore2018a, Marra2019a} or coupling to trivial zero-energy Andreev bound states (ZEABSs)\cite{Cayao2017a,Oladunjoye2019a}.
The situation is similar for the fractional Josephson effect, which can be probed by means of the Shapiro experiment, where odd Shapiro steps vanish \cite{Kitaev2001a, Fu2009a, Heck2011a, Badiane2011a, San-Jose2012a,Dominguez2012a,Pikulin2012a} or the Josephson radiation \cite{Deacon2017a,Laroche2019a}. Experimentally however, in most cases only few odd steps are suppressed when the driving frequency is low enough \cite{Rokhinson2012a, Wiedenmann2016a,Li2018a, Fischer2022a}, and only one contribution reports the lack of the first four odd steps\cite{Bocquillon2016a}. In this occasion, the signal can also be explained in terms of a topological state that coexists with trivial ones \cite{Dominguez2012a,Dominguez2017b,Pico2017a}, however, it is also possible that the behavior can be explained in terms of non-adiabatic transitions between Andreev bound states\cite{Yeyati2003a,San-Jose2012a,Dominguez2012a,Pikulin2012a, Houzet2013a, Virtanen2013a, Matthews2014a, Sau2017a}. Therefore, the need for detection schemes that are more susceptible to the triplet superconducting nature of the MBSs, such as the measurement of triplet correlations by coupling the topological superconductor to a spin-dependent current in a 3-terminal setup\cite{Haim2015a, Fleckenstein2021a} or the spin susceptibility of a Josephson junction are of utmost importance~\cite{Pakizer2021a}.

\begin{figure}[t]
\centering
\includegraphics*[width=8.0cm]{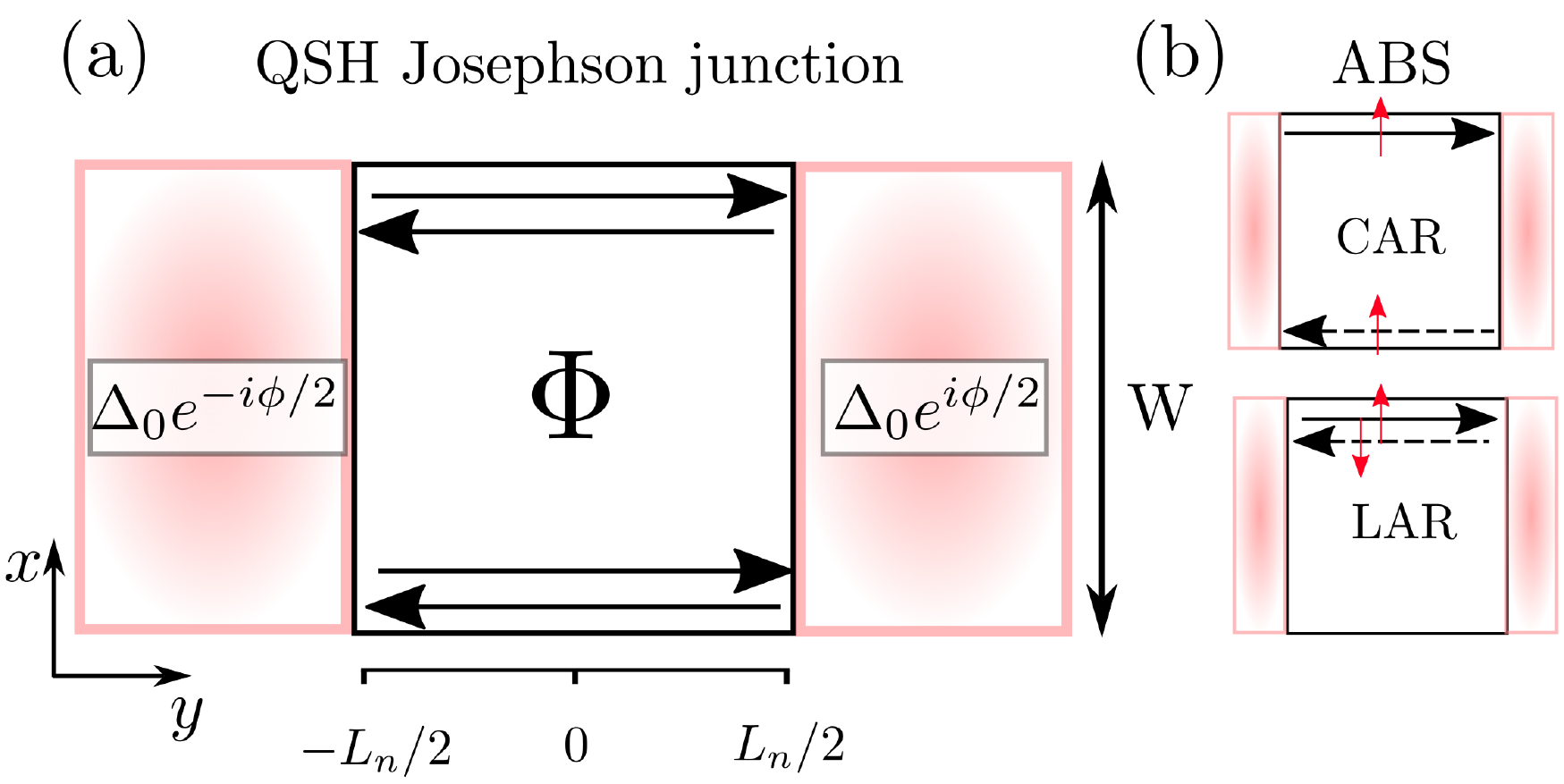}
\caption{[Panel (a)] Schematic representation of the planar quantum spin Hall Josephson junction. Here, pink and white areas represent superconducting and normal parts subjected to a magnetic flux $\Phi$. [Panel (b)] Possible Andreev bound states present in the setup. Due to spin-momentum locking, the crossed (local) Andreev bound states hold parallel (antiparallel) spin. Solid and dashed arrows represent electron- and hole-like helical edge states, respectively.
}\label{fig:QSH_ABS}
\end{figure}
 
In this work, we investigate signatures of the presence of MBSs at the NS interfaces that arise in the Fraunhofer pattern of a planar Josephson junction featuring quantum spin Hall edge states on the normal part, see Fig.~\ref{fig:QSH_ABS}.
Here, the spin-momentum locking of the helical edge states forces local and crossed Andreev reflections (LAR and CAR) to take different spin-symmetries, that is, $r^{he}_{\bar{s}s}$ for the LAR and $r^{he}_{s s}$ for the CAR. Note that in trivial junctions, the CAR is zero for a homogeneous   or effectively linear in momentum spin-orbit coupling. In contrast, in the presence of a MBS, LAR and CAR are of the same order. Interestingly, the presence of a magnetic flux can unravel both contributions since electrons accumulate a different magnetic flux when performing a LAR or a CAR process \cite{Schelter2012a,Haidekker2020a}.
In this way, we investigate the critical current of a Josephson junction as a function of an applied magnetic flux, also known as Fraunhofer pattern, expecting to observe an abrupt change on the profile when passing from the trivial to the topological regime. 
Since this detection scheme is performed in equilibrium, it removes the complications of parity conservation required in Fu and Kane's proposal~\cite{Fu2009a} and the possibility of non-adiabatic transitions that can appear in the Shapiro experiment or the Josephson radiation. Furthermore, using a scattering model we demonstrate that the change in Fraunhofer pattern profile is absent in the presence of accidental zero energy Andreev bound states if the direction of the Zeeman field is parallel to the helicity operator of the quantum spin Hall states. 

In the same spirit but with a different mechanism, other proposals suggest to use quantum Hall edge states to observe a change in the Fraunhofer pattern profile when MBSs are present\cite{Gavensky2020a}. 
Alternatively, one can study the change of periodicity in the conductance of a NSN junction interferometer fabricated on a 2D-topological insulator \cite{Benjamin2010a}. 
A recent experiment shows an abrupt $\pi$ shift in the Fraunhofer pattern of a Josephson junction in a SQUID loop configuration, formed by two parallel Al/InAs Josephson junctions\cite{Dartiailh2021a}. There, the phase shift is attributed to a topological phase transition produced by an external in-plane Zeeman field, but the mechanism for this transition is not further discussed. 

The paper is organized as follows. In Sec.~\ref{sec.TB} we introduce the proximitized BHZ model and the topological phase studied in this work. Then, in Sec.~\ref{sec.JJ} we explore the Fraunhofer pattern of the \emph{thin superconducting junction} where to expect a change in the Fraunhofer pattern periodicity. Finally, in Sec.~\ref{sec.scattering} we introduce an effective scattering model that captures the essential ingredients of the microscopic model and compute the resulting Fraunhofer pattern in the presence of MBSs and accidental zero energy crossings. Further information is given in several appendices.

\section{Microscopic model and transport formalism}\label{sec.TB}

In this section, we introduce the proximitized Bernevig-Hughes-Zhang (BHZ) model \cite{Bernevig2006a} and the transport formalism used to calculate the critical current of the Josephson junction. 

\subsection{Tight-binding Hamiltonian}

The proximitized Bogoliubov de Gennes (BdG) BHZ Hamiltonian is given by $H=(1/2)\int dxdy \Psi^\dagger \mathcal{H} \Psi$ with
\begin{equation}\label{eq:ham01}
\begin{array}{l}
\mathcal{H}=\left(
\begin{array}{cc}
                              \mathcal{H}_\text{e}-E_{F} & \hat{\Delta}(y) \\
                              \hat{\Delta}(y)^* & E_{F}-\mathcal{H}_\text{h} \\                             
                              \end{array}\right),
\end{array}
\end{equation}
with the Fermi energy $E_\text{F}$ and $\mathcal{H}_\text{h}=\mathcal{T}\mathcal{H}_\text{e}\mathcal{T}^{-1}$ is the Hamiltonian for holes, which is obtained from the one for electrons ($\mathcal{H}_\text{e}$) by performing a time-reversal transformation $\mathcal{T}=\it{i} s_y \sigma_0 \mathcal{C}$, with $\mathcal{C}$ being the complex conjugate operator. 
Here, $\hat{\Delta}(y)=\Delta_0(y) \exp[i \varphi(y)] \mathbb{1}_{4\times4}$ 
with the pairing potential $\Delta_0(y)$, which is finite and constant for $|y|>L_n/2$ and 0 otherwise, with $\varphi(y)=\phi/2$ $[\varphi(y)=-\phi/2]$ for $y>L_n/2$ ($y<-L_n/2$), see Fig.~\ref{fig:QSH_ABS}. 
This Hamiltonian is written in the electron-hole basis $\Psi(x,y)=(\psi_e,\psi_h)^t$, with
\begin{align}
&\psi_{e}(x,y)=(c_{E,\uparrow}, c_{H,\uparrow}, c_{E,\downarrow}, c_{H,\downarrow})_{(x,y)}^t,\label{psie}\\
&\psi_{h}(x,y)=(c^ \dagger_{E,\downarrow},c^ \dagger_{H,\downarrow},-c^ \dagger_{E,\uparrow},-c^ \dagger_{H,\uparrow} )_{(x,y)}^t, \label{psih}
\end{align}
where $c_{a,\sigma}^{(\dagger)}$ is the annihilation (creation) operator of an electron with orbital $a=E,~H$ and spin $\uparrow/\downarrow$ at position $(x,y)$.
The Pauli matrices $\sigma_i$ and $s_i$, span the orbital and spin degrees of freedom.

$\mathcal{H}_e=\mathcal{H}_0+\mathcal{H}_R+\mathcal{H}_D+\mathcal{H}_Z$ is the BHZ Hamiltonian, where
\begin{align}
 \mathcal{H}_0=&A(\hat{k}_x  \sigma_x s_z-\hat{k}_y  \sigma_y )+\xi(\hat{k})+M(\hat{k}) \sigma_z,\\
 \mathcal{H}_R=&\alpha\frac{\sigma_0+\sigma_z}{2}(\hat{k}_y  s_x-\hat{k}_x  s_y ),\label{eq.rashba}\\
 \mathcal{H}_D=&\delta_0\sigma_y s_y+ \delta_e\frac{\sigma_0+\sigma_z}{2}(\hat{k}_x  s_x-\hat{k}_y  s_y ) \label{eq.dress}\\
 &+\delta_h\frac{\sigma_0-\sigma_z}{2}(\hat{k}_x  s_x+\hat{k}_y  s_y ),\nonumber\\
 \mathcal{H}_Z=&B^e_\bot s_z(\sigma_0+\sigma_z )/2+B^h_\bot s_z(\sigma_0-\sigma_z )/2\\
 +&B_\parallel s_\theta \sigma_0,\nonumber
\end{align}
with $\xi(\hat{k})=C-D\hat{{\bf k}}^2$, $M(\hat{k})=M-B\hat{{\bf k}}^2$ and 
$A,B,C,D,M,$ are band structure parameters. Here, $2|M|$ gives approximately the energy band gap of the system and its sign determines the topological character of the Hamiltonian: $M>0$ ($M<0$), sets the Hamiltonian in the trivial (topological) insulating regime. Its topological character expresses through the emergence of helical quantum spin Hall edge states.
In the presence of time-reversal symmetry, these quantum spin Hall edge states propagate without backscattering even in the presence of disorder\cite{Bernevig2006a}. 

\begin{figure}[t]  
	\centering
	\includegraphics*[width=3.3in]{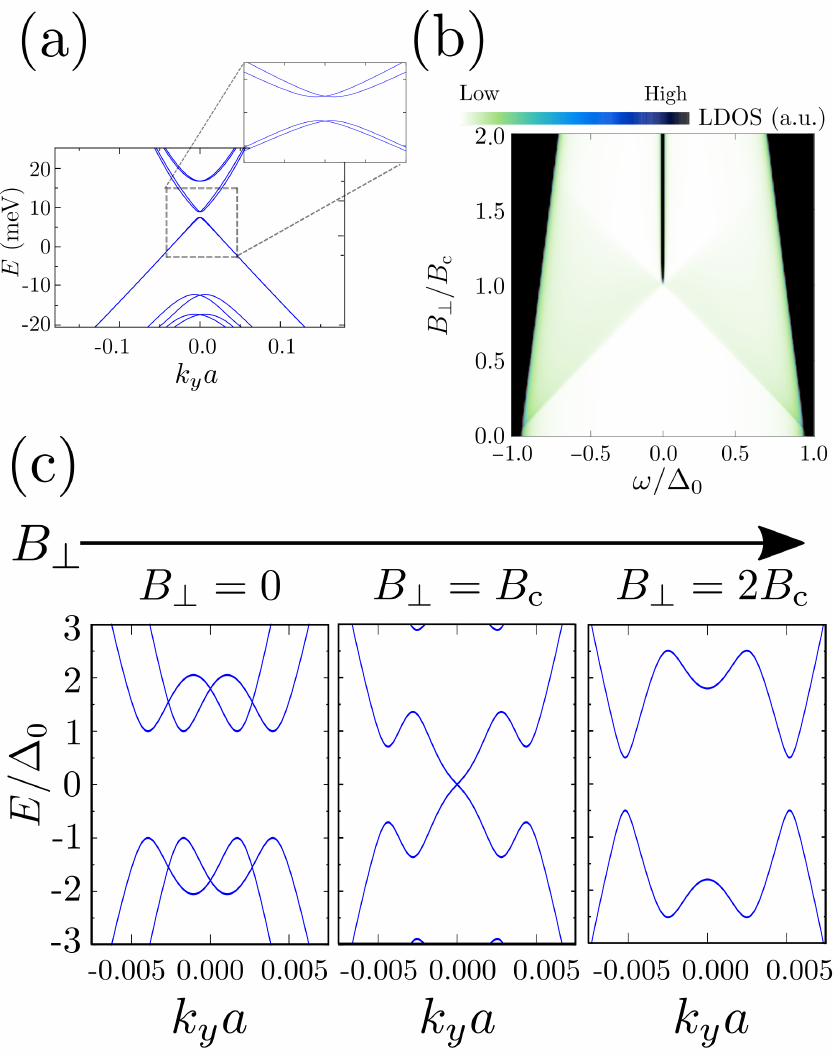}
	\caption{Panel (a), energy dispersion of the (normal) overlapped quantum spin Hall edge states in the thin limit, with ($C_N=0$). Inset: zoom within the gap opened by the overlap between the quantum spin Hall edge states. Panel (b), local density of states as a function of the energy $\omega$ and Zeeman energy $B_\perp$. Panel (c), BdG band spectrum inversion as a function of the Zeeman energy $B_\bot$, similar to the energy dispersion of the Rashba wire in Refs.~\onlinecite{Lutchyn2010a, Oreg2010a}. The width of the sample is W$=144\,$nm, the critical field $B_\text{c}\approx1.3\,$meV and $C_S=-9.2\,$meV.}\label{fig:E_dirac} 
\end{figure}

$\mathcal{H}_R$ is the Rashba spin-orbit coupling with $\alpha$ being the Rashba spin-orbit constant, which can be tuned by an external electric field\cite{Rothe2010}. 
$\mathcal{H}_D$ accounts for the bulk inversion asymmetry (BIA) contribution with $\delta_0, \delta_e,\delta_h,$ are bulk inversion asymmetry parameters, which are specific to the material under consideration. 
The effects of an external magnetic field ${\bf H}$ enter via the Peierls substitution, discussed below, and also via the Zeeman Hamiltonian $\mathcal{H}_Z$, with in-plane ${\bf H}_\parallel$ and out of plane ${\bf H}_\bot$ components. The corresponding Zeeman energies are given by $B_\parallel=g_{\parallel}\mu_B|{\bf H}_\parallel|/2$ and $B^{e/h}_\bot=g_\bot^{e/h}\mu_B|{\bf H}_\bot|/2$, with the parallel $g_\parallel\approx 2$ and perpendicular electron/hole band g-factor $g_\bot^{e/h}\approx 22.7(-1.21)$ and the Bohr magneton $\mu_B$. Further, we have introduced the in-plane spin-operator  $s_\theta=\cos \theta s_x+\sin \theta s_y$ with $\theta$ the in-plane polar angle. 
Guided by HgTe, we use different in-plane and out of plane g-factors\cite{Koenig2008a}. While $B_\parallel^e=B_\parallel^h$, $B_\bot^e\neq B_\bot^h$ because of different g-factors $|g_\bot^{h}/g_\bot^{e}|\ll 1$, and thus, we assume a negligible $g_\bot^{h}\approx 0$. Nevertheless, similar results can be found using $g_\bot^e=g_\bot^h$, as realized in InAs/GaSb wells\cite{Knez2011a}. 

Following standard finite difference methods, we  discretize the Hamiltonian on a 2D lattice, turning the continuum momenta $\hat{k}_{x/y} \Psi(x,y)=- {\it i} \partial_{x/y} \Psi(x,y)$, into  their discretized version, that is, $\partial_{x/y}\Psi_(x,y)\rightarrow \frac{1}{2a_{x/y}} (\Psi_{{\bf i_{x/y}}+\bf a_{x/y}}-\Psi_{{\bf i_{x/y}}-\bf a_{x/y}})$ and $\partial^2_{x/y} \Psi(x,y))\rightarrow \frac{1}{a^2_{x/y}} (\Psi_{{\bf i_{x/y}}+\bf a_{x/y}}-2\Psi_{{\bf i_{x/y}}}+\Psi_{{\bf i_{x/y}}-\bf a_{x/y}})$.
In the normal part of the junction $|y|<L_n/2$, we add a perpendicular magnetic field, whose orbital effects enter via the Peierls substitution $\hat{k}_{x,y}\rightarrow \hat{k}_{x,y}-(e/\hbar) A_{x,y}$, with the Landau gauge ${\bf A}=|{\bf H}_\bot|x \bm{e}_y $. Thus, $\psi_{e,\bf i_{y}}^\dagger\psi_{e,\bf i_{y}\pm \bf a_y}\rightarrow \exp [\pm{\it i}(\pi \Phi_p/\Phi_0)  i_{x}]\psi_{e,\bf i_{y}}^\dagger\psi_{e,\bf i_{y}\pm \bf a_y}$, with the flux quanta $\Phi_0=h/2e$ and the magnetic flux in a plaquette $\Phi_p= \mathcal{A}|{\bf H}_\bot|$ for $|y|<L_n/2$ and 0 otherwise, and with $\mathcal{A}=a_x a_y$ being the area of a single plaquet. Besides, we assume a negligible Zeeman contribution generated by the magnetic field responsible for the magnetic flux.

\subsection{Topological superconductivity}

There are different ways to induce Majorana bound states on the proximitized quantum spin Hall edges. The most direct one, a proximitized helical edge state \cite{Fu2008a}, is not relevant here because without conserving the parity, its corresponding Fraunhofer pattern gives no difference in the trivial and topological regimes. Here, we explore a different configuration closer to the Rashba quasi 1D well, where the proximitized BHZ model supports a chiral Majorana edge mode for a Zeeman field above the critical field $B_\bot>B_\text{c}$\cite{Weithofer2013a, Novik2020a} in the NS junction. In addition, in the quasi 1D limit a pair of proximitized quantum spin Hall edge channels can exhibit a superconducting topological phase when the width of the sample is smaller than the localization length of the quantum spin Hall edge states $\xi_\text{qsh}$, that is,  $\text{W}\lesssim \xi_\text{qsh}$~\cite{Fleckenstein2021a}.
To understand this scenario, we first represent the non-proximitized spectrum of a thin sample in Fig.~\ref{fig:E_dirac}(a), which develops a gap at the Dirac point. 
In the vicinity of the gap, the resulting spectrum exhibits the same structure as the Rashba wire, that is, parabolic bands shifted in $k$ by the spin-orbit coupling $\Delta k=\pm k_{so}$, see magnification in Fig.~\ref{fig:E_dirac}(a). Hence, it is not surprising to expect an inversion of the proximitized superconducting band produced by a Zeeman field applied perpendicularly to the effective spin-orbit field direction, see bottom panels in Fig.~\ref{fig:E_dirac}(c). As a result, two Majorana bound states appear at the extremes of the superconductor, which becomes visible imposing hard-wall boundary conditions. See LDOS at the NS interface in Fig.\ref{fig:E_dirac}(b)~\cite{Fleckenstein2021a}. 
It is important to remark that even though we need a finite overlap between opposite quantum spin Hall edge states on the superconducting part, the quantum spin Hall edge states placed on the normal part propagate without scattering to the opposite edge. This is possible by setting $C$ in the normal region, i.e.~$C_N$, in such a way that the Dirac point is away from the Fermi energy. 
Alternatively, one can use the same chemical potential as in the superconducting region, i.e.~$C_N = C_S$, with a different width $\text{W}_N> \text{W}_S$ on the normal part and superconducting parts. Although we use the former configuration $(C_N \neq C_S)$, the latter $(\text{W}_N> \text{W}_S)$ might be easier to accomplish experimentally since it does not involve a spatially dependent gate voltage control.

Experimentally, there are already two scenarios where non-proximitized quantum spin Hall modes of opposite edges overlap.
Firstly, the addition of an electric field can push the edge modes towards the bulk \cite{Strunz2020}. Secondly, materials like bismuthene on SiC develop topological line defects through which the quantum spin Hall edges propagating on opposite sides can overlap \cite{Stuehler2022a}.

\subsubsection*{Set of parameters}

Unless it is specified otherwise, we use the following set of parameters: $A=373\,$meV nm, $B=-857\,$meV nm$^2$, $D=-682\,$meV nm$^2$, $M=-10.0\,$meV. The parameters associated to spin-orbit coupling are $\alpha=0.0$, $\delta_0=1.6\,$meV, $\delta_e=-12.8\,$meV nm, $\delta_h=21.1\,$meV nm, $T=0.1\,$K and $\Delta_0=0.15\,$meV. In the tight-binding calculations, we use a discretization constant $a \equiv a_x=a_y=2.4\,$nm. We set the Fermi energy $E_F=0$ and the chemical potential on the normal and superconducting parts is tuned by $C_N$ and $C_S$, respectively.

In this work, we use the magnetic field for two different purposes: a Zeeman field to invert the superconducting gap and the magnetic flux to study the Fraunhofer pattern. Even though both fields are applied in the $z$-direction, it is possible to study the critical current as a function of both contributions independently because of the different order of magnitude of each contribution. Here, the Zeeman field is of the order of $B_\bot\sim\, 1$T, while the orbital field is a few mT.

\section{Fraunhofer pattern}\label{sec.JJ} 

\begin{figure}[t]
\centering
\includegraphics*[width=3.3in]{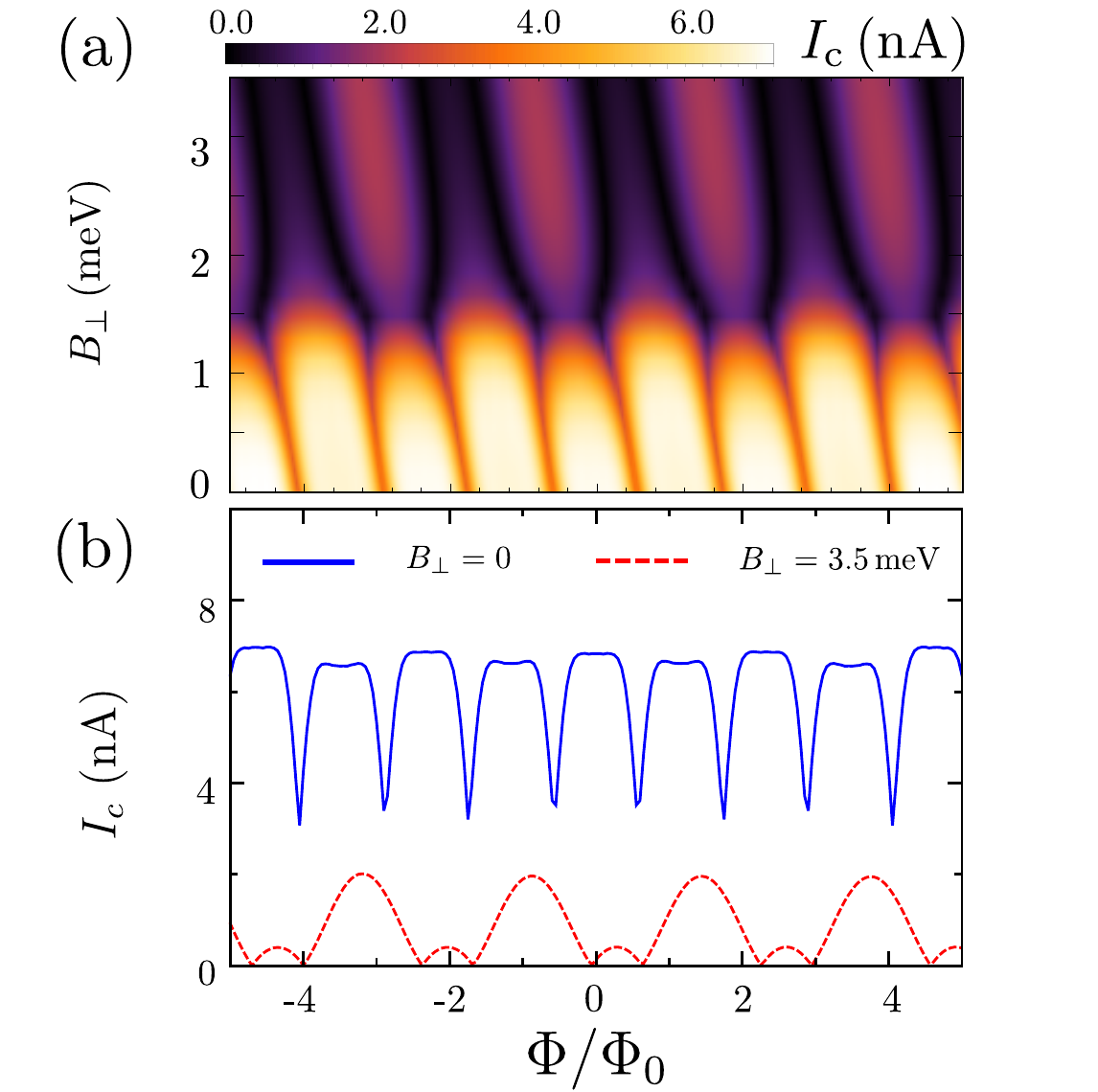}
\caption{Fraunhofer pattern for the thin superconducting junction: Panel (a), critical current as a function of the flux and perpendicular Zeeman field. The parameters of the model are $M_{S/N}=- 10\,$meV, W$=0.14\,\mu$m, $L_n=0.74\,\mu$m, $C_S=-9.25\,$meV,  $C_N=2.0\,$meV, $\Delta_0=0.15\,$meV. Panel (b), Fraunhofer pattern for two different line cuts $B_\bot=0.0$ (trivial regime) and $B_\bot=3.5\,$meV (topological regime).}\label{fig:Ic_inverted}
\end{figure}

The Fraunhofer pattern, i.e.~the critical current as a function of the magnetic flux,  contains important information about the spatial distribution of the supercurrent in a Josephson junction \cite{Dynes1971a}.
For example, when the normal part of the junction is dominated by bulk modes, the critical current oscillates and decays for an increasing magnetic flux. Whereas, when the supercurrent is carried along the edges (quantum spin Hall or quantum Hall states), the critical current oscillates with a period ($\Phi/\Phi_0$ or $2\Phi/\Phi_0$) without decaying. 
Using standard equilibrium Green's function methods introduced in App.~\ref{criticalGF}, we calculate the critical current of a Josephson junction with quantum spin Hall edges on the normal part in the absence and presence of Majorana bound states at the NS interfaces.

In the 2D limit, where chiral Majorana modes propagate at the boundary of the superconducting slabs, the Fraunhofer pattern does not exhibit a qualitative change when comparing the trivial and topological regimes, see Fig.~\ref{fig:Ic2D} in App.~\ref{app.wide}. 
This is caused by the dominant coupling between normal and proximitized quantum spin Hall edge states, which gives rise to a supercurrent contribution based on LAR, yielding negligible traces of the presence of the chiral Majorana contribution. Therefore, in order to observe a specific signature of the MBS in the Fraunhofer pattern profile, we need to reduce the supercurrent carried by local ABS so that we can enhance the coupling between the quantum spin Hall edge states and the central part of the superconducting slab, where the MBSs are located. To this aim, we induce MBSs directly on the proximitized quantum spin Hall edge states by considering a \emph{thin superconductor junction}. Alternatively, there is another way to increase the coupling between the helical modes and the MBSs, that involves a positive mass $M>0$ in the superconducting leads, having no helical edge states in the normal state. Since the results are analogous to the \emph{thin superconducting junction} we place this setup and its results to App.~\ref{app.massdomain}. In addition, we show there that it is also possible to observe the transition with an in-plane magnetic field $B_\parallel$.

In the trivial regime ($B_\bot<B_\text{c}$), the Fraunhofer pattern exhibits the usual SQUID pattern with period $p\sim \Phi_0$, see Figs.~\ref{fig:Ic_inverted} (a) and (b). In contrast to the wide superconductor junction, see App.~\ref{app.wide}, we observe a reduction of one lobe every $\Phi\approx 2 \Phi_0$. This beating pattern, also known as \emph{even-odd flux quanta effect}, occurs because the quantum spin Hall edges placed on opposite sides become partially coupled at the superconductor interface~\cite{Baxevanis2015a}, resulting into normal electron-electron and hole-hole reflections involving both edges. In this situation, particles can encircle the normal part several times before an electron-hole scattering event takes place, accumulating extra magnetic phase factors, and modifying the Fraunhofer pattern. 
Increasing the Zeeman field above the critical field $(B_\text{c}\sim1.5\,$meV), we observe the suppression of the originally ``odd'' lobes placed around $\Phi\approx (2n-1) \Phi_0$ (with $n$ being an integer), yielding a Fraunhofer pattern with a periodicity of $p\approx2\Phi_0$, see linecuts in Fig.~\ref{fig:Ic_inverted}(b). Numerical calculations show a similar change in periodicity for different set of parameters, such as, $C_S$, $\Delta_0$ as long as $B_\bot> B_\text{c}$, and thus, becoming a robust signature of the topological transition. Furthermore, we can observe a constant shift of the Fraunhofer pattern as a function of the Zeeman energy $B_\bot$, which can be gauged into the Peierls phase since the Zeeman field is parallel to the helicity operator of the quantum spin-Hall edges, see more details in App.~\ref{app.wide}. 

\begin{figure}[t]
\centering
\includegraphics*[width=3.4in]{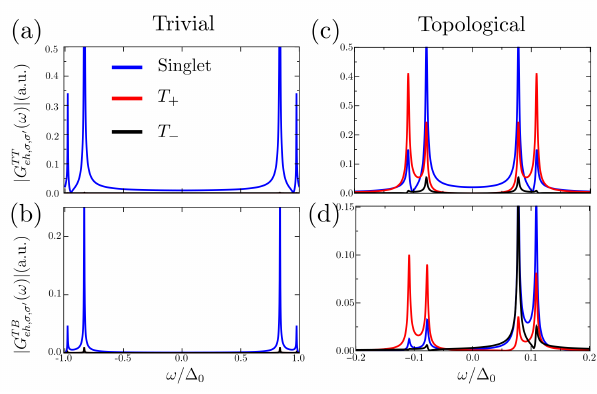} 
\caption{Intra and inter-edge GFs as a function of the frequency $\omega/\Delta$  for the trivial (a, b) and topological (c, d) regimes. The set of parameters is the same as in Fig.~\ref{fig:Ic_inverted}, with $\phi=0$ and (a) $B_\perp=0$, $\Phi/\Phi_0=0$ and (b) $B_\perp=3.5\,$meV, $\Phi/\Phi_0=-0.85$ which lead to the maximum critical current for each Zeeman fields.
}\label{fig:GFs}
\end{figure}

The abrupt change of the Fraunhofer periodicity can be understood from the new scattering channels introduced by the presence of MBSs, which enhance triplet electron-hole reflections with parallel spin, i.e.~$r_{ss}^{he}$. 
Due to spin-momentum locking, the ABSs involving $r_{ss}^{he}$ take place on opposite edges, while $r_{s\bar{s}}^{he}$ on the same edge. Due to this geometrical difference, the magnetic flux accumulated in each process is different. Crossed ABSs do not accumulate a magnetic flux, while local ABS accumulate the flux $\Phi$.
In the tunneling limit \cite{Haidekker2020a}, the critical current is given to lowest order in the particle tunneling by
\begin{align}
I_\text{c}\approx |I_{CAR}+I_{LAR}\cos(\pi\Phi/\Phi_0)| ,
\label{eq:analytics}
\end{align}
with $I_{CAR/LAR}$ the respective crossed and local critical currents. Using this simple equation, we can extract that the presence of MBSs at the NS boundaries generates a local and crossed supercurrents with approximately the same strength, which is in accordance with the electron-hole reflection coefficients of MBS \cite{Haim2015a} and the suppression of CAR in the trivial regime\cite{Adroguer2010a,Virtanen2013a,Reinthaler2013a}. 

\begin{figure}[t]
\centering
\includegraphics*[width=3.in]{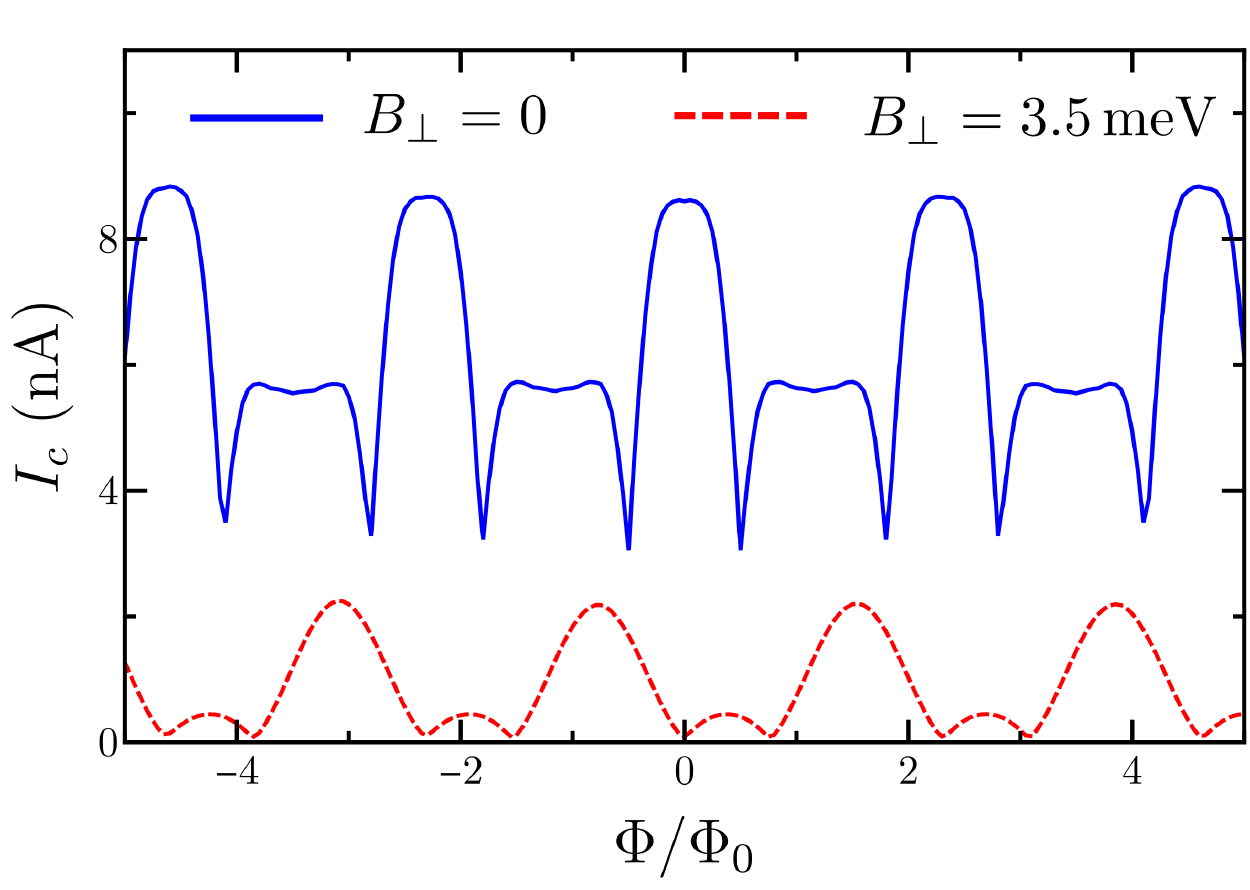} 
\caption{Fraunhofer pattern for the thin superconducting junction with a  disorder trench of $L_\text{trench}=150\,$nm and a disorder strength $\sigma=\pm 5\,$meV. The rest of the parameters are the same as in Fig.~\ref{fig:Ic_inverted}.}\label{fig:Icdisorder}
\end{figure}

{\it Intra and inter-edge GFs---} To support this interpretation, we compute the intra and inter-edge GFs defined respectively as,
\begin{align}
&G^{TT}_{eh,\sigma,\sigma'}(\omega)=\frac{2}{\text{W}}\int^{\text{W}}_{\text{W}/2} dx G_{eh,\sigma,\sigma'}(x,x,\omega) \\
&G^{TB}_{eh,\sigma,\sigma'}(\omega)=\frac{4}{\text{W}^2}\int^{\text{W}}_{\text{W}/2} dx \int^{\text{W}/2}_{0} dx' G_{eh,\sigma,\sigma'}(x,x',\omega) 
\end{align}
as a function of the frequency $\omega$, see Fig.~\ref{fig:GFs}. Here, the subscripts $T,B$ refer to top and bottom edges, respectively.
Note that these GFs are the result of tracing the orbital degrees of freedom and they are defined at the center of the junction $y=0$. Besides, we have used a superconducting phase difference $\phi=0$ and the magnetic flux $\Phi$ that gives rise to the maximal critical current for the given Zeeman field. We can observe that in the trivial regime ($B_\bot<B_\text{c}$), intra and inter-edge GFs contain only the singlet component $|S \rangle \sim |\uparrow_e \downarrow_h\rangle -|\downarrow_e \uparrow_h\rangle$, see panels~a and~c. 
In contrast, in the topological regime ($B_\bot>B_\text{c}$), we can observe that the triplet components are of the same order as the singlet one, see panels b and d.

\begin{figure}[t]
\centering
\includegraphics*[width=3.in]{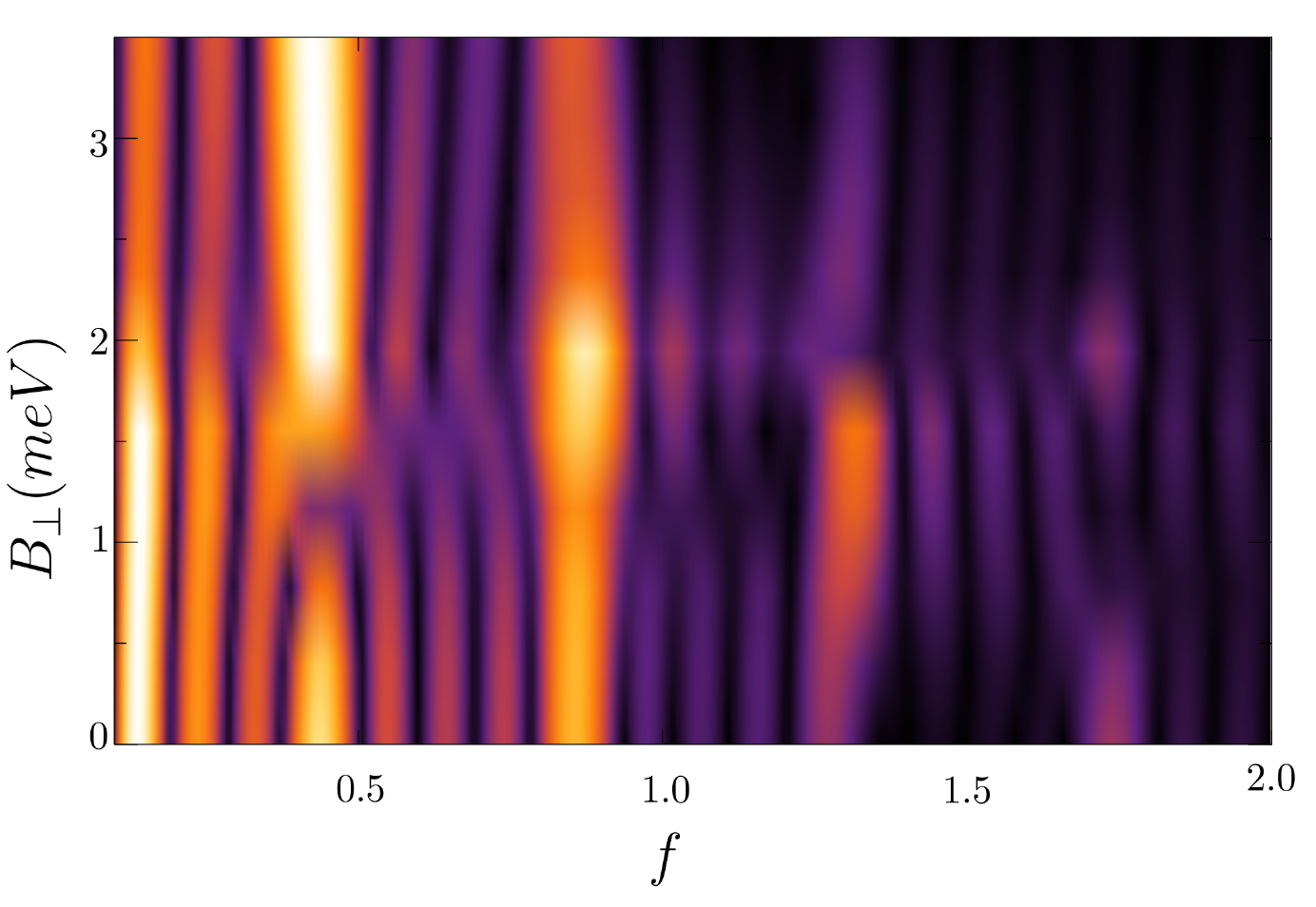} 
\caption{Absolute value of the Fourier transform of the critical current in the presence of disorder as shown in Fig.~\ref{fig:Ic_inverted} as a function of the characteristic frequency $f$ and the Zeeman field $B_\perp$.}\label{fig:Fourier}
\end{figure}

{\it Robustness against disorder---} Due to the presence of the magnetic field, time reversal symmetry is broken, and thus, the helical edge states are no longer protected against backscattering. Therefore, we check the robustness of the signature in the presence of disorder. To this aim, we set two stripes of length $L_\text{trench}$ and width $\text{W}$ at both NS interfaces, see Fig.~\ref{fig:setup}. This is consistent with an actual experiment, where the superconducting leads can introduce some disorder to the semiconductor system. 
In this way, we consider an on-site disorder potential that is randomly assigned in the range of $\pm\sigma$. 
We first note that for values of disorder strength $\sigma= 1\,$meV, the Fraunhofer pattern exhibits the same change of periodicity as the clean system, even when $L_\text{trench}$ extends over the whole setup ($\sim 0.74\,\mu$m), not shown. For larger disorder strenghts, $\sigma=5\,$meV, we observe no change in the Fraunhofer pattern relative to the clean system for $L_\text{trench}\lesssim 100\,$nm (not shown). However, increasing further $L_\text{trench}$ to 150$\,$nm the height of the odd lobes decays in the trivial regime, see two linecuts in Fig.~\ref{fig:Icdisorder}.
Although in this case the periodicity 
is strictly not doubled as a function of the Zeeman field, it is still possible to appreciate the impact of the Zeeman field on the Fraunhofer profile by performing the Fourier transform on the critical current, that is,
\begin{align}
I_\text{c}(f)=\int d\Phi e^{2\pi i f \Phi/\Phi_0} I_\text{c}(\Phi/\Phi_0).
\end{align}
In Fig.~\ref{fig:Fourier}, we represent the Fourier transform $I_\text{c}(f)$ as a function of the frequency $f$ and the Zeeman energy $B_\bot$. We can observe that in the trivial regime ($B_\bot< B_\text{c}$), $I_\text{c}(f)$ exhibits three maxima around $f\approx 0,\, 0.5$ and $1.0$, and smaller and narrower contributions for $0<f<1$. Here, the contribution around $f\approx 0$ is related to the constant background observed in Fig.~\ref{fig:Ic_inverted}, and the contributions with frequencies at $f<0.5$, correspond to the small oscillations observed on top of the odd lobes. Moreover, the contribution close to $f\approx 0.5$ is related to the \emph{even-odd flux quanta effect} mentioned above, whereas $f\approx 1$ is the characteristic frequency of a wide quantum spin-Hall Josephson junction.\footnote{Note that deviations of integer or semi-integer values of $f$ can be attributed to finite size effects.} For $B_\bot> B_\text{c}$, we can observe an abrupt change in $I_\text{c}(f)$, with a dominant component at $f\approx0.5$ resulting from the reduction of the odd lobes, which serves as an indication of the presence of MBSs as a function of the Zeeman field.

\section{Scattering model}\label{sec.scattering}

\begin{widetext}
 
\begin{figure}[t]
\centering 
\includegraphics*[width=3.9in]{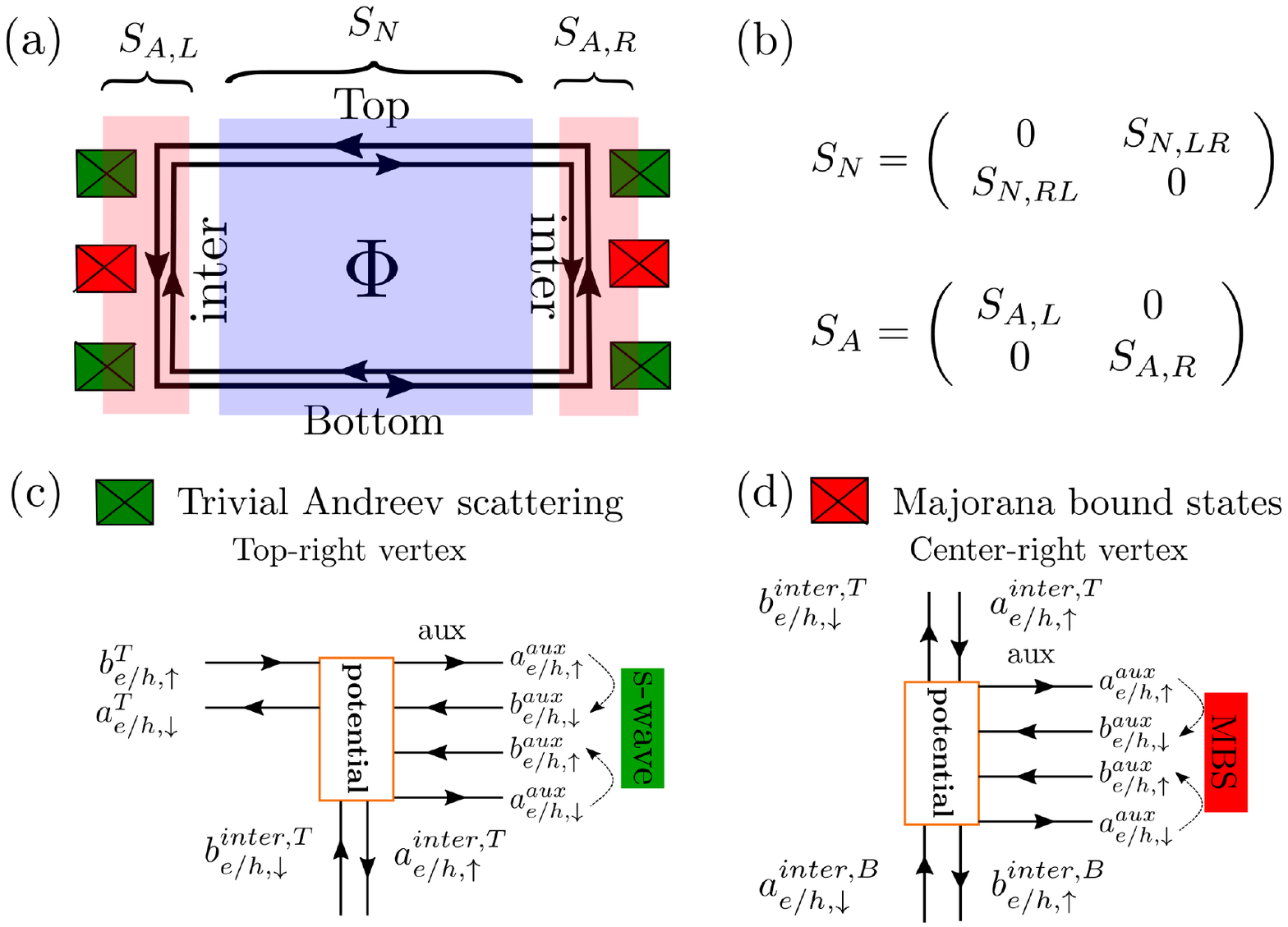}
\caption{Panel (a):
Josephson junction setup with helical modes confined in the normal region threaded by a magnetic flux $\Phi$. We highlight the normal and superconducting regions, where the $S$ matrices $S_N$ and $S_{A,L/R}$ act by light blue and red, respectively.
On the superconducting sides, the crossed boxes represent trivial (green) and topological (red) superconducting scattering centers, where the helical edge states perform electron-hole reflections. 
In panel (b), we show the structure of $S_N$ and $S_A$, which connect the incoming and outgoing $(a/b)$ scattering states on the left and right sides $(L/R)$, by means of $S_A (b_L,b_R)^t=(a_L,a_R)^t$ and $(b_L,b_R)^t=S_N(a_L,a_R)^t$. 
In panels~(c) and~(d), we detail the top-right (trivial) and center-right (topological) scattering processes. 
}\label{fig:Scattsetup}
\end{figure}

\end{widetext}

In order to have a deeper insight into the periodicity changes introduced in the Fraunhofer pattern by the presence of MBSs, we construct a phenomenological scattering model including the main ingredients participating in the setup, that is, helical modes propagating around the normal part in the presence of a magnetic flux, and two superconducting slabs accounting for scattering events with a bulk superconductor and MBS, see Fig.~\ref{fig:Scattsetup}. Furthermore, this model also allows us to consider accidental zero energy modes instead of MBSs and to study the resulting Fraunhofer pattern. 

The supercurrent at temperature $T$ can be written in terms of the $S$ matrices $S_A$ and $S_N$, namely, \cite{Beenakker1991a,Brouwer1997,Baxevanis2015a} 
\begin{align}\label{eq.supercurrent}
I=-kT \frac{2e}{\hbar}\frac{d}{d \phi} \sum_{n=0}^\infty \text{ln}~\text{det}[1-S_N(i\omega_n) S_A(i\omega_n)],
\end{align}
where the sum runs over the Matsubara frequencies $\omega_n=\pi k T (2n+1)$. Here, $S_N$ is the $S$ matrix that accounts for the propagation along the normal part, while $S_A$ accounts for the scattering events taking place at the left and right NS interfaces. 

\begin{figure}[tb]
\centering
\includegraphics*[width=3.1in]{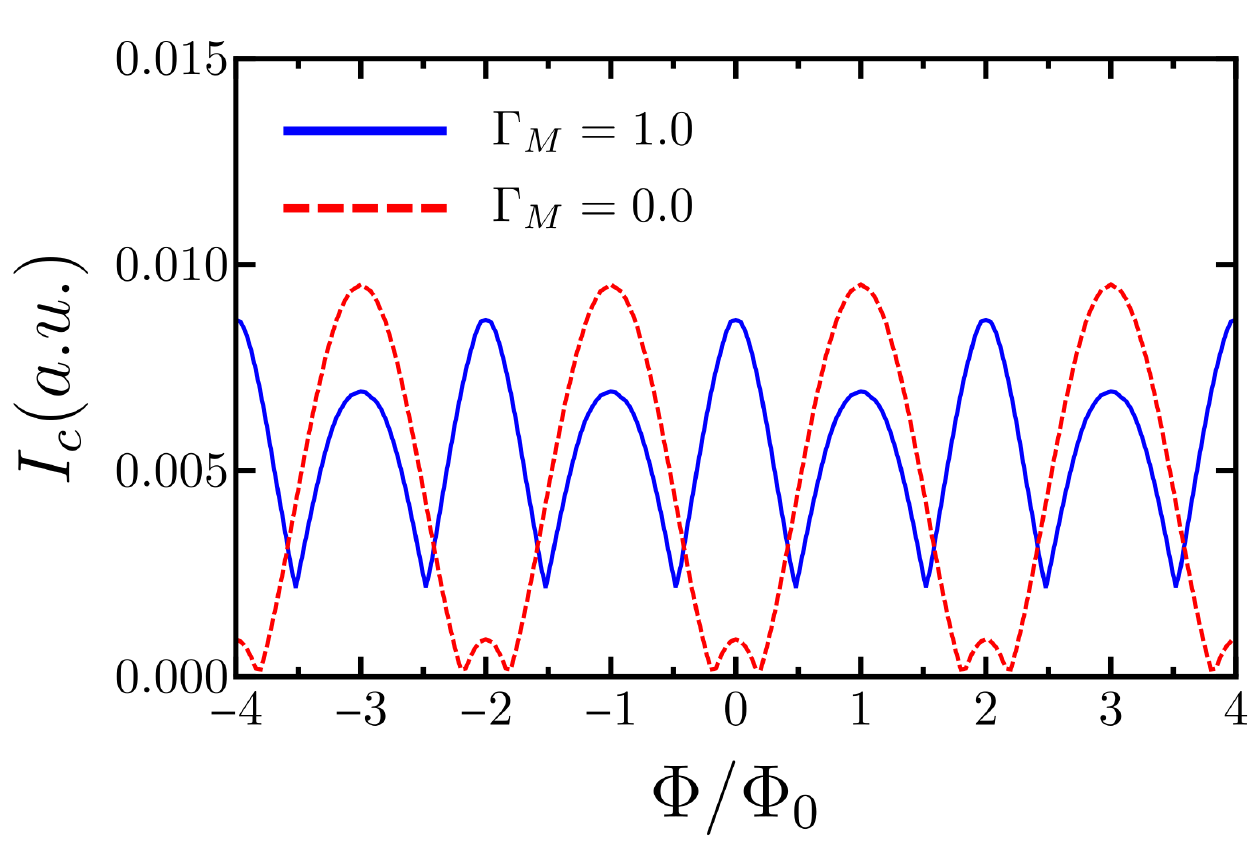} 
\caption{
Fraunhofer pattern obtained using the scattering model with MBSs scatterers for a deflection probability $\Gamma_d=0.2$ for two values of $\Gamma_M$: $1-\Gamma_M=0$ (solid blue) and $1-\Gamma_M=1$ (dashed red). 
We have used $t_{\uparrow,L}=t_{\downarrow,L}=2.5 \hbar v_F/\sqrt{L_n}$, $t_{\uparrow,R}=-t_{\downarrow,R}=2.5\hbar v_F/\sqrt{L_n}$, being $L_n$ the length of the normal part.
}\label{fig:Scatnontrivial}
\end{figure}

To model the processes occurring at each NS interface, we combine three $S$ matrices represented by crossed boxes in Fig.~\ref{fig:Scattsetup}, see more details in App.~\ref{app.scattering}. Two of them (green crossed boxes) capture the physics of the scattering events between the quantum spin Hall edges and the bulk superconductor, where we also account for finite electron-electron deflection probability ($\Gamma_d$) that connects quantum spin Hall edge states placed on opposite sides. Additionally, the scattering events produced by MBSs are modelled by the addition of a third scatterer (red crossed boxes) centered in between the bulk superconducting scatterers. Similarly as in the superconducting bulk scatterers, the intensity of the scattering processes with the MBSs scatterers is controlled by the parameter $\Gamma_M$: for $1-\Gamma_M=0$ particles propagate along the NS interface without scattering with the MBS, while for $1-\Gamma_M=1$ any particle propagating along the NS junction scatters with the MBS. 

As a first step, we recover a similar Fraunhofer pattern as the ones observed using the microscopic model for the trivial regime i.e.~$B_\bot=0$, see blue linecuts in Figs.~\ref{fig:Ic_inverted} and Fig.~\ref{fig:Scatnontrivial}. To do so, we set $\Gamma_M=1.0$, which sets a zero coupling to the MBSs and a finite $\Gamma_d>0$, which introduces finite deflection probability, connecting opposite edges with $\Gamma_d=0.2$. Here, the resulting Fraunhofer pattern exhibits the previously observed \emph{even-odd flux quanta effect} by the reduction of one of the lobe maxima every 2$\Phi_0$, see blue curves in Fig.~\ref{fig:Scatnontrivial}. Note that in contrast to the microscopic model, here the reduced lobes are always placed at odd multiples of $\Phi_0$. In general, the parity of the reduced lobes depends on the phase factors picked up during the deflection scattering processes \cite{Baxevanis2015a}. For simplicity, we have set all the phases picked up during the deflection to zero, and therefore, we cannot change the parity of the reduced lobes.

\subsection{Majorana bound states}  

Setting a finite scattering probability with the MBSs ($1-\Gamma_M>0$), introduces a finite electron-hole process with parallel spin \cite{He2014a}. Using the Mahaux-Weidenm\"uller formula we can obtain the reflection amplitudes of an isolated MBS, namely\cite{Fisher1981a,Iida1990a, Haim2015a}
\begin{align}
r^{ee}_{ss'} =\delta_{ss'}+\frac{2\pi \nu_0 t^*_{s}t_{s'}}{iE-\gamma},~~~~r^{he}_{ss'} =\frac{2\pi \nu_0 t_{s}t_{s'}}{iE-\gamma},
\label{eq.weidenm}
\end{align}
where $\gamma=2\pi \nu_0(|t_\uparrow|^2+|t_\downarrow|^2)$ and $t_\sigma$ is the effective coupling amplitude between the helical edge states and the Majorana bound states, see more details in App.~\ref{app.scattering}. As we have discussed above, these scattering processes have a strong impact on the Fraunhofer pattern because the resulting crossed ABS do not accumulate a net magnetic flux. 

Guided by the microscopic model, we assume $t_\sigma$ to be real\cite{Sticlet2012a,Prada2017a}. Concretely, we analyze the tunnel amplitude configuration:  $t_{\uparrow,L}=t_{\downarrow,L}$, $t_{\uparrow,R}=-t_{\downarrow,R}$, which reproduces approximately the Fraunhofer patterns observed using the microscopic model, see red dashed curves in Fig.~\ref{fig:Scatnontrivial}. 
Moreover, the form of the Fraunhofer pattern can only be recovered taking into account that the MBSs appear on both sides, thus, the change in the Fraunhofer profile also provides a non-local probe of a pair of separated  MBSs appearing at the left and right interfaces.

\begin{figure}[tb]
\centering
\includegraphics*[width=3.3in]{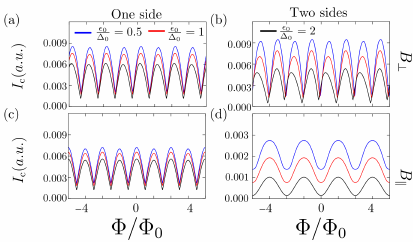}
\caption{
Fraunhofer pattern obtained using the scattering model with ZEABS scatterers for $B_{\parallel / \bot}=\sqrt{\Delta_0^2+\epsilon_0^2}$ and the same deflection probabilities as in Fig.~\ref{fig:Scatnontrivial}, i.e.~$\Gamma_d=0.2$, $\Gamma_M=0.0$. We set two different magnetic field directions $B_\bot$ (a and b) and $B_\parallel$ (c and d) and use either one (a and c) or two (b and d) ZEABS. We use in all panels three different values of the ratio $\epsilon_0/\Delta_0$, which controls indirectly the coupling between the quantum spin Hall edge states and the ZEABS, see more details in App.~\ref{app.scattering}.
}\label{figZEABS_Ic}
\end{figure}

\subsection{Zero energy Andreev bound states}

Accidental zero energy Andreev bound states (ZEABS) can arise in the presence of a Zeeman field when for example the NS interfaces of the proximitized region are less doped than the interior regions\cite{Fleckenstein2018a}. These states can give rise to similar signatures in the conductance as the MBSs, like a zero bias conductance peak. Thus, to test the robustness of the signatures observed in the Fraunhofer pattern, we replace the MBS $S$ matrix, by one resulting from the scattering with an accidental zero energy ABS (ZEABS) and study under which conditions we observe similar changes in the Fraunhofer pattern as the ones observed with the MBSs.

We model the ZEABS by means of the Hamiltonian\cite{Haim2015a} 
\begin{align}
H_\text{ZEABS}=\sum_{s=\uparrow,\downarrow} \epsilon_0 d_s^\dagger d_s +B_\xi d^\dagger_s d_{s'} (s_\xi)_{s,s'}+\Delta_0 d^\dagger_\uparrow d^\dagger_\downarrow+\text{h.c.},
\end{align}
where $\epsilon_0$ sets the doping of these regions and it is assumed to be smaller than the chemical potential of the bulk proximitized region. $\Delta_0$ is the pairing amplitude of this region and in general it can be different from the bulk $\Delta_0$. Furthermore, $B_{\xi}$ is the Zeeman energy of a field applied in the $\xi$ direction. 
This Hamiltonian exhibits zero energy states for $B_\xi =\sqrt{\Delta_0^2+\epsilon_0^2}$ independently of the Zeeman field direction. 
This region is coupled to the quantum spin Hall edge states by means of the tunnel Hamiltonian $H_T= \sum_s \omega_s \psi^\dagger_s(x_0) d_s+\text{h.c.}$, with $\psi(x)$ being the electronic field operator of the quantum spin Hall edges at position $x=x_0$, where the ZEABS is placed.
Moreover, the ratio $\epsilon_0/\Delta_0$ modifies effectively the tunnel couplings $\omega_{\uparrow/\downarrow}=3\hbar v_F/\sqrt{L_n}$ between the quantum spin Hall edges and the ZEABS. The reason for this reduction is that the larger $\epsilon_0$, the higher $B_\alpha$ we need to apply and the more spin-polarized the ZEABS become. See further details of the model in App.~\ref{app.scattering}.

As we have seen above, the presence of the accidental ZEABS depends on the specific conditions of the NS interfaces, like a spatial dependence of the chemical potential or pairing amplitude. Thus, it is possible that only one of the NS interfaces may develop a ZEABS. For this reason, we analyze two different scenarios depending on the number of ZEABS present in the setup, that is, whether only one or two NS interfaces develop a ZEABS \footnote{Naively, one could argue that these trivial states are present at zero energy only for some specific values of the Zeeman field, whereas the topological states should stick at zero energy, provided that the system is long enough. However, in an interacting scenario, the trivial states can be pinned at zero energy for a finite range of Zeeman fields because, in contrast with Majorana modes, these trivial states are charged and can interact with the electrostatic environment, changing the compressibility of the system once they are close to the Fermi level.}
\cite{Dominguez2017a, Escribano2018a}. Besides, we consider two different orientations of the Zeeman field: in-plane and an out of plane. 
In the latter case, we expect to observe no change in the Fraunhofer pattern because electron-hole reflections with parallel spin are zero\cite{Haim2015a}. In contrast, the ZEABS generated by an in-plane Zeeman field (independently of the direction) exhibit a finite electron-hole reflection coefficient with parallel spin and, as we will see, it can lead to similar Fraunhofer patterns as those shown by the MBSs. 

In Fig.~\ref{figZEABS_Ic}, we show a representative example of the Fraunhofer pattern generated with the scattering model using ZEABS. The main idea consists on checking whether it is possible to recover the Fraunhofer pattern observed in Fig.~\ref{fig:Scatnontrivial}, that is, the suppression of the half of the lobes by switching on the coupling to the MBSs, that is, for $\Gamma_M=1\rightarrow \Gamma_M=0$. 
To this aim, we use $\Gamma_d=0.2$, and analyze all four configurations of Zeeman fields and number of ZEABS ($\Gamma_M=0$). 
In panels (a) and (b) we show the results for the out of plane Zeeman field with one and two ZEABS, respectively, while in panels (c) and (d) we use an in-plane Zeeman field and again one and two ZEABS. 
In all panels, we represent the Fraunhofer pattern for three values of $\epsilon_0/\Delta_0=0.5,1,2$, which effectively change the coupling between the quantum spin Hall edges and the ZEABS. 
We observe that when only one ZEABS is present and independently of the Zeeman orientation, the relative height of the Fraunhofer pattern lobes does not change significantly with respect to the uncoupled case ($p\approx \Phi_0$). 
In addition, when ZEABS are present on both sides, the relative height of the lobes  remains unchanged relative to the trivial case $p\approx \Phi_0$ only for an out of plane Zeeman field $B_\bot$, see panel~(b). However, for an in-plane field [panel (d)] the Fraunhofer exhibits a periodicity of $p= 2\Phi_0$. Note that this in stark contrast to the case of MBSs, where the scattering matrix constructed from Eq.~\eqref{eq.weidenm} is independent  on the direction of the Zeeman field (as long as it is perpendicular to the spin-orbit field).

\section{Conclusions}

In conclusion, using tight-binding calculations based on the BHZ model, we predict that the presence of MBSs on both NS interfaces leads to a drastic change in the Fraunhofer pattern periodicity with respect to the trivial regime, where the half of the lobes become suppressed. We explain the change in periodicity in terms of the introduction of triplet electron-hole reflection amplitudes $r^{he}_{ss}$ enforced by the exotic Majorana bound states, which, when coupled to the spin-momentum locked quantum spin Hall edge states, leads to the coexistence of two different kinds of Andreev bound states: local and crossed, which are mediated by antiparallel and parallel spin electron-hole reflection amplitudes, respectively. Furthermore, this signature remains visible even in the presence of disorder strengths of $5\,$meV as long as the disorder trench is shorter than $L_{trench}\lesssim 100\,$nm. For larger disorder trenches $L_{trench}\geq  150\,$nm, the Fraunhofer pattern exhibits a larger even-odd effect in the trivial regime, making it more difficult to observe a change in periodicity as a function of the Zeeman energy. Nevertheless, making use of the Fourier transform, we observe an abrupt increment of the fractional frequencies for $B_\bot > B_\text{c}$.

Finally, we have developed a phenomenological scattering model with which we recover similar results as those observed in the microscopic model with MBS, supporting our physical picture of local and crossed Andreev reflections. Furthermore, we have used this model to explore the robustness of the detection scheme when accidental zero energy Andreev bound states (ZEABS) are present. We have found that when only one NS interface holds a ZEABS, the Fraunhofer pattern does not change its profile independently of the Zeeman orientation. Also when both NS interfaces hold a ZEABS the relative height of the Fraunhofer lobes remains unchanged for a perpendicular Zeeman field $B_\bot$. However, when both NS interfaces exhibit ZEABS generated by an in-plane Zeeman field $B_\parallel$, the resulting Fraunhofer pattern exhibits similarities to the topologically non-trivial case.

\appendix

\section{Supercurrent in a Josephson junction}\label{criticalGF}

The critical current $I_\text{c}$ is the maximum amount of current that the Josephson junction supports without generating any voltage. Given the geometry of a Josephson junction, this quantity characterizes its conducting properties and results from the self-adjustment of $\phi$, which minimizes the free energy and maximizes the supercurrent, that is 
\begin{align}
I_\text{c}=\text{Max}_\phi\{ I (\phi)\}.
\end{align}
here, $I(\phi)$ is the supercurrent of the Josephson junction. 

\begin{figure}[t]
\centering
\includegraphics*[width=6.0cm]{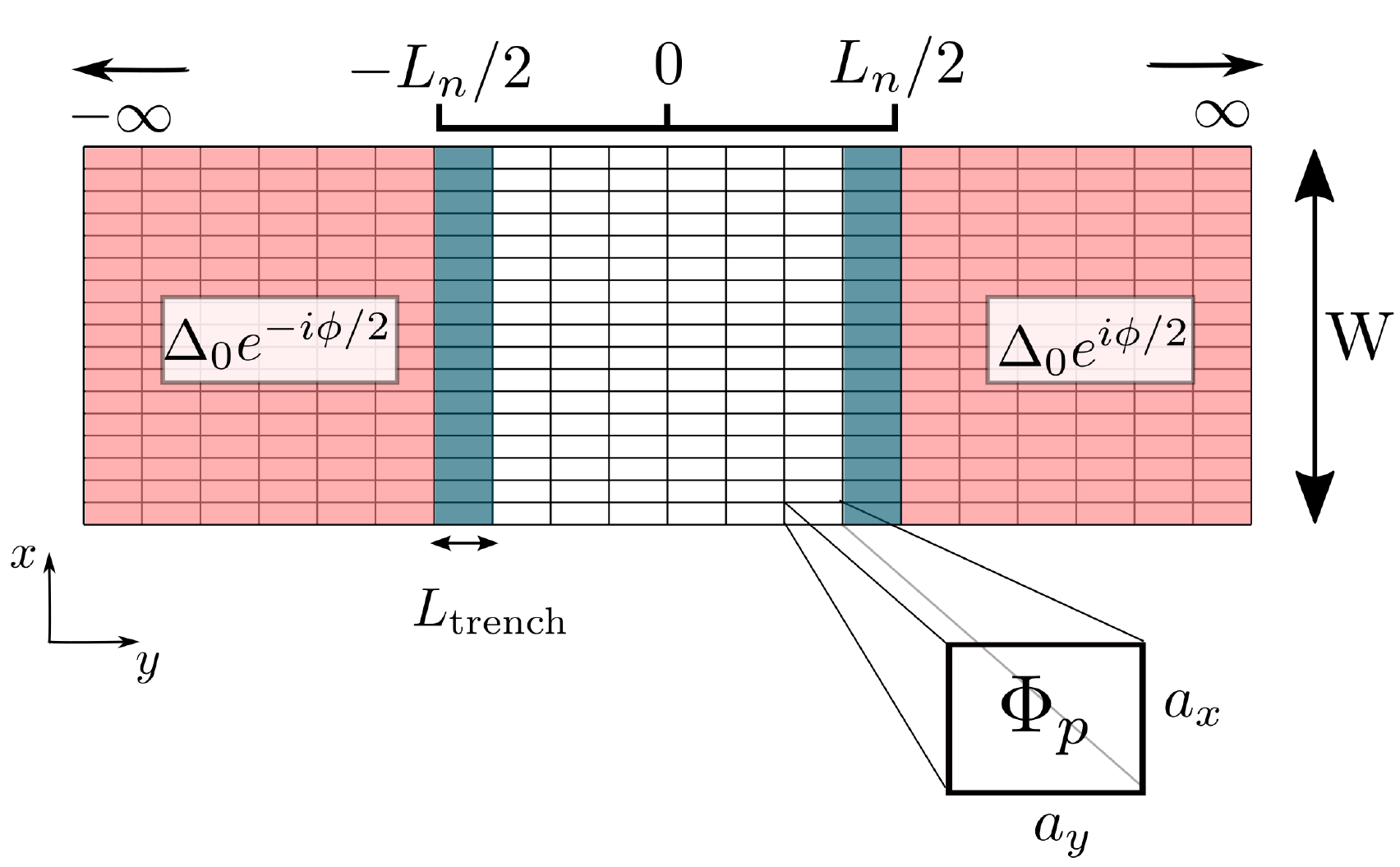}
\caption{Sketch of the discretized Josephson junction, with the mesh $a_x=a_y$. 
Here, we split the junction into two slabs to visualize the recursive inversion method, where the surface Green's functions $g_{LL}$ and $g_{RR}$ (blue areas) are coupled by $V_{LR/RL}$ and contain the information of the semi-infinite normal-superconducting leads, see further details in Sec.~\ref{criticalGF}. Besides, we introduce a magnetic flux $\Phi_p$ that threads each plaquette on the normal part of the junction. 
}\label{fig:setup}
\end{figure}

The stationary longitudinal supercurrent at a given position $y_0$ on the normal part can be expressed in terms of the Green's functions (GFs)\cite{Martin1994a,Yeyati1995a}, that is
\begin{align}\label{eq.statcurr}
I(\phi)= \frac{e}{\hbar}  \int_{-\infty}^\infty d\omega \text{Tr}_\text{W}\left\{[V_{LR} G^{+-}_{RL}(\omega)-V_{RL} G^{+-}_{LR}(\omega)]_e \right\}.
\end{align}
Here, the GFs $G^{+-}_{ij}(\omega)$ are the Fourier transform of the non-equilibrium GFs\cite{Keldysh1964a,Nambu1960a} 
$G^{+-}_{i,j}(t_+,t_-')=\langle T_\leftarrow [\Psi(x,y_i,t_+) \Psi^\dagger(x,y_j,t_-')] \rangle$, 
with the field operators $\Psi(x,y_i,t)$, introduced in Eqs.~\eqref{psie} and~\eqref{psih}, now written in the Heisenberg picture. Besides, $T_\leftarrow$ is the time-ordering operator and the subscripts $\pm$ refer to the upper and lower time branches of the Keldysh contour.
Here, the subindices $L,R$ denote the left and right sections relative to $y_0$, which is placed in the normal part at which we calculate the current, that is, $L=y_0-a/2,~R=y_0+a/2$ and $V_{ij}$ corresponds to the tunnel couplings that link the sites $i$ and $j$ in the $y$-direction. Due to current conservation, $I(\phi)$ is independent of the selected position and thus, we will not specify it. Furthermore, $V_{ij}$ and the GFs are represented by matrices of dimensions $8 \tilde{\text{W}}\times8\tilde{\text{W}}$ with $\tilde{\text{W}}=\text{Integer}[\text{W}/a]$, with W being the width of the Josephson junction, see Fig.~\ref{fig:setup}

In the absence of an applied bias voltage, the system is in equilibrium and thus, $G^{+-}$ is simply given by
\begin{align}
G^{+-}(\omega)=\left[ G^a(\omega)-G^r(\omega)\right] f(\omega),
\label{glesser}
\end{align}
where $f(\omega)$ is the equilibrium Fermi-Dirac distribution and ``$r$'', ``$a$'' denote the retarded and advanced GFs, respectively. Thus, using the Dyson equation we express the supercurrent in terms of the decoupled surface GFs $g_{LL}$ and $g_{RR}$, that is
\begin{align}
&G_{LR}^{r/a} =g_{LL}^{r/a} V_{LR} G_{RR}^{r/a},\\ 
&G_{RL}^{r/a} =G_{RR}^{r/a} V_{RL} g_{LL}^{r/a}\text{~~~and}\\
&G_{RR}^{r/a}=[(g_{RR}^{r/a})^{-1}-V_{RL}g_{LL}^{r/a}V_{LR}]^{-1},
\end{align}
valid for both retarded and advanced (r/a) GFs. The uncoupled GFs $g_{LL}$ and $g_{RR}$ contain information of the semi-infinite leads and the normal sectors from each side and are obtained using the recursive inversion method for the normal part together with an efficient method to calculate the surface GFs of semi-infinite superconducting leads\cite{Lopez1984a}. Then, these surface GFs are combined with the normal part using standard recursive methods, see for example Ref.~\onlinecite{Asano2001a}.

\section{Fraunhofer pattern in different configurations}

\subsection{Wide inverted junction $(M_{S,N}<0)$ with a perpendicular Zeeman field}\label{app.wide}

\begin{figure}[t]
\centering
\includegraphics*[width=3.33in]{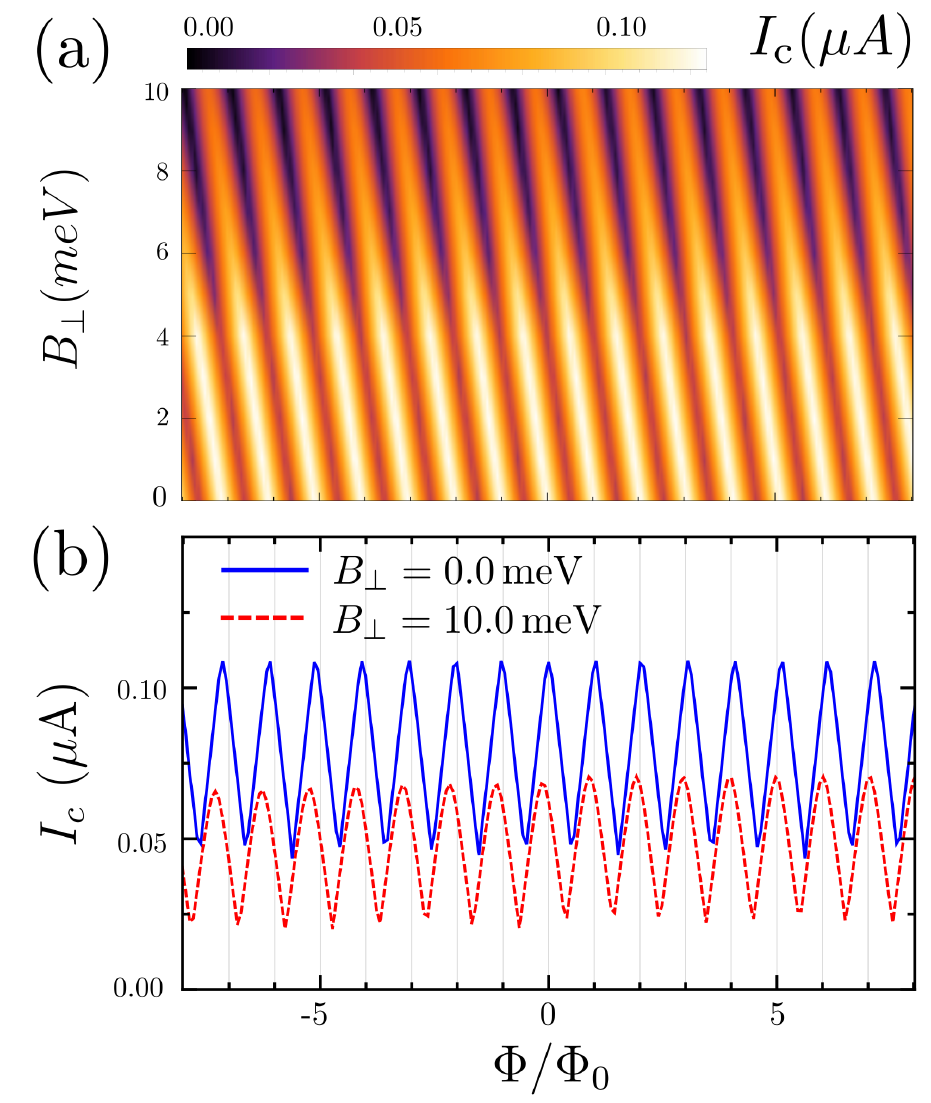} 
\caption{Panel (a), critical current as a function of the magnetic flux and Zeeman field $B_{\bot}$ for a proximitized Josephson junction, with $M=-8.0\,$meV for the whole junction. Panel (b), critical current for two different Zeeman fields corresponding to trivial and topological superconductors. The parameters used are $\Delta_0=4\,$meV, $W=1\,\mu$m, $L_n=0.35\,\mu$m, $C_N=-3\,$meV, $C_S=10\,$meV and lattice constant $a=5\,$nm.}\label{fig:Ic2D} 
\end{figure}

Using the microscopic BHZ model, we show the Fraunhofer pattern of a junction where the quantum spin Hall edge states on opposite sides are decoupled. In Fig.~\ref{fig:Ic2D}(a) we can observe that the critical current shows the usual oscillatory pattern exhibited by quantum spin Hall junctions, with periodicity $p=\Phi_0$ \cite{Hart2014a, Pribiag2015a, Bocquillon2016a}. In contrast to the case analyzed in the main text, the Fraunhofer pattern profile remains unchanged as a function of the Zeeman field $B_\bot$ even though the Zeeman field exceeds largely the critical field, i.e.~$B_\bot>B_\text{c}=0.3\,$meV, see also linecuts in panel (b). Due to the large coupling between the normal and proximitized quantum spin Hall edge states, the critical current is mainly given by LAR. Thus, the presence of MBSs located in the central regions of the NS junctions is not noticed by the edge states.

The shift observed in the Fraunhofer pattern as a function of the Zeeman field in Fig.~\ref{fig:Ic2D} comes from the fact that the Zeeman field is parallel to the helicity of the QSH edges and thus, it shifts the momentum in the same manner as the magnetic flux introduced by the Peierls substitution. To see this, we write the effective Hamiltonian for the QSH edge states with a Zeeman field, 
\begin{align}\label{eq.Heff}
 H_\text{eff}\sim \hbar v_f \hat{k}_y s_z \tau_z +B_\bot s_z,
\end{align}
being $s_z$ and $\tau_z$ the Pauli matrices describing the spin and the top/bottom edges. The addition of a magnetic flux enters via the Peierls substitution, shifting the momentum as $\hat{k}_y\rightarrow \hat{k}_y-e/\hbar A_y$, where $A_y$ changes the sign from the top to the bottom edge, and thus, $A_y\propto \tau_z$, which cancel the $\tau_z$ appearing in Eq.~\eqref{eq.Heff} and entering similarly as the Zeeman term. 

\subsection{Mass domain junction $M_N<0,~M_S>0$ with an in-plane Zeeman field $B_\parallel$}\label{app.massdomain}
An alternative way to enhance the coupling between the quantum spin Hall edge states propagating on the normal part and the MBSs placed at the NS interfaces is to consider $M_S>0$ on the superconducting part. In this configuration, the quantum spin Hall edge states propagate around the normal part and can scatter with MBSs when present. Experimentally, this setup requires a spatial control of the mass sign, which can be realized, for example, in InAs/GaSb wells\cite{Knez2011a}. Alternatively, one could form a Josephson junction by coupling laterally the superconductor \cite{Bai2020a} to a quantum spin Hall system. 
In addition, we show that the topological transition is also visible using an in-plane Zeeman field. Note that since the spin-orbit coupling depends on both momenta $k_x$ and $k_y$, it is not possible to invert the superconducting gap\cite{Alicea2010a}. However, this restriction can be overcome since the width of the junction W is small, and thus, it imposes a confinement energy given by $\sim \hbar^2 /2m^* \text{W}^2$, with $m^*$ being the effective mass of the band, is much larger than the spin-orbit energy $\sim \delta_{e,h}/\text{W}$, that couples different subbands.

\begin{figure}[t]
\centering
\includegraphics*[width=3.30in]{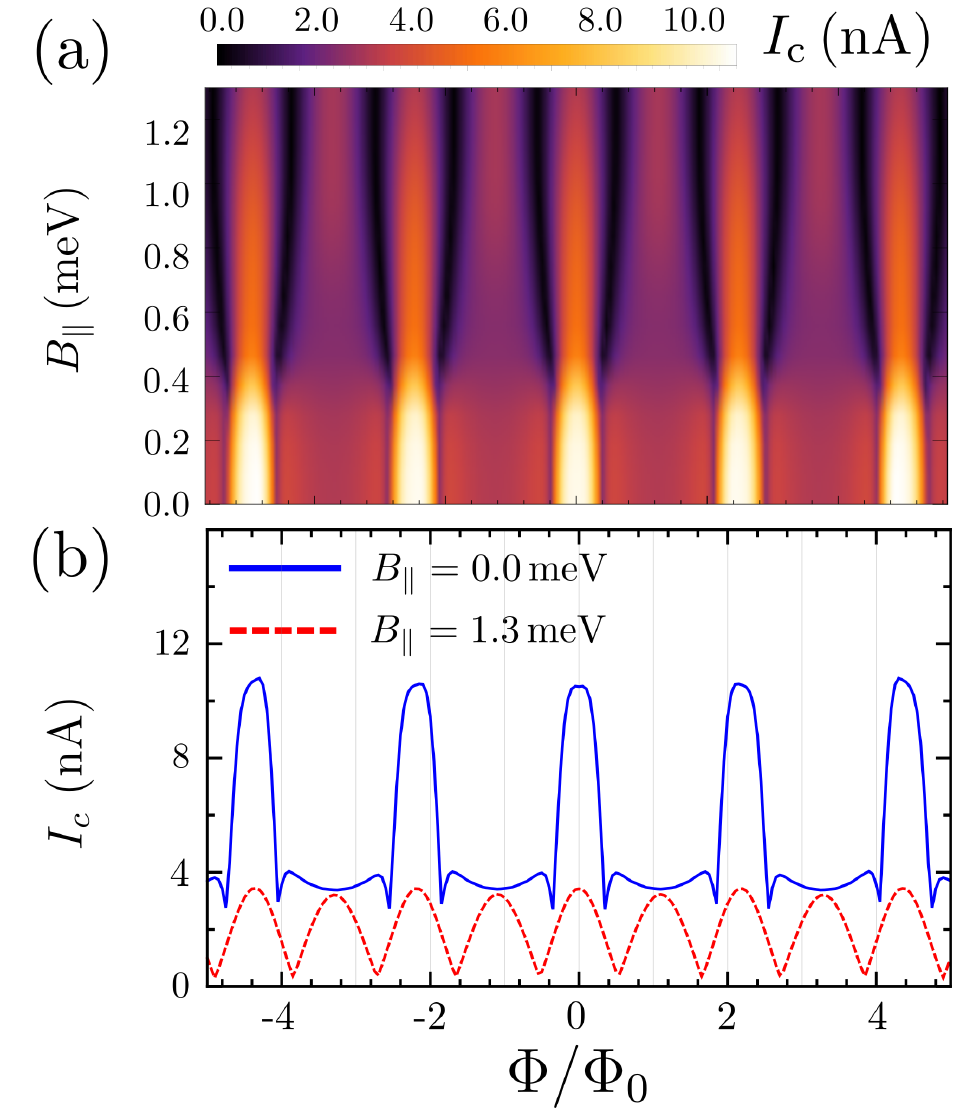} 
\caption{Fraunhofer pattern for the mass domain configuration: Panel (a) critical current as a function of the flux and parallel Zeeman field for the mass domain configuration. The parameters of the model are $M_{S/N}=\pm 10\,$meV, $C_S=12.0\,$meV, $C_N=2.0\,$meV, W$=0.24\,\mu$m and $L_n=0.4\,\mu$m. Panel (b): Fraunhofer pattern for two different linecuts $B_\parallel=0.0$ (trivial regime) and $B_\parallel=1.3\,$meV (topological regime).}\label{fig:Icbeating}
\end{figure}

\begin{figure}[t]
\centering
\includegraphics*[width=3.30in]{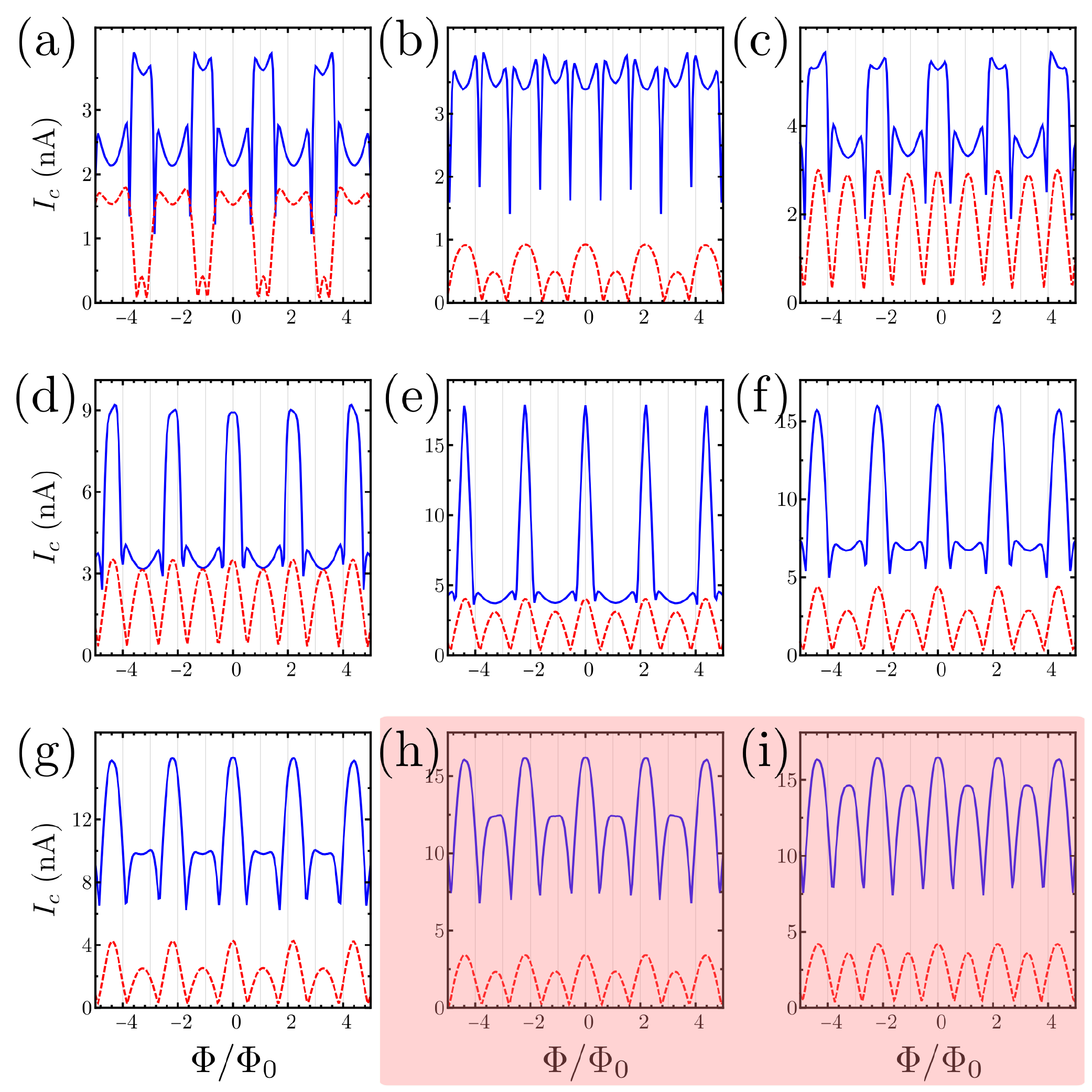}
\caption{Critical current for the mass domain configuration as a function of the magnetic flux, for two values of parallel Zeeman field $B_\parallel=0$ (blue) and 1.2\,meV (red). 
Different panels correspond to values of $C_S$, starting from $C_S=11.0\,$meV and increasing in steps of $\sim 0.3\,$meV from (a) to (i). There are two types of panels, shaded and non-shaded, corresponding to an even and odd number of occupied bands respectively. 
Thus, for finite magnetic field (red curves) only the non-shaded panels can develop MBSs, whereas for the shaded panels both blue and red curves correspond to the trivial case.
Besides, W$=0.24\,\mu$m and $L_n=0.4\,\mu$m.}\label{fig:Ic_masscn}
\end{figure}

\begin{figure}[t]
\centering
\includegraphics*[width=2.5in]{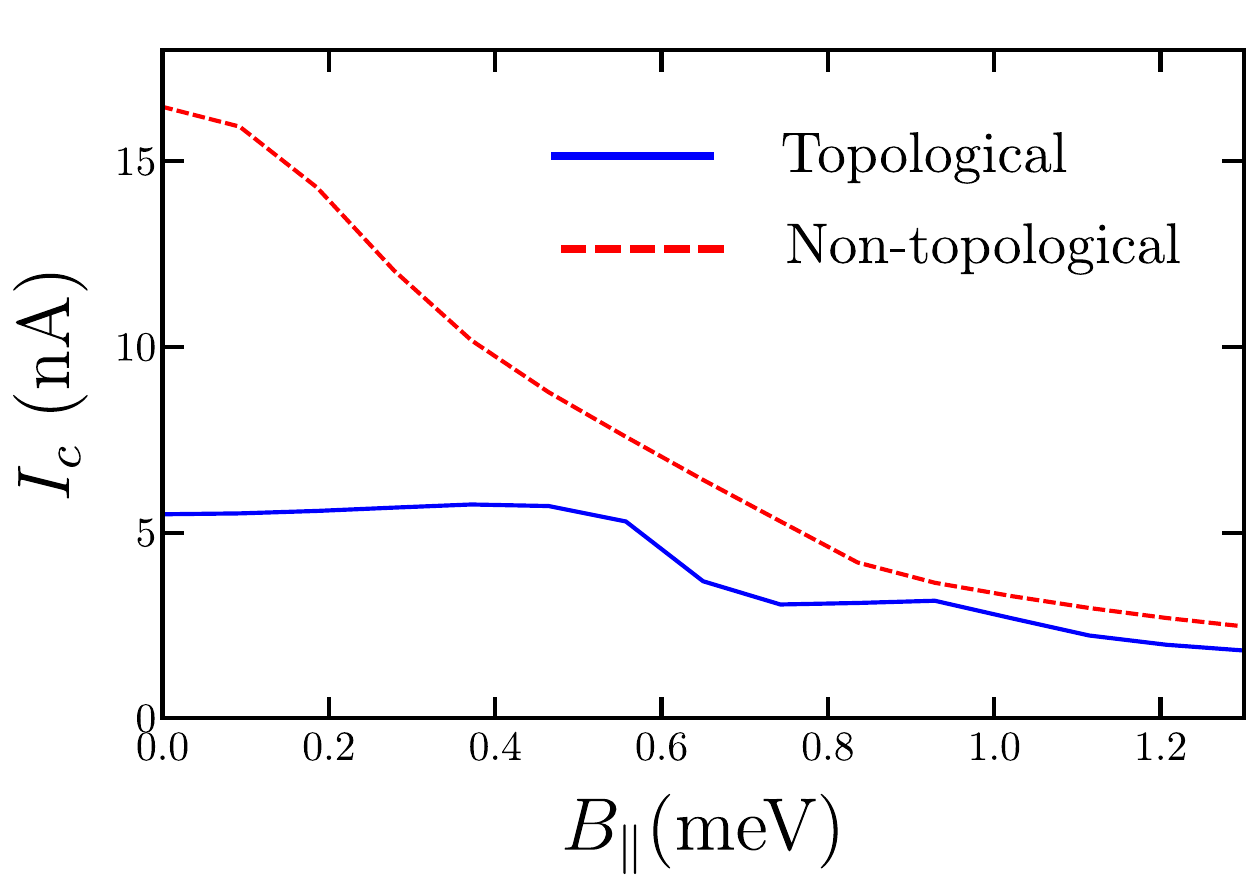} 
\caption{Critical current as a function of the Zeeman field $B_\parallel$ for $\Phi=0$. We have used the same set of parameters as in Fig.~\ref{fig:Ic_masscn}, panels (c) and (h) for the blue and red curves, respectively. The topological transition is evidenced by a sharp decrease of the critical current, see Ref.~\onlinecite{Cayao2017a}.
}\label{fig:Ictransition}
\end{figure}

In the trivial regime, the Fraunhofer pattern exhibits a doubled period with respect to the one obtained in the proximitized junction, i.e.~$p\approx2\Phi_0$, compare the blue linecuts in Figs.~\ref{fig:Icbeating}(b) with the SQUID pattern resulting from a junction with $M_{N,S}<0$ with only LAR processes in Fig.~\ref{fig:Ic2D}(b).
This scenario represents an extreme version of the \emph{even-odd flux quanta effect} observed in the \emph{thin superconducting junction}. Here, particles can encircle the normal part due to a finite Fermi velocity mismatch between the quantum spin Hall edge states and the superconducting part, which enhances the electron-electron reflection amplitude \cite{Mortensen1999a}. 
In the topological regime $(B_\parallel>B_{c})$, we again observe an abrupt change in the Fraunhofer pattern periodicity. In the particular case shown in Fig.~\ref{fig:Icbeating}, the periodicity is halfed to $p\approx \Phi_0$. However, the Fraunhofer pattern in the mass domain and the corresponding changes caused by the presence of a MBS do not exhibit a universal behavior. In contrast to the thin superconducting junction, here the coupling to the bulk s-wave superconductor is more visible yielding an extra source of LAR with respect to the electron-hole reflections provided by the MBS. The ratio between the two contributions is effectively controlled by $C_S$: the larger $k_F$, the more favored the LAR processes become. 
In Fig.~\ref{fig:Ic_masscn}, we show the Fraunhofer pattern for two values of the Zeeman field: $B_\parallel=0$ (blue) and $B_\parallel=1.2\,$meV (red) and nine values of the chemical potential on the superconducting part ($C_S$). Panels from~(a) to~(g) show a blue and a red curve corresponding to the trivial and topological regimes, respectively. The shaded panels (h) and (i) correspond to the trivial regime for both curves. In these two latter panels, an even number of superconducting bands are occupied and consequently above the critical Zeeman field, an even number of Majorana bound states form on each side, which overlap, gap out and yield a trivial superconductor\cite{Reuther2013a,Novik2020a}. This is known as the even/odd effect in the Majorana wells. Indeed, we can observe that the Fraunhofer pattern changes its profile for the six first panels (a)-(f), whereas close to the odd-even transition and within the even sectors, the Fraunhofer pattern does not exhibit a qualitative change in periodicity. In contrast to the \emph{thin junction} regime where the superconducting electrodes are composed by a single proximitized mode, here, the presence of extra modes can make the distinction between the trivial and topological regimes more difficult, see for example panels~c and~h, where blue and red curves exhibit a similar periodicity. 
To distinguish between both cases we study the critical current for $\Phi=0$ as a function of the Zeeman field $B_\parallel$, see Fig.~\ref{fig:Ictransition}. We can observe that in the case of a topological transition (blue curve), the critical current exhibits a sharp drop around $B_\parallel \sim B_\text{c}$, whereas the curve corresponding to the ``even'' case decreases continuously. This has been proposed as a detection scheme in Ref.~\onlinecite{Cayao2017a}. 

\section{Scattering model}\label{app.scattering}

In this section, we provide the explicit form of $S_N$ and $S_A$ used in Eq.~\eqref{eq.supercurrent}. 

This model is adapted from the one presented in Ref.~\onlinecite{Baxevanis2015a}, where we use helical propagation along the NS interfaces and add the possibility of scattering with MBSs or trivial ZEABSs. 

We start with $S_N$, which gives the relation between the incoming and outgoing modes relative to the central part of the junction. Thus, we write $(b_L,b_R)^t=S_N(a_L,a_R)^t$, where $a_{L(R)}$ and $b_{L(R)}$ are 4-component vectors containing left (right) incoming and outgoing scattering amplitudes on the top and bottom edges relative to the normal part of the junction, and are given by $a_L=(a_{e,\uparrow}^T,a_{e,\downarrow}^B,a_{h,\uparrow}^T,a_{h,\downarrow}^B)^t_L$, $a_R=(a_{e,\uparrow}^B,a_{e,\downarrow}^T,a_{h,\uparrow}^B,a_{h,\downarrow}^T)^t_R$, $b_L=(b_{e,\uparrow}^B,b_{e,\downarrow}^T,b_{h,\uparrow}^B,b_{h,\downarrow}^T)^t_L$, $b_R=(b_{e,\uparrow}^T,b_{e,\downarrow}^B,b_{h,\uparrow}^T,b_{h,\downarrow}^B)^t_R$, here $T/B$ stand for top and bottom edges. Note that the lack of some of the incoming and outgoing states comes from the spin-momentum locking of the edge states, e.g.~there is no left-top incoming down-spin state $a_{e/h,\downarrow}^T$. The form of $S_N$ is given by
\begin{equation}\label{eq:scatnormal}
S_N=\left(
\begin{array}{cc}
 0 & S_{N,LR} \\
 S_{N,RL} & 0 \\
 \end{array}
\right) 
\end{equation}
which is an off-diagonal matrix since $S_N$ connects the incoming of the L/R side with the outgoing scattering states of the R/L side. Here,
\begin{equation}
S_{N,LR}=\left(
\begin{array}{cccc}
  e^{i (\chi_b+\tilde{\Phi}/2)} & 0 & 0 & 0 \\
  0 & e^{i (\chi_t-\tilde{\Phi}/2)} & 0 & 0 \\
  0 & 0 & e^{i( \chi_b-\tilde{\Phi}/2)} & 0 \\
  0 & 0 & 0 & e^{i (\chi_t+\tilde{\Phi}/2)} \\
 \end{array}
\right) \nonumber
\end{equation}

\begin{equation}
S_{N,RL}=\left(
\begin{array}{cccc}
 e^{i (\chi_t+\tilde{\Phi}/2 )} & 0 & 0 & 0  \\
 0 & e^{i (\chi_b-\tilde{\Phi}/2)} & 0 & 0  \\
 0 & 0 & e^{i (\chi_t-\tilde{\Phi}/2 )} & 0 \\
 0 & 0 & 0 & e^{i (\chi_b+\tilde{\Phi}/2)}\\
\end{array}
\right) \nonumber
\end{equation}
being $\chi_{t/b}=E/E_{T,t/b}$ the dynamical phases accumulated along the top and bottom edges, $E_{T,t/b}$ the top/bottom Thouless energies and the dimensionless magnetic flux through the normal part $\tilde{\Phi}=\pi\Phi/\Phi_0=(e/\hbar)\int_0^{L_n} A_y|_{x=\text{W}} dy$ with ${\bf A}=|{\bf H}_\bot|x\bm{e}_y$. In the following we consider the short junction limit, i.e.~$\Delta_0\ll E_{T,t,b}= \hbar v_F /L_n$. However, the periodicity of the Fraunhofer pattern is similar in intermediate or the long junction limit. 

The $S$ matrix describing the scattering processes at the NS junctions is given by 
\begin{equation}\label{eq:SA}
S_A=\left(
\begin{array}{cc}
 S_{A,L} & 0 \\
 0 & S_{A,R} \\
\end{array}
\right) .
\end{equation}
% %
$S_{A,L/R}$ describes the scattering events taking place at the left/right interfaces and fulfills $S_A (b_L,b_R)^t=(a_L,a_R)^t$. Guided by the microscopic model, we include three different scattering processes:
\begin{itemize}
 \item Scattering with the bulk superconductor.
 \item Normal deflection between the top and bottom edges.
  \item Scattering with the MBS/ZEABS.
\end{itemize}
To do so, we model each NS interface by the combination ($\times$) of three $S$ matrices, that is, $S_{A,i}=S^T_{\text{green},i}\times S_{\text{red},i} \times S_{\text{green},i}^B$, with $i=L/R$. Here, $S^{T/B}_{\text{green},i}$ describes the scattering between the quantum spin Hall edges with a bulk superconductor that includes a deflection probability towards the opposite edge. 
Besides, $S_{\text{red},i}$ is the MBS scattering matrix on the $i$-side. 

We start introducing $S^{T/B}_{\text{green},i}$, which are represented by green crossed boxes in Fig.~\ref{fig:Scattsetup} and are obtained by the combination of two $S$ matrices, i.e.~$S^{T/B}_{\text{green},i}=S_{\text{pot},i}^{T/B}\times S_{\text{Bulk},i}$, where $S_{\text{pot},i}^{T/B}$ describes the scattering processes between the quantum spin Hall edges and a (ficticious) normal potential that introduces a deflection probability ($\Gamma_d$) towards the opposite edge.
For simplicity, we consider the same deflection probability on the top, bottom, left and right, and thus, the $S$ matrix becomes independent of the $T/B$ and $L/R$ labels, that is, $S_{\text{pot},L/R}^{T/B}\equiv S_{\text{pot}}$ and consequently $S_{\text{green},L/R}^{T/B}\equiv S_{\text{green},L/R}$. 
For the sake of simplicity, we model the potential scatterer by means of the simplest $S$ matrix that respects time-reversal symmetry and does not mix the spin degree of freedom, namely
\begin{equation}\label{eq:potential}
\begin{array}{l}
S_{\text{pot}}=
\left(
\begin{array}{cc}
                              e^{i \psi_{1}}\sqrt{\Gamma_d} & \sqrt{1-\Gamma_d}\\
                              \sqrt{1-\Gamma_d} & -e^{i \psi_{2}}\sqrt{\Gamma_d}\\
                              \end{array}\right),
\end{array}
\end{equation} 
where $\psi_{1}$ and $\psi_{2}$ are phases acquired after crossing the potential barriers. Furthermore, we set both phases to zero and also neglect the possibility of a spin-rotation due to the presence of spin-orbit coupling.
The scattering matrix connects the states 
$$(a_{edge},a_{aux})^t=S_{\text{pot}} (b_{edge},b_{aux})^t,$$ 
where ``aux'' refers to the auxiliary states placed between the potential and the superconductor and ``edge'' refers to the edge states propagating along the normal part, see Fig.~\ref{fig:Scattsetup}. 
Their specific components depend on the considered scattering center, for example, in Fig.~\ref{fig:Scattsetup} we illustrate the modes scattering on the top-right scattering center. 
Here, electron and hole components of the incoming top-right ``T'' (outgoing ``inter,T'') states with spin up fulfill $ (a_{e/h,\uparrow}^{inter, T},a_{e/h,\uparrow}^{aux})^t=S_{\text{pot}}(b_{e/h,\uparrow}^{T},b_{e/h,\uparrow}^{aux})^t$. A similar expression can be written for the spin-down components. Due to their helical character, the incoming and outgoing components with spin down are exchanged, that is, $(a_{e,\downarrow}^{T},a_{e,\downarrow}^{aux})^t=S_{\text{pot}} (b_{e,\downarrow}^{inter, T},b_{e,\downarrow}^{aux})^t$.

\begin{widetext}

So much for $S_{\text{pot}}$, we now introduce $S_{\text{Bulk},L/R}$, which describes the electron-hole reflection processes mediated by a trivial bulk superconductor and fulfills
\begin{equation}\label{eq:scat}
S_{\text{Bulk},L/R} \left(
\begin{array}{c}
                              a^{aux}_{e,\uparrow} \\
                              a^{aux}_{e,\downarrow}\\
                              a^{aux}_{h,\uparrow}\\
                              a^{aux}_{h,\downarrow}\\
                              \end{array}\right)
=\begin{array}{l}
\left(
\begin{array}{cccc}
                              r_{\uparrow,\uparrow}^{ee} & r_{\uparrow,\downarrow}^{ee} & r_{\uparrow,\uparrow}^{eh} & r_{\uparrow,\downarrow}^{eh}\\
                              r_{\downarrow,\uparrow}^{ee} & r_{\downarrow,\downarrow}^{ee} & r_{\downarrow,\uparrow}^{eh} & r_{\downarrow,\downarrow}^{eh}\\
                              r_{\uparrow,\uparrow}^{he} & r_{\uparrow,\downarrow}^{he} & r_{\uparrow,\uparrow}^{hh} & r_{\uparrow,\downarrow}^{hh}\\
                              r_{\downarrow,\uparrow}^{he} & r_{\downarrow,\downarrow}^{he} & r_{\downarrow,\uparrow}^{hh} & r_{\downarrow,\downarrow}^{hh}\\
                              \end{array}\right)_{L/R}
\end{array}
\left(
\begin{array}{c}
                              a^{aux}_{e,\uparrow} \\
                              a^{aux}_{e,\downarrow}\\
                              a^{aux}_{h,\uparrow}\\
                              a^{aux}_{h,\downarrow}\\
                              \end{array}\right)
=\left(
\begin{array}{c}
                              b^{aux}_{e,\uparrow} \\
                              b^{aux}_{e,\downarrow}\\
                              b^{aux}_{h,\uparrow}\\
                              b^{aux}_{h,\downarrow}\\
                              \end{array}\right),
\end{equation}
which relates the incoming $a^{aux}_{e/h,\uparrow \downarrow}$ and the outgoing $b^{aux}_{e/h,\uparrow \downarrow}$ scattering amplitudes in the ``aux'' region.
Using the Andreev approximation, the right reflection coefficients are given by $r^{he}_{\downarrow \uparrow}(E)=(r^{eh}_{\downarrow \uparrow}(-E))^*= \exp[i (\chi-\phi/2)]$ and $r^{he}_{\uparrow \downarrow}(E)=(r^{eh}_{\uparrow \downarrow}(-E))^*=- \exp[i (\chi-\phi/2)]$, with $\chi=-\text{arccos}(E/\Delta_0)$, 
while the left-reflection coefficients are obtained from the right coefficients by substituting $\phi \rightarrow -\phi$. 
An alternative way to model $S_{\text{green},i}$ is to relax the conditions that lead to the Andreev approximation. Away from this limit one can naturally obtain electron-electron reflection amplitudes which lead to a finite deflection probability~\cite{Chen2011a, Reinthaler2013a, Novik2014a, Pientka2017a}. 

To proceed further, we combine $S_{\text{pot}}$ in Eq.~\eqref{eq:potential} with the corresponding superconducting $S$ matrix Eq.~\eqref{eq:scat} eliminating the ``aux'' modes. In the trivial case, the resulting matrix is given by
\begin{equation}\label{eq:weidenmuller}
S_{\text{green},L/R}=\frac{1}{1-e^{2 i \chi } \Gamma_d}\left(
\begin{array}{cccc}
 \sqrt{\Gamma_d } (1-e^{2 i \chi}) & 0 & 0 & (1-\Gamma_d ) e^{i(\chi\mp \phi/2)} \\
 0 & \sqrt{\Gamma_d } (1-e^{2 i \chi }) & -(1-\Gamma_d ) e^{i(\chi \mp \phi/2)}  & 0 \\
 0 & -(1-\Gamma_d ) e^{i (\chi\pm \phi/2 )}  & \sqrt{\Gamma_d } (1-e^{2 i \chi }) & 0 \\
 (1-\Gamma_d ) e^{i( \chi\pm \phi/2 )} & 0 & 0 & \sqrt{\Gamma_d } (1-e^{2 i \chi }) \\
\end{array}
\right).
\end{equation}

\end{widetext}

The $S$ matrices $S_{\text{red},i}$, represented by red crossed boxes in Fig.~\ref{fig:Scattsetup}, are obtained in a similar fashion as $S_{\text{green},i}$, that is, $S_{\text{red},i}=S_{\text{pot},M}\times S_{\text{MBS},i}$. Here, we combine two $S$ matrices, one accounting for the probability of particles propagating along the NS junction without scattering with the MBS, $\Gamma_M=1$, and the MBS $S$ matrix.
While $S_{\text{pot},M}$ is obtained from Eq.~\ref{eq:potential} using the replacement $\Gamma_d\rightarrow \Gamma_M$.

We now introduce the matrix $S_{\text{MBS},i}$, which accounts for the scattering with MBSs or ZEABSs. In both cases, we use the Weidenm\"uller formula,\cite{Fisher1981a,Iida1990a} \cite{Haim2015a}
\begin{equation}\label{eq:weid}
\begin{array}{l} 
\left(
\begin{array}{cc}
                              r^{ee} & r^{eh} \\
                              r^{he} & r^{hh}\\                             
                              \end{array}\right)=\mathbb{1}-2\pi \nu_0 i W^\dagger (E+i \pi \nu_0 W W^\dagger)^{-1}W,
\end{array}
\end{equation}
with the density of states of the quantum spin Hall edges $\nu_0=1/\pi \hbar v_F $ and $W$ accounting for the tunnel amplitudes between the MBSs/ZEABSs and the quantum spin Hall edge states.

{\bf Majorana bound states}--- In the case of MBSs $W=(t_\uparrow,t_\downarrow,t_\uparrow^*,t_\downarrow^*)$, yielding the reflection coefficients
\begin{align}
r^{ee}_{ss'} =\delta_{ss'}+\frac{2\pi \nu_0 t^*_{s}t_{s'}}{iE-\gamma},~~~~r^{he}_{ss'} =\frac{2\pi \nu_0 t_{s}t_{s'}}{iE-\gamma},
\end{align}
where $\gamma=2\pi \nu_0(|t_\uparrow|^2+|t_\downarrow|^2)$ and $t_\sigma$ is the effective coupling amplitude between the helical edge states and the Majorana bound states. 
The analytical expressions for $S_{\text{red},L/R}$ is cumbersome and therefore, we obtain it only numerically.
The same expressions can be rederived without using explicitly the Weidenm\"uller formula following Refs.~\onlinecite{Pikulin2016a,Fleckenstein2021a}.

\begin{figure}[t]
\centering
\includegraphics*[width=3.30in]{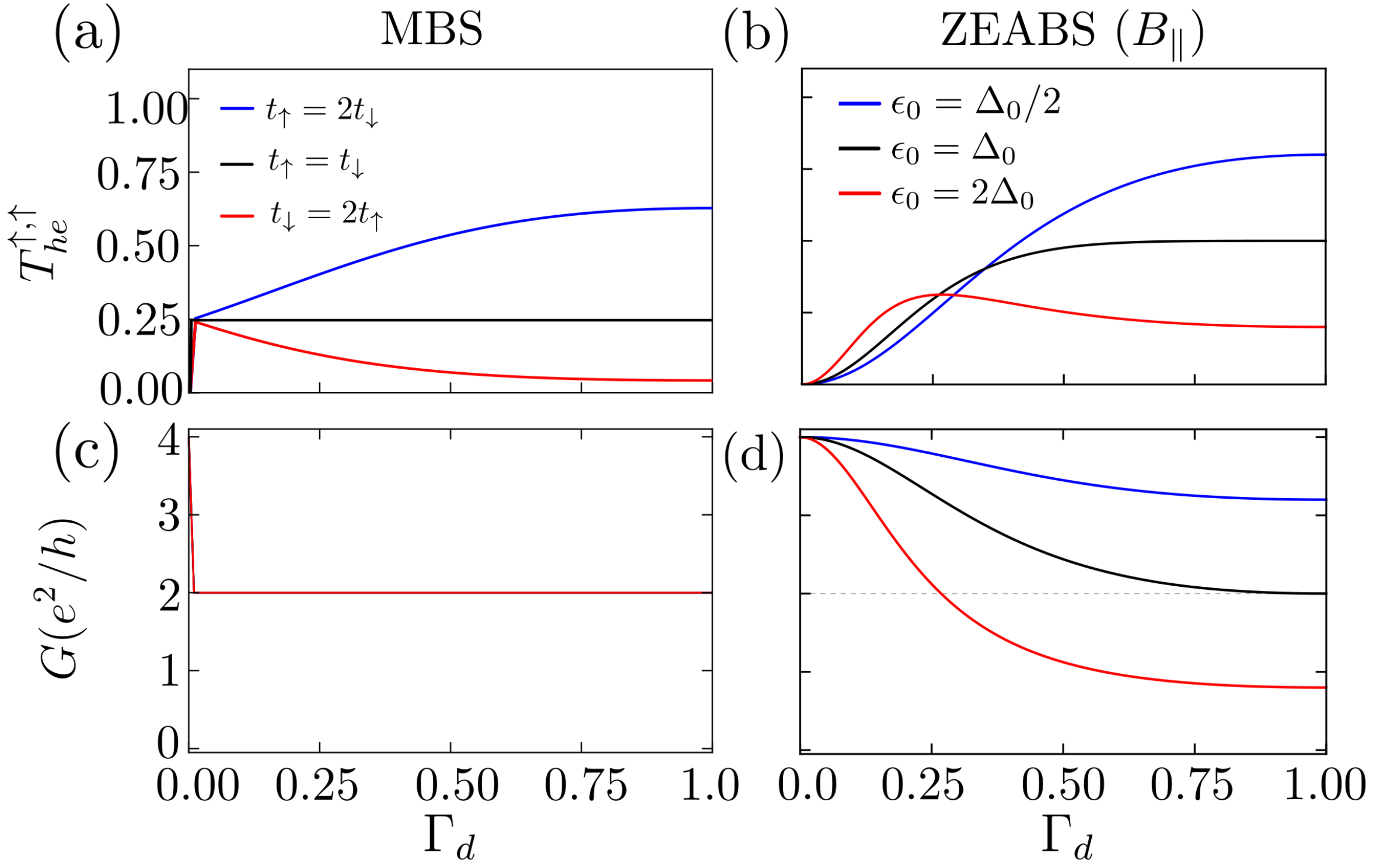}
\caption{Panels~(a) and~(b): Interedge electron-hole reflection probability $T_{he}^{\uparrow,\uparrow}$ as a function of the deflection probability $\Gamma_d$ for the MBS (a) and the ZEABS (b), with $\Gamma_M=0$ and $E=0$. We have used three different tunnel amplitude configurations $t_{\uparrow}=2t_{\downarrow}$, $t_{\uparrow}=t_{\downarrow}$, $t_{\downarrow}=2t_{\uparrow}$ in the MBS case and $\epsilon_0=\Delta_0/2$, $\epsilon_0=\Delta_0$ and $\epsilon_0=2\Delta_0$ for the ZEABS case. 
Panels~(c) and~(d): zero bias conductance as a function of $\Gamma_d$ for the MBS and ZEABS with the same parameters as in panels~(a) and~(b).
}\label{fig:transmissioneh}
\end{figure}

{\bf Zero energy bound states}--- In the same way, we can obtain the $S$ matrix describing the scattering processes between the quantum spin Hall edges and an accidental ZEABS. To do so, we use an effective model describing a proximitized region ($\Delta_0$) that can develop a zero energy state under a Zeeman term ($B_{\parallel/\bot}$), namely \cite{Haim2015a} 
\begin{align}
H_\text{ZEABS}=&\sum_{s=\uparrow,\downarrow} \epsilon_0 d_s^\dagger d_s +\Delta_0 d^\dagger_\uparrow d^\dagger_\downarrow+\text{h.c.}\nonumber\\
&+B_\parallel d^\dagger_s d_{s'} (s_\theta)_{s,s'}+B_\bot d^\dagger_s d_{s'} (s_z)_{s,s'}, 
\end{align}
here $d^{(\dagger)}$ are fermionic destruction (creation) operators and $\epsilon_0$ is the energy of the particle. This region is coupled to the quantum spin Hall edge states by means of the tunnel Hamiltonian $H_T= \sum_s \omega_s \psi^\dagger_s(x_0) d_s+\text{h.c.}$, with $\psi(x)$ being the electronic field operator of the quantum spin Hall edges at position $x=x_0$, where the ZEABS is placed. 

{\it In-plane magnetic field}--- For a finite in-plane field $B_\parallel \neq 0$ and a zero perpendicular Zeeman field $B_\bot= 0$, we diagonalize the Hamiltonian $H_\text{ZEABS}=(\sqrt{\Delta_0^2+\epsilon_0^2}-B_\parallel)a^\dagger a+(\sqrt{\Delta_0^2+\epsilon_0^2}+B_\parallel)b^\dagger b+\text{const}$. For $B_\parallel = \sqrt{\Delta_0^2+\epsilon_0^2}$, the Hamiltonian develops a zero energy state and the spectrum has a low and high energy sector. In this way, we project $H_T$ onto the low energy sector by means of the transformation 
\begin{align}
&d_\uparrow\approx -\sin(\alpha) a^\dagger -\cos(\alpha) a,\\ 
&d_\downarrow\approx -\sin(\alpha) a^\dagger +\cos(\alpha) a,
\end{align}
with the parametrization 
\begin{align}
&\sin(\alpha)=\frac{\sqrt{\Delta_0^2+\epsilon_0(\epsilon_0-\sqrt{\Delta_0^2+\epsilon_0^2})}}{2\sqrt{\Delta_0^2+\epsilon_0^2}},\label{eq.sin}\\ 
&\cos(\alpha)=\frac{\sqrt{\Delta_0^2+\epsilon_0(\epsilon_0+\sqrt{\Delta_0^2+\epsilon_0^2})}}{2\sqrt{\Delta_0^2+\epsilon_0^2}}\label{eq.cos}.
\end{align}
Thus, in the low energy sector, the tunnel Hamiltonian reads,
\begin{equation}
H_T=\sum_k \begin{array}{l} 
\left(
\begin{array}{cc}
                              a^\dagger, & a\\
                              \end{array}\right)W_\parallel
\end{array}
\begin{array}{l} \left(
\begin{array}{c}
                               \psi_{k,\uparrow} \\
                               \psi_{k,\downarrow} \\
                               \psi_{k,\uparrow}^\dagger \\
                               \psi_{k,\downarrow}^\dagger \\
                               \end{array}\right)
\end{array}
\end{equation}
with 
\begin{equation}\label{eq:weidtrivpar}
W_\parallel=\begin{array}{l} 
\left(
\begin{array}{cccc}
                              -t_{2,\uparrow} & t_{1,\downarrow} & t_{1,\uparrow}^* & t_{2,\downarrow}^* \\
                              -t_{1,\uparrow} & -t_{2,\downarrow} & t_{2,\uparrow}^* & -t_{1,\downarrow}^*  \\
                              \end{array}\right),
\end{array}
\end{equation}
and the tunnel amplitudes $t_{1[2],\uparrow}=\omega_\uparrow^* \sin(\alpha) [\cos(\alpha)]$ and $t_{1[2],\downarrow}=\omega_\downarrow^* \cos(\alpha) [\sin(\alpha)]$. Now, $W_\parallel$ enters into the Weidenm\"uller formula~\eqref{eq:weid}, so we obtain the reflection coefficients. In this case, the analytical expressions for $r^{ee}_{ss'}$ and $r^{he}_{ss'}$ are cumbersome, thus, we restrict ourselves to evaluate it numerically. In general, we observe a finite electron-hole reflection amplitude with parallel spin. Thus, we expect to observe a similar (but not equal) behavior in the Fraunhofer pattern as with MBSs.

{\it Perpendicular magnetic field}--- When the Zeeman field is applied in $z$-direction, that is, $B_\parallel=0$ and $B_\bot \neq 0$ the gap also closes at  $B_\bot = \sqrt{\Delta_0^2+\epsilon_0^2}$ and the low energy sector allows us to write 
\begin{align}
&d_\uparrow\approx -\sin(\alpha) a^\dagger \\ 
&d_\downarrow\approx \cos(\alpha) a
\end{align}
with the parametrization 
\begin{align}
&\sin(\alpha)=\frac{\Delta_0}{\sqrt{2\Delta_0^2+2\epsilon_0(\epsilon_0+\sqrt{\Delta_0^2+\epsilon_0^2})}}\label{eq.sinz}\\ 
&\cos(\alpha)=\frac{\Delta_0}{\sqrt{2\Delta_0^2+2\epsilon_0(\epsilon_0-\sqrt{\Delta_0^2+\epsilon_0^2})}}\label{eq.cosz}.
\end{align}
yielding 
\begin{equation}\label{eq:weidtrivbot}
W_\bot=\begin{array}{l} 
\left(
\begin{array}{cccc}
                              0 & \tilde{t}_{\downarrow} & \tilde{t}_{\uparrow}^* & 0 \\
                              -\tilde{t}_{\uparrow} & 0 & 0 & -\tilde{t}_{\downarrow}^*  \\
                              \end{array}\right),
\end{array}
\end{equation}
with $\tilde{t}_{\uparrow}\equiv \sin(\alpha) \omega_\uparrow^*$ and $\tilde{t}_{\downarrow}\equiv \cos(\alpha) \omega_\downarrow^*$.

In this case, the electron-hole reflection coefficients can be obtained analytically\cite{Haim2015a}, namely, 
\begin{align}
r^{he}=2\pi\nu_0\frac{ \tilde{t}_\uparrow \tilde{t}_\downarrow }{i E-\gamma/2}\sigma_x, 
\end{align}
where in contrast to the in-plane ZEABS, here $r^{he}_{\uparrow \uparrow}=r^{he}_{\downarrow \downarrow}=0$. Thus, we expect to observe a completely different behavior as compared to the in-plane fields.

\subsection{Interedge transmission probability and total conductance}

To have more insight into the reflection properties of the combined $S$ matrix $S_{A,i}\,$, we calculate numerically the inter-edge electron-hole reflection probability $T^{\uparrow  \uparrow}_{he}$ and the zero bias conductance at zero temperature, given by
\begin{align}
G=\frac{e^2}{h}\left(2- \text{Tr}[S_{ee}^\dagger S_{ee}]+\text{Tr}[S_{eh}^\dagger S_{eh}]\right),  
\end{align}
with $\Gamma_M=0$ and $E=0$. 

Here, we consider two different scenarios, the first one with a MBS and the second one with an accidental zero energy ABSs with an in-plane magnetic field. In both cases we restrict to the case where all tunnel amplitudes are real $t_\sigma\in \mathbb{R}$.

We start with the MBS case, whose results are shown in Fig.~\ref{fig:transmissioneh}(a,c).
For zero deflection probability $\Gamma_d=0$, we observe that electrons are locally reflected from the bulk superconductor, yielding $T^{\uparrow \uparrow}_{he}=0$ and $G=4e^2/h$.
Then, increasing $\Gamma_d$, the conductance becomes constant and equal to $G=2e^2/h$ independent on the tunnel couplings $t_{\sigma}$, whereas the transmission $T^{\uparrow \uparrow}_{he}$ changes in the following manner: 
For $t_\uparrow =2 t_\downarrow$, with $t_{\uparrow/\downarrow}\in $ Reals $T^{\uparrow \uparrow}_{he}$ increases linearly as a function of $\Gamma_d$, reaching a plateau for large $\Gamma_d$.  Similarly, for $t_\downarrow =2 t_\uparrow$, $T^{\uparrow \uparrow}_{he}$ decreases linearly as a function of $\Gamma_d$ until it reaches a plateau. In all cases, when $\Gamma_d\approx 1$ the transmission is mainly given by the Majorana electron-hole reflection coefficient, that is, $T^{\uparrow \uparrow}_{he}\approx|r^{he}_{\uparrow\uparrow}|^2=\frac{|t_\uparrow|^4}{|t_\uparrow^2+t_\downarrow^2|^2}$.

In the case of zero energy Andreev bound states, $T^{\uparrow \uparrow}_{he}$ is in general finite for $\Gamma_d>0$, due to finite $r_{\uparrow\uparrow}^{he}$ introduced by the zero energy scatterers. In contrast to the MBS scenario, here, the conductance $G$ exhibits any value between 4 and 0, depending on the used set of tunnel couplings $t_{i,\sigma}$, see Fig.~\ref{fig:transmissioneh}(b,d). These couplings can be tuned as a function of the microscopic parameters $\epsilon_0$ and $\Delta_0$, which modify $\sin(\alpha)$ and $\cos(\alpha)$  in Eqs.~\eqref{eq.sin} and \eqref{eq.cos}.

\begin{acknowledgments}
We thank D. Breunig and B. Scharf for helpful discussions. F.D.\ and P.R.\ gratefully acknowledge funding by the Deutsche Forschungsgemeinschaft (DFG, German Research Foundation) within the framework of Germany's Excellence Strategy -- EXC-2123 QuantumFrontiers -- 390837967. E. G. N. gratefully acknowledges funding by SFB 1143 (Project ID 247310070).
\end{acknowledgments}

%\bibliography{QH_S}

\begin{thebibliography}{94}%
\makeatletter
\providecommand \@ifxundefined [1]{%
 \@ifx{#1\undefined}
}%
\providecommand \@ifnum [1]{%
 \ifnum #1\expandafter \@firstoftwo
 \else \expandafter \@secondoftwo
 \fi
}%
\providecommand \@ifx [1]{%
 \ifx #1\expandafter \@firstoftwo
 \else \expandafter \@secondoftwo
 \fi
}%
\providecommand \natexlab [1]{#1}%
\providecommand \enquote  [1]{``#1''}%
\providecommand \bibnamefont  [1]{#1}%
\providecommand \bibfnamefont [1]{#1}%
\providecommand \citenamefont [1]{#1}%
\providecommand \href@noop [0]{\@secondoftwo}%
\providecommand \href [0]{\begingroup \@sanitize@url \@href}%
\providecommand \@href[1]{\@@startlink{#1}\@@href}%
\providecommand \@@href[1]{\endgroup#1\@@endlink}%
\providecommand \@sanitize@url [0]{\catcode `\\12\catcode `\$12\catcode
  `\&12\catcode `\#12\catcode `\^12\catcode `\_12\catcode `\%12\relax}%
\providecommand \@@startlink[1]{}%
\providecommand \@@endlink[0]{}%
\providecommand \url  [0]{\begingroup\@sanitize@url \@url }%
\providecommand \@url [1]{\endgroup\@href {#1}{\urlprefix }}%
\providecommand \urlprefix  [0]{URL }%
\providecommand \Eprint [0]{\href }%
\providecommand \doibase [0]{http://dx.doi.org/}%
\providecommand \selectlanguage [0]{\@gobble}%
\providecommand \bibinfo  [0]{\@secondoftwo}%
\providecommand \bibfield  [0]{\@secondoftwo}%
\providecommand \translation [1]{[#1]}%
\providecommand \BibitemOpen [0]{}%
\providecommand \bibitemStop [0]{}%
\providecommand \bibitemNoStop [0]{.\EOS\space}%
\providecommand \EOS [0]{\spacefactor3000\relax}%
\providecommand \BibitemShut  [1]{\csname bibitem#1\endcsname}%
\let\auto@bib@innerbib\@empty
%</preamble>
\bibitem [{\citenamefont {Majorana}(1937)}]{Majorana1937a}%
  \BibitemOpen
  \bibfield  {author} {\bibinfo {author} {\bibfnamefont {E.}~\bibnamefont
  {Majorana}},\ }\href {\doibase 10.1007/BF02961314} {\bibfield  {journal}
  {\bibinfo  {journal} {Nuovo Cimento}\ }\textbf {\bibinfo {volume} {14}},\
  \bibinfo {pages} {171} (\bibinfo {year} {1937})}\BibitemShut {NoStop}%
\bibitem [{\citenamefont {Read}\ and\ \citenamefont {Green}(2000)}]{Read2000a}%
  \BibitemOpen
  \bibfield  {author} {\bibinfo {author} {\bibfnamefont {N.}~\bibnamefont
  {Read}}\ and\ \bibinfo {author} {\bibfnamefont {D.}~\bibnamefont {Green}},\
  }\href {\doibase 10.1103/PhysRevB.61.10267} {\bibfield  {journal} {\bibinfo
  {journal} {Phys. Rev. B}\ }\textbf {\bibinfo {volume} {61}},\ \bibinfo
  {pages} {10267} (\bibinfo {year} {2000})}\BibitemShut {NoStop}%
\bibitem [{\citenamefont {Ivanov}(2001)}]{Ivanov2001a}%
  \BibitemOpen
  \bibfield  {author} {\bibinfo {author} {\bibfnamefont {D.~A.}\ \bibnamefont
  {Ivanov}},\ }\href {\doibase 10.1103/PhysRevLett.86.268} {\bibfield
  {journal} {\bibinfo  {journal} {Phys. Rev. Lett.}\ }\textbf {\bibinfo
  {volume} {86}},\ \bibinfo {pages} {268} (\bibinfo {year} {2001})}\BibitemShut
  {NoStop}%
\bibitem [{\citenamefont {Kitaev}(2001)}]{Kitaev2001a}%
  \BibitemOpen
  \bibfield  {author} {\bibinfo {author} {\bibfnamefont {A.~Y.}\ \bibnamefont
  {Kitaev}},\ }\href {http://stacks.iop.org/1063-7869/44/i=10S/a=S29}
  {\bibfield  {journal} {\bibinfo  {journal} {Physics-Uspekhi}\ }\textbf
  {\bibinfo {volume} {44}},\ \bibinfo {pages} {131} (\bibinfo {year}
  {2001})}\BibitemShut {NoStop}%
\bibitem [{\citenamefont {Fu}\ and\ \citenamefont {Kane}(2009)}]{Fu2009a}%
  \BibitemOpen
  \bibfield  {author} {\bibinfo {author} {\bibfnamefont {L.}~\bibnamefont
  {Fu}}\ and\ \bibinfo {author} {\bibfnamefont {C.~L.}\ \bibnamefont {Kane}},\
  }\href {\doibase 10.1103/PhysRevB.79.161408} {\bibfield  {journal} {\bibinfo
  {journal} {Phys. Rev. B}\ }\textbf {\bibinfo {volume} {79}},\ \bibinfo
  {pages} {161408} (\bibinfo {year} {2009})}\BibitemShut {NoStop}%
\bibitem [{\citenamefont {Lutchyn}\ \emph {et~al.}(2010)\citenamefont
  {Lutchyn}, \citenamefont {Sau},\ and\ \citenamefont
  {Das~Sarma}}]{Lutchyn2010a}%
  \BibitemOpen
  \bibfield  {author} {\bibinfo {author} {\bibfnamefont {R.~M.}\ \bibnamefont
  {Lutchyn}}, \bibinfo {author} {\bibfnamefont {J.~D.}\ \bibnamefont {Sau}}, \
  and\ \bibinfo {author} {\bibfnamefont {S.}~\bibnamefont {Das~Sarma}},\ }\href
  {\doibase 10.1103/PhysRevLett.105.077001} {\bibfield  {journal} {\bibinfo
  {journal} {Phys. Rev. Lett.}\ }\textbf {\bibinfo {volume} {105}},\ \bibinfo
  {pages} {077001} (\bibinfo {year} {2010})}\BibitemShut {NoStop}%
\bibitem [{\citenamefont {Oreg}\ \emph {et~al.}(2010)\citenamefont {Oreg},
  \citenamefont {Refael},\ and\ \citenamefont {von Oppen}}]{Oreg2010a}%
  \BibitemOpen
  \bibfield  {author} {\bibinfo {author} {\bibfnamefont {Y.}~\bibnamefont
  {Oreg}}, \bibinfo {author} {\bibfnamefont {G.}~\bibnamefont {Refael}}, \ and\
  \bibinfo {author} {\bibfnamefont {F.}~\bibnamefont {von Oppen}},\ }\href
  {\doibase 10.1103/PhysRevLett.105.177002} {\bibfield  {journal} {\bibinfo
  {journal} {Phys. Rev. Lett.}\ }\textbf {\bibinfo {volume} {105}},\ \bibinfo
  {pages} {177002} (\bibinfo {year} {2010})}\BibitemShut {NoStop}%
\bibitem [{\citenamefont {Law}\ \emph {et~al.}(2009)\citenamefont {Law},
  \citenamefont {Lee},\ and\ \citenamefont {Ng}}]{Law2009a}%
  \BibitemOpen
  \bibfield  {author} {\bibinfo {author} {\bibfnamefont {K.~T.}\ \bibnamefont
  {Law}}, \bibinfo {author} {\bibfnamefont {P.~A.}\ \bibnamefont {Lee}}, \ and\
  \bibinfo {author} {\bibfnamefont {T.~K.}\ \bibnamefont {Ng}},\ }\href
  {\doibase 10.1103/PhysRevLett.103.237001} {\bibfield  {journal} {\bibinfo
  {journal} {Phys. Rev. Lett.}\ }\textbf {\bibinfo {volume} {103}},\ \bibinfo
  {pages} {237001} (\bibinfo {year} {2009})}\BibitemShut {NoStop}%
\bibitem [{\citenamefont {Nayak}\ \emph {et~al.}(2008)\citenamefont {Nayak},
  \citenamefont {Simon}, \citenamefont {Stern}, \citenamefont {Freedman},\ and\
  \citenamefont {Das~Sarma}}]{Nayak2008a}%
  \BibitemOpen
  \bibfield  {author} {\bibinfo {author} {\bibfnamefont {C.}~\bibnamefont
  {Nayak}}, \bibinfo {author} {\bibfnamefont {S.~H.}\ \bibnamefont {Simon}},
  \bibinfo {author} {\bibfnamefont {A.}~\bibnamefont {Stern}}, \bibinfo
  {author} {\bibfnamefont {M.}~\bibnamefont {Freedman}}, \ and\ \bibinfo
  {author} {\bibfnamefont {S.}~\bibnamefont {Das~Sarma}},\ }\href {\doibase
  10.1103/RevModPhys.80.1083} {\bibfield  {journal} {\bibinfo  {journal} {Rev.
  Mod. Phys.}\ }\textbf {\bibinfo {volume} {80}},\ \bibinfo {pages} {1083}
  (\bibinfo {year} {2008})}\BibitemShut {NoStop}%
\bibitem [{\citenamefont {Alicea}(2012)}]{Alicea2012a}%
  \BibitemOpen
  \bibfield  {author} {\bibinfo {author} {\bibfnamefont {J.}~\bibnamefont
  {Alicea}},\ }\href@noop {} {\bibfield  {journal} {\bibinfo  {journal} {Rep.
  Prog. Phys.}\ }\textbf {\bibinfo {volume} {75}},\ \bibinfo {pages} {076501}
  (\bibinfo {year} {2012})}\BibitemShut {NoStop}%
\bibitem [{\citenamefont {Beenakker}(2013)}]{Beenakker2013a}%
  \BibitemOpen
  \bibfield  {author} {\bibinfo {author} {\bibfnamefont {C.}~\bibnamefont
  {Beenakker}},\ }\href {\doibase 10.1146/annurev-conmatphys-030212-184337}
  {\bibfield  {journal} {\bibinfo  {journal} {Annual Review of Condensed Matter
  Physics}\ }\textbf {\bibinfo {volume} {4}},\ \bibinfo {pages} {113} (\bibinfo
  {year} {2013})}\BibitemShut {NoStop}%
\bibitem [{\citenamefont {Aguado}(2017)}]{Aguado2017a}%
  \BibitemOpen
  \bibfield  {author} {\bibinfo {author} {\bibfnamefont {R.}~\bibnamefont
  {Aguado}},\ }\href {\doibase 10.1393/ncr/i2017-10141-9} {\bibfield  {journal}
  {\bibinfo  {journal} {Riv. Nuovo Cimento}\ }\textbf {\bibinfo {volume}
  {40}},\ \bibinfo {pages} {523} (\bibinfo {year} {2017})}\BibitemShut
  {NoStop}%
\bibitem [{\citenamefont {Schuray}\ \emph {et~al.}(2020)\citenamefont
  {Schuray}, \citenamefont {Frombach}, \citenamefont {Park},\ and\
  \citenamefont {Recher}}]{Schuray2020a}%
  \BibitemOpen
  \bibfield  {author} {\bibinfo {author} {\bibfnamefont {A.}~\bibnamefont
  {Schuray}}, \bibinfo {author} {\bibfnamefont {D.}~\bibnamefont {Frombach}},
  \bibinfo {author} {\bibfnamefont {S.}~\bibnamefont {Park}}, \ and\ \bibinfo
  {author} {\bibfnamefont {P.}~\bibnamefont {Recher}},\ }\href {\doibase
  10.1140/epjst/e2019-900150-7} {\bibfield  {journal} {\bibinfo  {journal} {The
  European Physical Journal Special Topics}\ }\textbf {\bibinfo {volume}
  {229}},\ \bibinfo {pages} {593} (\bibinfo {year} {2020})}\BibitemShut
  {NoStop}%
\bibitem [{\citenamefont {Prada}\ \emph {et~al.}(2020)\citenamefont {Prada},
  \citenamefont {San-Jose}, \citenamefont {de~Moor}, \citenamefont {Geresdi},
  \citenamefont {Lee}, \citenamefont {Klinovaja}, \citenamefont {Loss},
  \citenamefont {Nygard}, \citenamefont {Aguado},\ and\ \citenamefont
  {Kouwenhoven}}]{Prada2020a}%
  \BibitemOpen
  \bibfield  {author} {\bibinfo {author} {\bibfnamefont {E.}~\bibnamefont
  {Prada}}, \bibinfo {author} {\bibfnamefont {P.}~\bibnamefont {San-Jose}},
  \bibinfo {author} {\bibfnamefont {M.~W.~A.}\ \bibnamefont {de~Moor}},
  \bibinfo {author} {\bibfnamefont {A.}~\bibnamefont {Geresdi}}, \bibinfo
  {author} {\bibfnamefont {E.~J.~H.}\ \bibnamefont {Lee}}, \bibinfo {author}
  {\bibfnamefont {J.}~\bibnamefont {Klinovaja}}, \bibinfo {author}
  {\bibfnamefont {D.}~\bibnamefont {Loss}}, \bibinfo {author} {\bibfnamefont
  {J.}~\bibnamefont {Nygard}}, \bibinfo {author} {\bibfnamefont
  {R.}~\bibnamefont {Aguado}}, \ and\ \bibinfo {author} {\bibfnamefont {L.~P.}\
  \bibnamefont {Kouwenhoven}},\ }\href {\doibase 10.1038/s42254-020-0228-y}
  {\bibfield  {journal} {\bibinfo  {journal} {Nat. Rev. Phys.}\ }\textbf
  {\bibinfo {volume} {2}},\ \bibinfo {pages} {575} (\bibinfo {year}
  {2020})}\BibitemShut {NoStop}%
\bibitem [{\citenamefont {Flensberg}\ \emph {et~al.}(2021)\citenamefont
  {Flensberg}, \citenamefont {von Oppen},\ and\ \citenamefont
  {Stern}}]{Flensberg2021a}%
  \BibitemOpen
  \bibfield  {author} {\bibinfo {author} {\bibfnamefont {K.}~\bibnamefont
  {Flensberg}}, \bibinfo {author} {\bibfnamefont {F.}~\bibnamefont {von
  Oppen}}, \ and\ \bibinfo {author} {\bibfnamefont {A.}~\bibnamefont {Stern}},\
  }\href {\doibase 10.1038/s41578-021-00336-6} {\bibfield  {journal} {\bibinfo
  {journal} {Nat. Rev. Mat.}\ }\textbf {\bibinfo {volume} {6}},\ \bibinfo
  {pages} {944} (\bibinfo {year} {2021})}\BibitemShut {NoStop}%
\bibitem [{\citenamefont {Prada}\ \emph {et~al.}(2012)\citenamefont {Prada},
  \citenamefont {San-Jose},\ and\ \citenamefont {Aguado}}]{Prada2012a}%
  \BibitemOpen
  \bibfield  {author} {\bibinfo {author} {\bibfnamefont {E.}~\bibnamefont
  {Prada}}, \bibinfo {author} {\bibfnamefont {P.}~\bibnamefont {San-Jose}}, \
  and\ \bibinfo {author} {\bibfnamefont {R.}~\bibnamefont {Aguado}},\ }\href
  {\doibase 10.1103/PhysRevB.86.180503} {\bibfield  {journal} {\bibinfo
  {journal} {Phys. Rev. B}\ }\textbf {\bibinfo {volume} {86}},\ \bibinfo
  {pages} {180503} (\bibinfo {year} {2012})}\BibitemShut {NoStop}%
\bibitem [{\citenamefont {Li}\ \emph {et~al.}(2012)\citenamefont {Li},
  \citenamefont {Fleury},\ and\ \citenamefont {B\"uttiker}}]{Li2012a}%
  \BibitemOpen
  \bibfield  {author} {\bibinfo {author} {\bibfnamefont {J.}~\bibnamefont
  {Li}}, \bibinfo {author} {\bibfnamefont {G.}~\bibnamefont {Fleury}}, \ and\
  \bibinfo {author} {\bibfnamefont {M.}~\bibnamefont {B\"uttiker}},\ }\href
  {\doibase 10.1103/PhysRevB.85.125440} {\bibfield  {journal} {\bibinfo
  {journal} {Phys. Rev. B}\ }\textbf {\bibinfo {volume} {85}},\ \bibinfo
  {pages} {125440} (\bibinfo {year} {2012})}\BibitemShut {NoStop}%
\bibitem [{\citenamefont {Mourik}\ \emph {et~al.}(2012)\citenamefont {Mourik},
  \citenamefont {Zuo}, \citenamefont {Frolov}, \citenamefont {Plissard},
  \citenamefont {Bakkers},\ and\ \citenamefont {Kouwenhoven}}]{Mourik2012a}%
  \BibitemOpen
  \bibfield  {author} {\bibinfo {author} {\bibfnamefont {V.}~\bibnamefont
  {Mourik}}, \bibinfo {author} {\bibfnamefont {K.}~\bibnamefont {Zuo}},
  \bibinfo {author} {\bibfnamefont {S.~M.}\ \bibnamefont {Frolov}}, \bibinfo
  {author} {\bibfnamefont {S.~R.}\ \bibnamefont {Plissard}}, \bibinfo {author}
  {\bibfnamefont {E.~P. A.~M.}\ \bibnamefont {Bakkers}}, \ and\ \bibinfo
  {author} {\bibfnamefont {L.~P.}\ \bibnamefont {Kouwenhoven}},\ }\href
  {\doibase 10.1126/science.1222360} {\bibfield  {journal} {\bibinfo  {journal}
  {Science}\ }\textbf {\bibinfo {volume} {336}},\ \bibinfo {pages} {1003}
  (\bibinfo {year} {2012})}\BibitemShut {NoStop}%
\bibitem [{\citenamefont {Deng}\ \emph {et~al.}(2012)\citenamefont {Deng},
  \citenamefont {Yu}, \citenamefont {Huang}, \citenamefont {Larsson},
  \citenamefont {Caroff},\ and\ \citenamefont {Xu}}]{Deng2012a}%
  \BibitemOpen
  \bibfield  {author} {\bibinfo {author} {\bibfnamefont {M.~T.}\ \bibnamefont
  {Deng}}, \bibinfo {author} {\bibfnamefont {C.~L.}\ \bibnamefont {Yu}},
  \bibinfo {author} {\bibfnamefont {G.~Y.}\ \bibnamefont {Huang}}, \bibinfo
  {author} {\bibfnamefont {M.}~\bibnamefont {Larsson}}, \bibinfo {author}
  {\bibfnamefont {P.}~\bibnamefont {Caroff}}, \ and\ \bibinfo {author}
  {\bibfnamefont {H.~Q.}\ \bibnamefont {Xu}},\ }\href {\doibase
  10.1021/nl303758w} {\bibfield  {journal} {\bibinfo  {journal} {Nano Letters}\
  }\textbf {\bibinfo {volume} {12}},\ \bibinfo {pages} {6414} (\bibinfo {year}
  {2012})}\BibitemShut {NoStop}%
\bibitem [{\citenamefont {Das}\ \emph {et~al.}(2012)\citenamefont {Das},
  \citenamefont {Ronen}, \citenamefont {Most}, \citenamefont {Oreg},
  \citenamefont {Heiblum},\ and\ \citenamefont {Shtrikman}}]{Das2012a}%
  \BibitemOpen
  \bibfield  {author} {\bibinfo {author} {\bibfnamefont {A.}~\bibnamefont
  {Das}}, \bibinfo {author} {\bibfnamefont {Y.}~\bibnamefont {Ronen}}, \bibinfo
  {author} {\bibfnamefont {Y.}~\bibnamefont {Most}}, \bibinfo {author}
  {\bibfnamefont {Y.}~\bibnamefont {Oreg}}, \bibinfo {author} {\bibfnamefont
  {M.}~\bibnamefont {Heiblum}}, \ and\ \bibinfo {author} {\bibfnamefont
  {H.}~\bibnamefont {Shtrikman}},\ }\href {\doibase 10.1038/nphys2479}
  {\bibfield  {journal} {\bibinfo  {journal} {Nat. Phys.}\ }\textbf {\bibinfo
  {volume} {8}},\ \bibinfo {pages} {887} (\bibinfo {year} {2012})}\BibitemShut
  {NoStop}%
\bibitem [{\citenamefont {Nichele}\ \emph {et~al.}(2017)\citenamefont
  {Nichele}, \citenamefont {Drachmann}, \citenamefont {Whiticar}, \citenamefont
  {O'Farrell}, \citenamefont {Suominen}, \citenamefont {Fornieri},
  \citenamefont {Wang}, \citenamefont {Gardner}, \citenamefont {Thomas},
  \citenamefont {Hatke}, \citenamefont {Krogstrup}, \citenamefont {Manfra},
  \citenamefont {Flensberg},\ and\ \citenamefont {Marcus}}]{Nichele2017a}%
  \BibitemOpen
  \bibfield  {author} {\bibinfo {author} {\bibfnamefont {F.}~\bibnamefont
  {Nichele}}, \bibinfo {author} {\bibfnamefont {A.~C.~C.}\ \bibnamefont
  {Drachmann}}, \bibinfo {author} {\bibfnamefont {A.~M.}\ \bibnamefont
  {Whiticar}}, \bibinfo {author} {\bibfnamefont {E.~C.~T.}\ \bibnamefont
  {O'Farrell}}, \bibinfo {author} {\bibfnamefont {H.~J.}\ \bibnamefont
  {Suominen}}, \bibinfo {author} {\bibfnamefont {A.}~\bibnamefont {Fornieri}},
  \bibinfo {author} {\bibfnamefont {T.}~\bibnamefont {Wang}}, \bibinfo {author}
  {\bibfnamefont {G.~C.}\ \bibnamefont {Gardner}}, \bibinfo {author}
  {\bibfnamefont {C.}~\bibnamefont {Thomas}}, \bibinfo {author} {\bibfnamefont
  {A.~T.}\ \bibnamefont {Hatke}}, \bibinfo {author} {\bibfnamefont
  {P.}~\bibnamefont {Krogstrup}}, \bibinfo {author} {\bibfnamefont {M.~J.}\
  \bibnamefont {Manfra}}, \bibinfo {author} {\bibfnamefont {K.}~\bibnamefont
  {Flensberg}}, \ and\ \bibinfo {author} {\bibfnamefont {C.~M.}\ \bibnamefont
  {Marcus}},\ }\href {\doibase 10.1103/PhysRevLett.119.136803} {\bibfield
  {journal} {\bibinfo  {journal} {Phys. Rev. Lett.}\ }\textbf {\bibinfo
  {volume} {119}},\ \bibinfo {pages} {136803} (\bibinfo {year}
  {2017})}\BibitemShut {NoStop}%
\bibitem [{\citenamefont {Zhang}\ \emph {et~al.}(2021)\citenamefont {Zhang},
  \citenamefont {de~Moor}, \citenamefont {Bommer}, \citenamefont {Xu},
  \citenamefont {Wang}, \citenamefont {van Loo}, \citenamefont {Liu},
  \citenamefont {Gazibegovic}, \citenamefont {Logan}, \citenamefont {Car},
  \citenamefont {het Veld}, \citenamefont {van Veldhoven}, \citenamefont
  {Koelling}, \citenamefont {Verheijen}, \citenamefont {Pendharkar},
  \citenamefont {Pennachio}, \citenamefont {Shojaei}, \citenamefont {Lee},
  \citenamefont {Palmstr\"om}, \citenamefont {Bakkers}, \citenamefont
  {Das~Sarma},\ and\ \citenamefont {Kouwenhoven}}]{Hao2021a}%
  \BibitemOpen
  \bibfield  {author} {\bibinfo {author} {\bibfnamefont {H.}~\bibnamefont
  {Zhang}}, \bibinfo {author} {\bibfnamefont {M.~W.~A.}\ \bibnamefont
  {de~Moor}}, \bibinfo {author} {\bibfnamefont {J.~D.~S.}\ \bibnamefont
  {Bommer}}, \bibinfo {author} {\bibfnamefont {D.}~\bibnamefont {Xu}}, \bibinfo
  {author} {\bibfnamefont {G.}~\bibnamefont {Wang}}, \bibinfo {author}
  {\bibfnamefont {N.}~\bibnamefont {van Loo}}, \bibinfo {author} {\bibfnamefont
  {C.-X.}\ \bibnamefont {Liu}}, \bibinfo {author} {\bibfnamefont
  {S.}~\bibnamefont {Gazibegovic}}, \bibinfo {author} {\bibfnamefont {J.~A.}\
  \bibnamefont {Logan}}, \bibinfo {author} {\bibfnamefont {D.}~\bibnamefont
  {Car}}, \bibinfo {author} {\bibfnamefont {R.~L.~M.~O.}\ \bibnamefont {het
  Veld}}, \bibinfo {author} {\bibfnamefont {P.~J.}\ \bibnamefont {van
  Veldhoven}}, \bibinfo {author} {\bibfnamefont {S.}~\bibnamefont {Koelling}},
  \bibinfo {author} {\bibfnamefont {M.~A.}\ \bibnamefont {Verheijen}}, \bibinfo
  {author} {\bibfnamefont {M.}~\bibnamefont {Pendharkar}}, \bibinfo {author}
  {\bibfnamefont {D.~J.}\ \bibnamefont {Pennachio}}, \bibinfo {author}
  {\bibfnamefont {B.}~\bibnamefont {Shojaei}}, \bibinfo {author} {\bibfnamefont
  {J.~S.}\ \bibnamefont {Lee}}, \bibinfo {author} {\bibfnamefont {C.~J.}\
  \bibnamefont {Palmstr\"om}}, \bibinfo {author} {\bibfnamefont {E.~P.~A.~M.}\
  \bibnamefont {Bakkers}}, \bibinfo {author} {\bibfnamefont {S.}~\bibnamefont
  {Das~Sarma}}, \ and\ \bibinfo {author} {\bibfnamefont {L.~P.}\ \bibnamefont
  {Kouwenhoven}},\ }\href@noop {} {\bibfield  {journal} {\bibinfo  {journal}
  {arXiv e-prints}\ ,\ \bibinfo {eid} {arXiv:2101.11456}} (\bibinfo {year}
  {2021})},\ \Eprint {http://arxiv.org/abs/2101.11456} {arXiv:2101.11456
  [cond-mat.mes-hall]} \BibitemShut {NoStop}%
\bibitem [{\citenamefont {Fleckenstein}\ \emph {et~al.}(2018)\citenamefont
  {Fleckenstein}, \citenamefont {Dom\'{\i}nguez}, \citenamefont
  {Traverso~Ziani},\ and\ \citenamefont {Trauzettel}}]{Fleckenstein2018a}%
  \BibitemOpen
  \bibfield  {author} {\bibinfo {author} {\bibfnamefont {C.}~\bibnamefont
  {Fleckenstein}}, \bibinfo {author} {\bibfnamefont {F.}~\bibnamefont
  {Dom\'{\i}nguez}}, \bibinfo {author} {\bibfnamefont {N.}~\bibnamefont
  {Traverso~Ziani}}, \ and\ \bibinfo {author} {\bibfnamefont {B.}~\bibnamefont
  {Trauzettel}},\ }\href {\doibase 10.1103/PhysRevB.97.155425} {\bibfield
  {journal} {\bibinfo  {journal} {Phys. Rev. B}\ }\textbf {\bibinfo {volume}
  {97}},\ \bibinfo {pages} {155425} (\bibinfo {year} {2018})}\BibitemShut
  {NoStop}%
\bibitem [{\citenamefont {Moore}\ \emph {et~al.}(2018)\citenamefont {Moore},
  \citenamefont {Zeng}, \citenamefont {Stanescu},\ and\ \citenamefont
  {Tewari}}]{Moore2018a}%
  \BibitemOpen
  \bibfield  {author} {\bibinfo {author} {\bibfnamefont {C.}~\bibnamefont
  {Moore}}, \bibinfo {author} {\bibfnamefont {C.}~\bibnamefont {Zeng}},
  \bibinfo {author} {\bibfnamefont {T.~D.}\ \bibnamefont {Stanescu}}, \ and\
  \bibinfo {author} {\bibfnamefont {S.}~\bibnamefont {Tewari}},\ }\href
  {\doibase 10.1103/PhysRevB.98.155314} {\bibfield  {journal} {\bibinfo
  {journal} {Phys. Rev. B}\ }\textbf {\bibinfo {volume} {98}},\ \bibinfo
  {pages} {155314} (\bibinfo {year} {2018})}\BibitemShut {NoStop}%
\bibitem [{\citenamefont {Marra}\ and\ \citenamefont
  {Nitta}(2019)}]{Marra2019a}%
  \BibitemOpen
  \bibfield  {author} {\bibinfo {author} {\bibfnamefont {P.}~\bibnamefont
  {Marra}}\ and\ \bibinfo {author} {\bibfnamefont {M.}~\bibnamefont {Nitta}},\
  }\href {\doibase 10.1103/PhysRevB.100.220502} {\bibfield  {journal} {\bibinfo
   {journal} {Phys. Rev. B}\ }\textbf {\bibinfo {volume} {100}},\ \bibinfo
  {pages} {220502} (\bibinfo {year} {2019})}\BibitemShut {NoStop}%
\bibitem [{\citenamefont {Cayao}\ \emph {et~al.}(2017)\citenamefont {Cayao},
  \citenamefont {San-Jose}, \citenamefont {Black-Schaffer}, \citenamefont
  {Aguado},\ and\ \citenamefont {Prada}}]{Cayao2017a}%
  \BibitemOpen
  \bibfield  {author} {\bibinfo {author} {\bibfnamefont {J.}~\bibnamefont
  {Cayao}}, \bibinfo {author} {\bibfnamefont {P.}~\bibnamefont {San-Jose}},
  \bibinfo {author} {\bibfnamefont {A.~M.}\ \bibnamefont {Black-Schaffer}},
  \bibinfo {author} {\bibfnamefont {R.}~\bibnamefont {Aguado}}, \ and\ \bibinfo
  {author} {\bibfnamefont {E.}~\bibnamefont {Prada}},\ }\href {\doibase
  10.1103/PhysRevB.96.205425} {\bibfield  {journal} {\bibinfo  {journal} {Phys.
  Rev. B}\ }\textbf {\bibinfo {volume} {96}},\ \bibinfo {pages} {205425}
  (\bibinfo {year} {2017})}\BibitemShut {NoStop}%
\bibitem [{\citenamefont {Awoga}\ \emph {et~al.}(2019)\citenamefont {Awoga},
  \citenamefont {Cayao},\ and\ \citenamefont
  {Black-Schaffer}}]{Oladunjoye2019a}%
  \BibitemOpen
  \bibfield  {author} {\bibinfo {author} {\bibfnamefont {O.~A.}\ \bibnamefont
  {Awoga}}, \bibinfo {author} {\bibfnamefont {J.}~\bibnamefont {Cayao}}, \ and\
  \bibinfo {author} {\bibfnamefont {A.~M.}\ \bibnamefont {Black-Schaffer}},\
  }\href {\doibase 10.1103/PhysRevLett.123.117001} {\bibfield  {journal}
  {\bibinfo  {journal} {Phys. Rev. Lett.}\ }\textbf {\bibinfo {volume} {123}},\
  \bibinfo {pages} {117001} (\bibinfo {year} {2019})}\BibitemShut {NoStop}%
\bibitem [{\citenamefont {van Heck}\ \emph {et~al.}(2011)\citenamefont {van
  Heck}, \citenamefont {Hassler}, \citenamefont {Akhmerov},\ and\ \citenamefont
  {Beenakker}}]{Heck2011a}%
  \BibitemOpen
  \bibfield  {author} {\bibinfo {author} {\bibfnamefont {B.}~\bibnamefont {van
  Heck}}, \bibinfo {author} {\bibfnamefont {F.}~\bibnamefont {Hassler}},
  \bibinfo {author} {\bibfnamefont {A.~R.}\ \bibnamefont {Akhmerov}}, \ and\
  \bibinfo {author} {\bibfnamefont {C.~W.~J.}\ \bibnamefont {Beenakker}},\
  }\href {\doibase 10.1103/PhysRevB.84.180502} {\bibfield  {journal} {\bibinfo
  {journal} {Phys. Rev. B}\ }\textbf {\bibinfo {volume} {84}},\ \bibinfo
  {pages} {180502(R)} (\bibinfo {year} {2011})}\BibitemShut {NoStop}%
\bibitem [{\citenamefont {Badiane}\ \emph {et~al.}(2011)\citenamefont
  {Badiane}, \citenamefont {Houzet},\ and\ \citenamefont
  {Meyer}}]{Badiane2011a}%
  \BibitemOpen
  \bibfield  {author} {\bibinfo {author} {\bibfnamefont {D.~M.}\ \bibnamefont
  {Badiane}}, \bibinfo {author} {\bibfnamefont {M.}~\bibnamefont {Houzet}}, \
  and\ \bibinfo {author} {\bibfnamefont {J.~S.}\ \bibnamefont {Meyer}},\
  }\href@noop {} {\bibfield  {journal} {\bibinfo  {journal} {Phys. Rev. Lett.}\
  }\textbf {\bibinfo {volume} {107}},\ \bibinfo {pages} {177002} (\bibinfo
  {year} {2011})}\BibitemShut {NoStop}%
\bibitem [{\citenamefont {San-Jose}\ \emph {et~al.}(2012)\citenamefont
  {San-Jose}, \citenamefont {Prada},\ and\ \citenamefont
  {Aguado}}]{San-Jose2012a}%
  \BibitemOpen
  \bibfield  {author} {\bibinfo {author} {\bibfnamefont {P.}~\bibnamefont
  {San-Jose}}, \bibinfo {author} {\bibfnamefont {E.}~\bibnamefont {Prada}}, \
  and\ \bibinfo {author} {\bibfnamefont {R.}~\bibnamefont {Aguado}},\ }\href
  {\doibase 10.1103/PhysRevLett.108.257001} {\bibfield  {journal} {\bibinfo
  {journal} {Phys. Rev. Lett.}\ }\textbf {\bibinfo {volume} {108}},\ \bibinfo
  {pages} {257001} (\bibinfo {year} {2012})}\BibitemShut {NoStop}%
\bibitem [{\citenamefont {Dom\'inguez}\ \emph {et~al.}(2012)\citenamefont
  {Dom\'inguez}, \citenamefont {Hassler},\ and\ \citenamefont
  {Platero}}]{Dominguez2012a}%
  \BibitemOpen
  \bibfield  {author} {\bibinfo {author} {\bibfnamefont {F.}~\bibnamefont
  {Dom\'inguez}}, \bibinfo {author} {\bibfnamefont {F.}~\bibnamefont
  {Hassler}}, \ and\ \bibinfo {author} {\bibfnamefont {G.}~\bibnamefont
  {Platero}},\ }\href {\doibase 10.1103/PhysRevB.86.140503} {\bibfield
  {journal} {\bibinfo  {journal} {Phys. Rev. B}\ }\textbf {\bibinfo {volume}
  {86}},\ \bibinfo {pages} {140503} (\bibinfo {year} {2012})}\BibitemShut
  {NoStop}%
\bibitem [{\citenamefont {Pikulin}\ and\ \citenamefont
  {Nazarov}(2012)}]{Pikulin2012a}%
  \BibitemOpen
  \bibfield  {author} {\bibinfo {author} {\bibfnamefont {D.~I.}\ \bibnamefont
  {Pikulin}}\ and\ \bibinfo {author} {\bibfnamefont {Y.~V.}\ \bibnamefont
  {Nazarov}},\ }\href@noop {} {\bibfield  {journal} {\bibinfo  {journal} {Phys.
  Rev. B}\ }\textbf {\bibinfo {volume} {86}},\ \bibinfo {pages} {140504(R)}
  (\bibinfo {year} {2012})}\BibitemShut {NoStop}%
\bibitem [{\citenamefont {Deacon}\ \emph {et~al.}(2017)\citenamefont {Deacon},
  \citenamefont {Wiedenmann}, \citenamefont {Bocquillon}, \citenamefont
  {Dom\'{\i}nguez}, \citenamefont {Klapwijk}, \citenamefont {Leubner},
  \citenamefont {Br\"une}, \citenamefont {Hankiewicz}, \citenamefont {Tarucha},
  \citenamefont {Ishibashi}, \citenamefont {Buhmann},\ and\ \citenamefont
  {Molenkamp}}]{Deacon2017a}%
  \BibitemOpen
  \bibfield  {author} {\bibinfo {author} {\bibfnamefont {R.~S.}\ \bibnamefont
  {Deacon}}, \bibinfo {author} {\bibfnamefont {J.}~\bibnamefont {Wiedenmann}},
  \bibinfo {author} {\bibfnamefont {E.}~\bibnamefont {Bocquillon}}, \bibinfo
  {author} {\bibfnamefont {F.}~\bibnamefont {Dom\'{\i}nguez}}, \bibinfo
  {author} {\bibfnamefont {T.~M.}\ \bibnamefont {Klapwijk}}, \bibinfo {author}
  {\bibfnamefont {P.}~\bibnamefont {Leubner}}, \bibinfo {author} {\bibfnamefont
  {C.}~\bibnamefont {Br\"une}}, \bibinfo {author} {\bibfnamefont {E.~M.}\
  \bibnamefont {Hankiewicz}}, \bibinfo {author} {\bibfnamefont
  {S.}~\bibnamefont {Tarucha}}, \bibinfo {author} {\bibfnamefont
  {K.}~\bibnamefont {Ishibashi}}, \bibinfo {author} {\bibfnamefont
  {H.}~\bibnamefont {Buhmann}}, \ and\ \bibinfo {author} {\bibfnamefont
  {L.~W.}\ \bibnamefont {Molenkamp}},\ }\href {\doibase
  10.1103/PhysRevX.7.021011} {\bibfield  {journal} {\bibinfo  {journal} {Phys.
  Rev. X}\ }\textbf {\bibinfo {volume} {7}},\ \bibinfo {pages} {021011}
  (\bibinfo {year} {2017})}\BibitemShut {NoStop}%
\bibitem [{\citenamefont {Laroche}\ \emph {et~al.}(2019)\citenamefont
  {Laroche}, \citenamefont {Bouman}, \citenamefont {van Woerkom}, \citenamefont
  {Proutski}, \citenamefont {Murthy}, \citenamefont {Pikulin}, \citenamefont
  {Nayak}, \citenamefont {van Gulik}, \citenamefont {Nygard}, \citenamefont
  {Krogstrup}, \citenamefont {Kouwenhoven},\ and\ \citenamefont
  {Geresdi}}]{Laroche2019a}%
  \BibitemOpen
  \bibfield  {author} {\bibinfo {author} {\bibfnamefont {D.}~\bibnamefont
  {Laroche}}, \bibinfo {author} {\bibfnamefont {D.}~\bibnamefont {Bouman}},
  \bibinfo {author} {\bibfnamefont {D.~J.}\ \bibnamefont {van Woerkom}},
  \bibinfo {author} {\bibfnamefont {A.}~\bibnamefont {Proutski}}, \bibinfo
  {author} {\bibfnamefont {C.}~\bibnamefont {Murthy}}, \bibinfo {author}
  {\bibfnamefont {D.~I.}\ \bibnamefont {Pikulin}}, \bibinfo {author}
  {\bibfnamefont {C.}~\bibnamefont {Nayak}}, \bibinfo {author} {\bibfnamefont
  {R.~J.~J.}\ \bibnamefont {van Gulik}}, \bibinfo {author} {\bibfnamefont
  {J.}~\bibnamefont {Nygard}}, \bibinfo {author} {\bibfnamefont
  {P.}~\bibnamefont {Krogstrup}}, \bibinfo {author} {\bibfnamefont {L.~P.}\
  \bibnamefont {Kouwenhoven}}, \ and\ \bibinfo {author} {\bibfnamefont
  {A.}~\bibnamefont {Geresdi}},\ }\href {\doibase 10.1038/s41467-018-08161-2}
  {\bibfield  {journal} {\bibinfo  {journal} {Nat. Comm.}\ }\textbf {\bibinfo
  {volume} {10}},\ \bibinfo {pages} {245} (\bibinfo {year} {2019})}\BibitemShut
  {NoStop}%
\bibitem [{\citenamefont {Rokhinson}\ \emph {et~al.}(2012)\citenamefont
  {Rokhinson}, \citenamefont {Liu},\ and\ \citenamefont
  {Furdyna}}]{Rokhinson2012a}%
  \BibitemOpen
  \bibfield  {author} {\bibinfo {author} {\bibfnamefont {L.~P.}\ \bibnamefont
  {Rokhinson}}, \bibinfo {author} {\bibfnamefont {X.}~\bibnamefont {Liu}}, \
  and\ \bibinfo {author} {\bibfnamefont {J.~K.}\ \bibnamefont {Furdyna}},\
  }\href {\doibase 10.1038/nphys2429} {\bibfield  {journal} {\bibinfo
  {journal} {Nat. Phys.}\ }\textbf {\bibinfo {volume} {1038}},\ \bibinfo
  {pages} {2429} (\bibinfo {year} {2012})}\BibitemShut {NoStop}%
\bibitem [{\citenamefont {Wiedenmann}\ \emph {et~al.}(2016)\citenamefont
  {Wiedenmann}, \citenamefont {Bocquillon}, \citenamefont {Deacon},
  \citenamefont {Hartinger}, \citenamefont {Herrmann}, \citenamefont
  {Klapwijk}, \citenamefont {Maier}, \citenamefont {Ames}, \citenamefont
  {Br{\"u}ne}, \citenamefont {Gould}, \citenamefont {Oiwa}, \citenamefont
  {Ishibashi}, \citenamefont {Tarucha}, \citenamefont {Buhmann},\ and\
  \citenamefont {Molenkamp}}]{Wiedenmann2016a}%
  \BibitemOpen
  \bibfield  {author} {\bibinfo {author} {\bibfnamefont {J.}~\bibnamefont
  {Wiedenmann}}, \bibinfo {author} {\bibfnamefont {E.}~\bibnamefont
  {Bocquillon}}, \bibinfo {author} {\bibfnamefont {R.~S.}\ \bibnamefont
  {Deacon}}, \bibinfo {author} {\bibfnamefont {S.}~\bibnamefont {Hartinger}},
  \bibinfo {author} {\bibfnamefont {O.}~\bibnamefont {Herrmann}}, \bibinfo
  {author} {\bibfnamefont {T.~M.}\ \bibnamefont {Klapwijk}}, \bibinfo {author}
  {\bibfnamefont {L.}~\bibnamefont {Maier}}, \bibinfo {author} {\bibfnamefont
  {C.}~\bibnamefont {Ames}}, \bibinfo {author} {\bibfnamefont {C.}~\bibnamefont
  {Br{\"u}ne}}, \bibinfo {author} {\bibfnamefont {C.}~\bibnamefont {Gould}},
  \bibinfo {author} {\bibfnamefont {A.}~\bibnamefont {Oiwa}}, \bibinfo {author}
  {\bibfnamefont {K.}~\bibnamefont {Ishibashi}}, \bibinfo {author}
  {\bibfnamefont {S.}~\bibnamefont {Tarucha}}, \bibinfo {author} {\bibfnamefont
  {H.}~\bibnamefont {Buhmann}}, \ and\ \bibinfo {author} {\bibfnamefont
  {L.~W.}\ \bibnamefont {Molenkamp}},\ }\href {\doibase 10.1038/ncomms10303}
  {\bibfield  {journal} {\bibinfo  {journal} {Nature Communications}\ }\textbf
  {\bibinfo {volume} {7}},\ \bibinfo {pages} {10303} (\bibinfo {year}
  {2016})}\BibitemShut {NoStop}%
\bibitem [{\citenamefont {Li}\ \emph {et~al.}(2018)\citenamefont {Li},
  \citenamefont {de~Boer}, \citenamefont {de~Ronde}, \citenamefont
  {Ramankutty}, \citenamefont {van Heumen}, \citenamefont {Huang},
  \citenamefont {de~Visser}, \citenamefont {Golubov}, \citenamefont {Golden},\
  and\ \citenamefont {Brinkman}}]{Li2018a}%
  \BibitemOpen
  \bibfield  {author} {\bibinfo {author} {\bibfnamefont {C.}~\bibnamefont
  {Li}}, \bibinfo {author} {\bibfnamefont {J.~C.}\ \bibnamefont {de~Boer}},
  \bibinfo {author} {\bibfnamefont {B.}~\bibnamefont {de~Ronde}}, \bibinfo
  {author} {\bibfnamefont {S.~V.}\ \bibnamefont {Ramankutty}}, \bibinfo
  {author} {\bibfnamefont {E.}~\bibnamefont {van Heumen}}, \bibinfo {author}
  {\bibfnamefont {Y.}~\bibnamefont {Huang}}, \bibinfo {author} {\bibfnamefont
  {A.}~\bibnamefont {de~Visser}}, \bibinfo {author} {\bibfnamefont {A.~A.}\
  \bibnamefont {Golubov}}, \bibinfo {author} {\bibfnamefont {M.~S.}\
  \bibnamefont {Golden}}, \ and\ \bibinfo {author} {\bibfnamefont
  {A.}~\bibnamefont {Brinkman}},\ }\href {\doibase 10.1038/s41563-018-0158-6}
  {\bibfield  {journal} {\bibinfo  {journal} {Nat. Materials}\ }\textbf
  {\bibinfo {volume} {17}},\ \bibinfo {pages} {880} (\bibinfo {year}
  {2018})}\BibitemShut {NoStop}%
\bibitem [{\citenamefont {Fischer}\ \emph {et~al.}(2022)\citenamefont
  {Fischer}, \citenamefont {Pic\'o-Cort\'es}, \citenamefont {Himmler},
  \citenamefont {Platero}, \citenamefont {Grifoni}, \citenamefont {Kozlov},
  \citenamefont {Mikhailov}, \citenamefont {Dvoretsky}, \citenamefont
  {Strunk},\ and\ \citenamefont {Weiss}}]{Fischer2022a}%
  \BibitemOpen
  \bibfield  {author} {\bibinfo {author} {\bibfnamefont {R.}~\bibnamefont
  {Fischer}}, \bibinfo {author} {\bibfnamefont {J.}~\bibnamefont
  {Pic\'o-Cort\'es}}, \bibinfo {author} {\bibfnamefont {W.}~\bibnamefont
  {Himmler}}, \bibinfo {author} {\bibfnamefont {G.}~\bibnamefont {Platero}},
  \bibinfo {author} {\bibfnamefont {M.}~\bibnamefont {Grifoni}}, \bibinfo
  {author} {\bibfnamefont {D.~A.}\ \bibnamefont {Kozlov}}, \bibinfo {author}
  {\bibfnamefont {N.~N.}\ \bibnamefont {Mikhailov}}, \bibinfo {author}
  {\bibfnamefont {S.~A.}\ \bibnamefont {Dvoretsky}}, \bibinfo {author}
  {\bibfnamefont {C.}~\bibnamefont {Strunk}}, \ and\ \bibinfo {author}
  {\bibfnamefont {D.}~\bibnamefont {Weiss}},\ }\href {\doibase
  10.1103/PhysRevResearch.4.013087} {\bibfield  {journal} {\bibinfo  {journal}
  {Phys. Rev. Research}\ }\textbf {\bibinfo {volume} {4}},\ \bibinfo {pages}
  {013087} (\bibinfo {year} {2022})}\BibitemShut {NoStop}%
\bibitem [{\citenamefont {Bocquillon}\ \emph {et~al.}(2016)\citenamefont
  {Bocquillon}, \citenamefont {Deacon}, \citenamefont {Wiedenmann},
  \citenamefont {P.}, \citenamefont {Klapwijk}, \citenamefont {Br{\"u}ne},
  \citenamefont {Ishibashi}, \citenamefont {Buhmann},\ and\ \citenamefont
  {Molenkamp}}]{Bocquillon2016a}%
  \BibitemOpen
  \bibfield  {author} {\bibinfo {author} {\bibfnamefont {E.}~\bibnamefont
  {Bocquillon}}, \bibinfo {author} {\bibfnamefont {R.~S.}\ \bibnamefont
  {Deacon}}, \bibinfo {author} {\bibfnamefont {J.}~\bibnamefont {Wiedenmann}},
  \bibinfo {author} {\bibfnamefont {L.}~\bibnamefont {P.}}, \bibinfo {author}
  {\bibnamefont {Klapwijk}}, \bibinfo {author} {\bibfnamefont {C.}~\bibnamefont
  {Br{\"u}ne}}, \bibinfo {author} {\bibfnamefont {K.}~\bibnamefont
  {Ishibashi}}, \bibinfo {author} {\bibfnamefont {H.}~\bibnamefont {Buhmann}},
  \ and\ \bibinfo {author} {\bibfnamefont {L.~W.}\ \bibnamefont {Molenkamp}},\
  }\href {\doibase 10.1038/nnano.2016.159} {\bibfield  {journal} {\bibinfo
  {journal} {Nat. Nano.}\ }\textbf {\bibinfo {volume} {12}},\ \bibinfo {pages}
  {137} (\bibinfo {year} {2016})}\BibitemShut {NoStop}%
\bibitem [{\citenamefont {Dom\'inguez}\ \emph
  {et~al.}(2017{\natexlab{a}})\citenamefont {Dom\'inguez}, \citenamefont
  {Kashuba}, \citenamefont {Bocquillon}, \citenamefont {Wiedenmann},
  \citenamefont {Deacon}, \citenamefont {Klapwijk}, \citenamefont {Platero},
  \citenamefont {Molenkamp}, \citenamefont {Trauzettel},\ and\ \citenamefont
  {Hankiewicz}}]{Dominguez2017b}%
  \BibitemOpen
  \bibfield  {author} {\bibinfo {author} {\bibfnamefont {F.}~\bibnamefont
  {Dom\'inguez}}, \bibinfo {author} {\bibfnamefont {O.}~\bibnamefont
  {Kashuba}}, \bibinfo {author} {\bibfnamefont {E.}~\bibnamefont {Bocquillon}},
  \bibinfo {author} {\bibfnamefont {J.}~\bibnamefont {Wiedenmann}}, \bibinfo
  {author} {\bibfnamefont {R.~S.}\ \bibnamefont {Deacon}}, \bibinfo {author}
  {\bibfnamefont {T.~M.}\ \bibnamefont {Klapwijk}}, \bibinfo {author}
  {\bibfnamefont {G.}~\bibnamefont {Platero}}, \bibinfo {author} {\bibfnamefont
  {L.~W.}\ \bibnamefont {Molenkamp}}, \bibinfo {author} {\bibfnamefont
  {B.}~\bibnamefont {Trauzettel}}, \ and\ \bibinfo {author} {\bibfnamefont
  {E.~M.}\ \bibnamefont {Hankiewicz}},\ }\href {\doibase
  10.1103/PhysRevB.95.195430} {\bibfield  {journal} {\bibinfo  {journal} {Phys.
  Rev. B}\ }\textbf {\bibinfo {volume} {95}},\ \bibinfo {pages} {195430}
  (\bibinfo {year} {2017}{\natexlab{a}})}\BibitemShut {NoStop}%
\bibitem [{\citenamefont {Pic\'o-Cort\'es}\ \emph {et~al.}(2017)\citenamefont
  {Pic\'o-Cort\'es}, \citenamefont {Dom\'{\i}nguez},\ and\ \citenamefont
  {Platero}}]{Pico2017a}%
  \BibitemOpen
  \bibfield  {author} {\bibinfo {author} {\bibfnamefont {J.}~\bibnamefont
  {Pic\'o-Cort\'es}}, \bibinfo {author} {\bibfnamefont {F.}~\bibnamefont
  {Dom\'{\i}nguez}}, \ and\ \bibinfo {author} {\bibfnamefont {G.}~\bibnamefont
  {Platero}},\ }\href {\doibase 10.1103/PhysRevB.96.125438} {\bibfield
  {journal} {\bibinfo  {journal} {Phys. Rev. B}\ }\textbf {\bibinfo {volume}
  {96}},\ \bibinfo {pages} {125438} (\bibinfo {year} {2017})}\BibitemShut
  {NoStop}%
\bibitem [{\citenamefont {Yeyati}\ \emph {et~al.}(2003)\citenamefont {Yeyati},
  \citenamefont {Mart\'{\i}n-Rodero},\ and\ \citenamefont
  {Vecino}}]{Yeyati2003a}%
  \BibitemOpen
  \bibfield  {author} {\bibinfo {author} {\bibfnamefont {A.~L.}\ \bibnamefont
  {Yeyati}}, \bibinfo {author} {\bibfnamefont {A.}~\bibnamefont
  {Mart\'{\i}n-Rodero}}, \ and\ \bibinfo {author} {\bibfnamefont
  {E.}~\bibnamefont {Vecino}},\ }\href {\doibase 10.1103/PhysRevLett.91.266802}
  {\bibfield  {journal} {\bibinfo  {journal} {Phys. Rev. Lett.}\ }\textbf
  {\bibinfo {volume} {91}},\ \bibinfo {pages} {266802} (\bibinfo {year}
  {2003})}\BibitemShut {NoStop}%
\bibitem [{\citenamefont {Houzet}\ \emph {et~al.}(2013)\citenamefont {Houzet},
  \citenamefont {Meyer}, \citenamefont {Badiane},\ and\ \citenamefont
  {Glazman}}]{Houzet2013a}%
  \BibitemOpen
  \bibfield  {author} {\bibinfo {author} {\bibfnamefont {M.}~\bibnamefont
  {Houzet}}, \bibinfo {author} {\bibfnamefont {J.~S.}\ \bibnamefont {Meyer}},
  \bibinfo {author} {\bibfnamefont {D.~M.}\ \bibnamefont {Badiane}}, \ and\
  \bibinfo {author} {\bibfnamefont {L.~I.}\ \bibnamefont {Glazman}},\ }\href
  {\doibase 10.1103/PhysRevLett.111.046401} {\bibfield  {journal} {\bibinfo
  {journal} {Phys. Rev. Lett.}\ }\textbf {\bibinfo {volume} {111}},\ \bibinfo
  {pages} {046401} (\bibinfo {year} {2013})}\BibitemShut {NoStop}%
\bibitem [{\citenamefont {Virtanen}\ and\ \citenamefont
  {Recher}(2013)}]{Virtanen2013a}%
  \BibitemOpen
  \bibfield  {author} {\bibinfo {author} {\bibfnamefont {P.}~\bibnamefont
  {Virtanen}}\ and\ \bibinfo {author} {\bibfnamefont {P.}~\bibnamefont
  {Recher}},\ }\href {\doibase 10.1103/PhysRevB.88.144507} {\bibfield
  {journal} {\bibinfo  {journal} {Phys. Rev. B}\ }\textbf {\bibinfo {volume}
  {88}},\ \bibinfo {pages} {144507} (\bibinfo {year} {2013})}\BibitemShut
  {NoStop}%
\bibitem [{\citenamefont {Matthews}\ \emph {et~al.}(2014)\citenamefont
  {Matthews}, \citenamefont {Ribeiro},\ and\ \citenamefont
  {Garc\'{\i}a-Garc\'{\i}a}}]{Matthews2014a}%
  \BibitemOpen
  \bibfield  {author} {\bibinfo {author} {\bibfnamefont {P.}~\bibnamefont
  {Matthews}}, \bibinfo {author} {\bibfnamefont {P.}~\bibnamefont {Ribeiro}}, \
  and\ \bibinfo {author} {\bibfnamefont {A.~M.}\ \bibnamefont
  {Garc\'{\i}a-Garc\'{\i}a}},\ }\href {\doibase 10.1103/PhysRevLett.112.247001}
  {\bibfield  {journal} {\bibinfo  {journal} {Phys. Rev. Lett.}\ }\textbf
  {\bibinfo {volume} {112}},\ \bibinfo {pages} {247001} (\bibinfo {year}
  {2014})}\BibitemShut {NoStop}%
\bibitem [{\citenamefont {Sau}\ and\ \citenamefont
  {Setiawan}(2017)}]{Sau2017a}%
  \BibitemOpen
  \bibfield  {author} {\bibinfo {author} {\bibfnamefont {J.~D.}\ \bibnamefont
  {Sau}}\ and\ \bibinfo {author} {\bibfnamefont {F.}~\bibnamefont {Setiawan}},\
  }\href {\doibase 10.1103/PhysRevB.95.060501} {\bibfield  {journal} {\bibinfo
  {journal} {Phys. Rev. B}\ }\textbf {\bibinfo {volume} {95}},\ \bibinfo
  {pages} {060501} (\bibinfo {year} {2017})}\BibitemShut {NoStop}%
\bibitem [{\citenamefont {Haim}\ \emph {et~al.}(2015)\citenamefont {Haim},
  \citenamefont {Berg}, \citenamefont {von Oppen},\ and\ \citenamefont
  {Oreg}}]{Haim2015a}%
  \BibitemOpen
  \bibfield  {author} {\bibinfo {author} {\bibfnamefont {A.}~\bibnamefont
  {Haim}}, \bibinfo {author} {\bibfnamefont {E.}~\bibnamefont {Berg}}, \bibinfo
  {author} {\bibfnamefont {F.}~\bibnamefont {von Oppen}}, \ and\ \bibinfo
  {author} {\bibfnamefont {Y.}~\bibnamefont {Oreg}},\ }\href {\doibase
  10.1103/PhysRevLett.114.166406} {\bibfield  {journal} {\bibinfo  {journal}
  {Phys. Rev. Lett.}\ }\textbf {\bibinfo {volume} {114}},\ \bibinfo {pages}
  {166406} (\bibinfo {year} {2015})}\BibitemShut {NoStop}%
\bibitem [{\citenamefont {Fleckenstein}\ \emph {et~al.}(2021)\citenamefont
  {Fleckenstein}, \citenamefont {Ziani}, \citenamefont {Calzona}, \citenamefont
  {Sassetti},\ and\ \citenamefont {Trauzettel}}]{Fleckenstein2021a}%
  \BibitemOpen
  \bibfield  {author} {\bibinfo {author} {\bibfnamefont {C.}~\bibnamefont
  {Fleckenstein}}, \bibinfo {author} {\bibfnamefont {N.~T.}\ \bibnamefont
  {Ziani}}, \bibinfo {author} {\bibfnamefont {A.}~\bibnamefont {Calzona}},
  \bibinfo {author} {\bibfnamefont {M.}~\bibnamefont {Sassetti}}, \ and\
  \bibinfo {author} {\bibfnamefont {B.}~\bibnamefont {Trauzettel}},\ }\href
  {\doibase 10.1103/PhysRevB.103.125303} {\bibfield  {journal} {\bibinfo
  {journal} {Phys. Rev. B}\ }\textbf {\bibinfo {volume} {103}},\ \bibinfo
  {pages} {125303} (\bibinfo {year} {2021})}\BibitemShut {NoStop}%
\bibitem [{\citenamefont {Pakizer}\ and\ \citenamefont
  {Matos-Abiague}(2021)}]{Pakizer2021a}%
  \BibitemOpen
  \bibfield  {author} {\bibinfo {author} {\bibfnamefont {J.~D.}\ \bibnamefont
  {Pakizer}}\ and\ \bibinfo {author} {\bibfnamefont {A.}~\bibnamefont
  {Matos-Abiague}},\ }\href {\doibase 10.1103/PhysRevB.104.L100506} {\bibfield
  {journal} {\bibinfo  {journal} {Phys. Rev. B}\ }\textbf {\bibinfo {volume}
  {104}},\ \bibinfo {pages} {L100506} (\bibinfo {year} {2021})}\BibitemShut
  {NoStop}%
\bibitem [{\citenamefont {Schelter}\ \emph {et~al.}(2012)\citenamefont
  {Schelter}, \citenamefont {Trauzettel},\ and\ \citenamefont
  {Recher}}]{Schelter2012a}%
  \BibitemOpen
  \bibfield  {author} {\bibinfo {author} {\bibfnamefont {J.}~\bibnamefont
  {Schelter}}, \bibinfo {author} {\bibfnamefont {B.}~\bibnamefont
  {Trauzettel}}, \ and\ \bibinfo {author} {\bibfnamefont {P.}~\bibnamefont
  {Recher}},\ }\href {\doibase 10.1103/PhysRevLett.108.106603} {\bibfield
  {journal} {\bibinfo  {journal} {Phys. Rev. Lett.}\ }\textbf {\bibinfo
  {volume} {108}},\ \bibinfo {pages} {106603} (\bibinfo {year}
  {2012})}\BibitemShut {NoStop}%
\bibitem [{\citenamefont {Haidekker~Galambos}\ \emph
  {et~al.}(2020)\citenamefont {Haidekker~Galambos}, \citenamefont {Hoffman},
  \citenamefont {Recher}, \citenamefont {Klinovaja},\ and\ \citenamefont
  {Loss}}]{Haidekker2020a}%
  \BibitemOpen
  \bibfield  {author} {\bibinfo {author} {\bibfnamefont {T.}~\bibnamefont
  {Haidekker~Galambos}}, \bibinfo {author} {\bibfnamefont {S.}~\bibnamefont
  {Hoffman}}, \bibinfo {author} {\bibfnamefont {P.}~\bibnamefont {Recher}},
  \bibinfo {author} {\bibfnamefont {J.}~\bibnamefont {Klinovaja}}, \ and\
  \bibinfo {author} {\bibfnamefont {D.}~\bibnamefont {Loss}},\ }\href {\doibase
  10.1103/PhysRevLett.125.157701} {\bibfield  {journal} {\bibinfo  {journal}
  {Phys. Rev. Lett.}\ }\textbf {\bibinfo {volume} {125}},\ \bibinfo {pages}
  {157701} (\bibinfo {year} {2020})}\BibitemShut {NoStop}%
\bibitem [{\citenamefont {Peralta~Gavensky}\ \emph {et~al.}(2020)\citenamefont
  {Peralta~Gavensky}, \citenamefont {Usaj},\ and\ \citenamefont
  {Balseiro}}]{Gavensky2020a}%
  \BibitemOpen
  \bibfield  {author} {\bibinfo {author} {\bibfnamefont {L.}~\bibnamefont
  {Peralta~Gavensky}}, \bibinfo {author} {\bibfnamefont {G.}~\bibnamefont
  {Usaj}}, \ and\ \bibinfo {author} {\bibfnamefont {C.~A.}\ \bibnamefont
  {Balseiro}},\ }\href {\doibase 10.1103/PhysRevResearch.2.033218} {\bibfield
  {journal} {\bibinfo  {journal} {Phys. Rev. Res.}\ }\textbf {\bibinfo {volume}
  {2}},\ \bibinfo {pages} {033218} (\bibinfo {year} {2020})}\BibitemShut
  {NoStop}%
\bibitem [{\citenamefont {Benjamin}\ and\ \citenamefont
  {Pachos}(2010)}]{Benjamin2010a}%
  \BibitemOpen
  \bibfield  {author} {\bibinfo {author} {\bibfnamefont {C.}~\bibnamefont
  {Benjamin}}\ and\ \bibinfo {author} {\bibfnamefont {J.~K.}\ \bibnamefont
  {Pachos}},\ }\href {\doibase 10.1103/PhysRevB.81.085101} {\bibfield
  {journal} {\bibinfo  {journal} {Phys. Rev. B}\ }\textbf {\bibinfo {volume}
  {81}},\ \bibinfo {pages} {085101} (\bibinfo {year} {2010})}\BibitemShut
  {NoStop}%
\bibitem [{\citenamefont {Dartiailh}\ \emph {et~al.}(2021)\citenamefont
  {Dartiailh}, \citenamefont {Mayer}, \citenamefont {Yuan}, \citenamefont
  {Wickramasinghe}, \citenamefont {Matos-Abiague}, \citenamefont {\ifmmode
  \check{Z}\else \v{Z}\fi{}uti\ifmmode~\acute{c}\else \'{c}\fi{}},\ and\
  \citenamefont {Shabani}}]{Dartiailh2021a}%
  \BibitemOpen
  \bibfield  {author} {\bibinfo {author} {\bibfnamefont {M.~C.}\ \bibnamefont
  {Dartiailh}}, \bibinfo {author} {\bibfnamefont {W.}~\bibnamefont {Mayer}},
  \bibinfo {author} {\bibfnamefont {J.}~\bibnamefont {Yuan}}, \bibinfo {author}
  {\bibfnamefont {K.~S.}\ \bibnamefont {Wickramasinghe}}, \bibinfo {author}
  {\bibfnamefont {A.}~\bibnamefont {Matos-Abiague}}, \bibinfo {author}
  {\bibfnamefont {I.}~\bibnamefont {\ifmmode \check{Z}\else
  \v{Z}\fi{}uti\ifmmode~\acute{c}\else \'{c}\fi{}}}, \ and\ \bibinfo {author}
  {\bibfnamefont {J.}~\bibnamefont {Shabani}},\ }\href {\doibase
  10.1103/PhysRevLett.126.036802} {\bibfield  {journal} {\bibinfo  {journal}
  {Phys. Rev. Lett.}\ }\textbf {\bibinfo {volume} {126}},\ \bibinfo {pages}
  {036802} (\bibinfo {year} {2021})}\BibitemShut {NoStop}%
\bibitem [{\citenamefont {Bernevig}\ \emph {et~al.}(2006)\citenamefont
  {Bernevig}, \citenamefont {Hughes},\ and\ \citenamefont
  {Zhang}}]{Bernevig2006a}%
  \BibitemOpen
  \bibfield  {author} {\bibinfo {author} {\bibfnamefont {B.~A.}\ \bibnamefont
  {Bernevig}}, \bibinfo {author} {\bibfnamefont {T.~L.}\ \bibnamefont
  {Hughes}}, \ and\ \bibinfo {author} {\bibfnamefont {S.-C.}\ \bibnamefont
  {Zhang}},\ }\href {\doibase 10.1126/science.1133734} {\bibfield  {journal}
  {\bibinfo  {journal} {Science}\ }\textbf {\bibinfo {volume} {314}},\ \bibinfo
  {pages} {1757} (\bibinfo {year} {2006})},\ \Eprint
  {http://arxiv.org/abs/https://www.science.org/doi/pdf/10.1126/science.1133734}
  {https://www.science.org/doi/pdf/10.1126/science.1133734} \BibitemShut
  {NoStop}%
\bibitem [{\citenamefont {Rothe}\ \emph {et~al.}(2010)\citenamefont {Rothe},
  \citenamefont {Reinthaler}, \citenamefont {Liu}, \citenamefont {Molenkamp},
  \citenamefont {Zhang},\ and\ \citenamefont {Hankiewicz}}]{Rothe2010}%
  \BibitemOpen
  \bibfield  {author} {\bibinfo {author} {\bibfnamefont {D.~G.}\ \bibnamefont
  {Rothe}}, \bibinfo {author} {\bibfnamefont {R.~W.}\ \bibnamefont
  {Reinthaler}}, \bibinfo {author} {\bibfnamefont {C.-X.}\ \bibnamefont {Liu}},
  \bibinfo {author} {\bibfnamefont {L.~W.}\ \bibnamefont {Molenkamp}}, \bibinfo
  {author} {\bibfnamefont {S.-C.}\ \bibnamefont {Zhang}}, \ and\ \bibinfo
  {author} {\bibfnamefont {E.~M.}\ \bibnamefont {Hankiewicz}},\ }\href
  {\doibase 10.1088/1367-2630/12/6/065012} {\bibfield  {journal} {\bibinfo
  {journal} {New Journal of Physics}\ }\textbf {\bibinfo {volume} {12}},\
  \bibinfo {pages} {065012} (\bibinfo {year} {2010})}\BibitemShut {NoStop}%
\bibitem [{\citenamefont {K\"onig}\ \emph {et~al.}(2008)\citenamefont
  {K\"onig}, \citenamefont {Buhmann}, \citenamefont {W.~Molenkamp},
  \citenamefont {Hughes}, \citenamefont {Liu}, \citenamefont {Qi},\ and\
  \citenamefont {Zhang}}]{Koenig2008a}%
  \BibitemOpen
  \bibfield  {author} {\bibinfo {author} {\bibfnamefont {M.}~\bibnamefont
  {K\"onig}}, \bibinfo {author} {\bibfnamefont {H.}~\bibnamefont {Buhmann}},
  \bibinfo {author} {\bibfnamefont {L.}~\bibnamefont {W.~Molenkamp}}, \bibinfo
  {author} {\bibfnamefont {T.}~\bibnamefont {Hughes}}, \bibinfo {author}
  {\bibfnamefont {C.-X.}\ \bibnamefont {Liu}}, \bibinfo {author} {\bibfnamefont
  {X.-L.}\ \bibnamefont {Qi}}, \ and\ \bibinfo {author} {\bibfnamefont {S.-C.}\
  \bibnamefont {Zhang}},\ }\href {\doibase 10.1143/JPSJ.77.031007} {\bibfield
  {journal} {\bibinfo  {journal} {Journal of the Physical Society of Japan}\
  }\textbf {\bibinfo {volume} {77}},\ \bibinfo {pages} {031007} (\bibinfo
  {year} {2008})},\ \Eprint
  {http://arxiv.org/abs/https://doi.org/10.1143/JPSJ.77.031007}
  {https://doi.org/10.1143/JPSJ.77.031007} \BibitemShut {NoStop}%
\bibitem [{\citenamefont {Knez}\ \emph {et~al.}(2011)\citenamefont {Knez},
  \citenamefont {Du},\ and\ \citenamefont {Sullivan}}]{Knez2011a}%
  \BibitemOpen
  \bibfield  {author} {\bibinfo {author} {\bibfnamefont {I.}~\bibnamefont
  {Knez}}, \bibinfo {author} {\bibfnamefont {R.-R.}\ \bibnamefont {Du}}, \ and\
  \bibinfo {author} {\bibfnamefont {G.}~\bibnamefont {Sullivan}},\ }\href
  {\doibase 10.1103/PhysRevLett.107.136603} {\bibfield  {journal} {\bibinfo
  {journal} {Phys. Rev. Lett.}\ }\textbf {\bibinfo {volume} {107}},\ \bibinfo
  {pages} {136603} (\bibinfo {year} {2011})}\BibitemShut {NoStop}%
\bibitem [{\citenamefont {Fu}\ and\ \citenamefont {Kane}(2008)}]{Fu2008a}%
  \BibitemOpen
  \bibfield  {author} {\bibinfo {author} {\bibfnamefont {L.}~\bibnamefont
  {Fu}}\ and\ \bibinfo {author} {\bibfnamefont {C.~L.}\ \bibnamefont {Kane}},\
  }\href {\doibase 10.1103/PhysRevLett.100.096407} {\bibfield  {journal}
  {\bibinfo  {journal} {Phys. Rev. Lett.}\ }\textbf {\bibinfo {volume} {100}},\
  \bibinfo {pages} {096407} (\bibinfo {year} {2008})}\BibitemShut {NoStop}%
\bibitem [{\citenamefont {Weithofer}\ and\ \citenamefont
  {Recher}(2013)}]{Weithofer2013a}%
  \BibitemOpen
  \bibfield  {author} {\bibinfo {author} {\bibfnamefont {L.}~\bibnamefont
  {Weithofer}}\ and\ \bibinfo {author} {\bibfnamefont {P.}~\bibnamefont
  {Recher}},\ }\href {\doibase 10.1088/1367-2630/15/8/085008} {\bibfield
  {journal} {\bibinfo  {journal} {New Journal of Physics}\ }\textbf {\bibinfo
  {volume} {15}},\ \bibinfo {pages} {085008} (\bibinfo {year}
  {2013})}\BibitemShut {NoStop}%
\bibitem [{\citenamefont {Novik}\ \emph {et~al.}(2020)\citenamefont {Novik},
  \citenamefont {Trauzettel},\ and\ \citenamefont {Recher}}]{Novik2020a}%
  \BibitemOpen
  \bibfield  {author} {\bibinfo {author} {\bibfnamefont {E.~G.}\ \bibnamefont
  {Novik}}, \bibinfo {author} {\bibfnamefont {B.}~\bibnamefont {Trauzettel}}, \
  and\ \bibinfo {author} {\bibfnamefont {P.}~\bibnamefont {Recher}},\ }\href
  {\doibase 10.1103/PhysRevB.101.235308} {\bibfield  {journal} {\bibinfo
  {journal} {Phys. Rev. B}\ }\textbf {\bibinfo {volume} {101}},\ \bibinfo
  {pages} {235308} (\bibinfo {year} {2020})}\BibitemShut {NoStop}%
\bibitem [{\citenamefont {Strunz}\ \emph {et~al.}(2020)\citenamefont {Strunz},
  \citenamefont {Wiedenmann}, \citenamefont {Fleckenstein}, \citenamefont
  {Lunczer}, \citenamefont {Beugeling}, \citenamefont {M\"uller}, \citenamefont
  {Shekhar}, \citenamefont {Ziani}, \citenamefont {Shamim}, \citenamefont
  {Kleinlein}, \citenamefont {Buhmann}, \citenamefont {Trauzettel},\ and\
  \citenamefont {Molenkamp}}]{Strunz2020}%
  \BibitemOpen
  \bibfield  {author} {\bibinfo {author} {\bibfnamefont {J.}~\bibnamefont
  {Strunz}}, \bibinfo {author} {\bibfnamefont {J.}~\bibnamefont {Wiedenmann}},
  \bibinfo {author} {\bibfnamefont {C.}~\bibnamefont {Fleckenstein}}, \bibinfo
  {author} {\bibfnamefont {L.}~\bibnamefont {Lunczer}}, \bibinfo {author}
  {\bibfnamefont {W.}~\bibnamefont {Beugeling}}, \bibinfo {author}
  {\bibfnamefont {V.~L.}\ \bibnamefont {M\"uller}}, \bibinfo {author}
  {\bibfnamefont {P.}~\bibnamefont {Shekhar}}, \bibinfo {author} {\bibfnamefont
  {N.~T.}\ \bibnamefont {Ziani}}, \bibinfo {author} {\bibfnamefont
  {S.}~\bibnamefont {Shamim}}, \bibinfo {author} {\bibfnamefont
  {J.}~\bibnamefont {Kleinlein}}, \bibinfo {author} {\bibfnamefont
  {H.}~\bibnamefont {Buhmann}}, \bibinfo {author} {\bibfnamefont
  {B.}~\bibnamefont {Trauzettel}}, \ and\ \bibinfo {author} {\bibfnamefont
  {L.~W.}\ \bibnamefont {Molenkamp}},\ }\href {\doibase
  10.1038/s41567-019-0692-4} {\bibfield  {journal} {\bibinfo  {journal} {Nature
  Physics}\ }\textbf {\bibinfo {volume} {16}},\ \bibinfo {pages} {83} (\bibinfo
  {year} {2020})}\BibitemShut {NoStop}%
\bibitem [{\citenamefont {{St\"uhler}}\ \emph {et~al.}(2022)\citenamefont
  {{St\"uhler}}, \citenamefont {{Kowalewski}}, \citenamefont {{Reis}},
  \citenamefont {{Jungblut}}, \citenamefont {{Dominguez}}, \citenamefont
  {{Scharf}}, \citenamefont {{Li}}, \citenamefont {{Sch\"afer}}, \citenamefont
  {{Hankiewicz}},\ and\ \citenamefont {{Claessen}}}]{Stuehler2022a}%
  \BibitemOpen
  \bibfield  {author} {\bibinfo {author} {\bibfnamefont {R.}~\bibnamefont
  {{St\"uhler}}}, \bibinfo {author} {\bibfnamefont {A.}~\bibnamefont
  {{Kowalewski}}}, \bibinfo {author} {\bibfnamefont {F.}~\bibnamefont
  {{Reis}}}, \bibinfo {author} {\bibfnamefont {D.}~\bibnamefont {{Jungblut}}},
  \bibinfo {author} {\bibfnamefont {F.}~\bibnamefont {{Dominguez}}}, \bibinfo
  {author} {\bibfnamefont {B.}~\bibnamefont {{Scharf}}}, \bibinfo {author}
  {\bibfnamefont {G.}~\bibnamefont {{Li}}}, \bibinfo {author} {\bibfnamefont
  {J.}~\bibnamefont {{Sch\"afer}}}, \bibinfo {author} {\bibfnamefont {E.~M.}\
  \bibnamefont {{Hankiewicz}}}, \ and\ \bibinfo {author} {\bibfnamefont
  {R.}~\bibnamefont {{Claessen}}},\ }\href {\doibase
  10.1038/s41467-022-30996-z} {\bibfield  {journal} {\bibinfo  {journal} {Nat.
  Comm.}\ }\textbf {\bibinfo {volume} {13}},\ \bibinfo {pages} {3480} (\bibinfo
  {year} {2022})}\BibitemShut {NoStop}%
\bibitem [{\citenamefont {Dynes}\ and\ \citenamefont
  {Fulton}(1971)}]{Dynes1971a}%
  \BibitemOpen
  \bibfield  {author} {\bibinfo {author} {\bibfnamefont {R.~C.}\ \bibnamefont
  {Dynes}}\ and\ \bibinfo {author} {\bibfnamefont {T.~A.}\ \bibnamefont
  {Fulton}},\ }\href {\doibase 10.1103/PhysRevB.3.3015} {\bibfield  {journal}
  {\bibinfo  {journal} {Phys. Rev. B}\ }\textbf {\bibinfo {volume} {3}},\
  \bibinfo {pages} {3015} (\bibinfo {year} {1971})}\BibitemShut {NoStop}%
\bibitem [{\citenamefont {Baxevanis}\ \emph {et~al.}(2015)\citenamefont
  {Baxevanis}, \citenamefont {Ostroukh},\ and\ \citenamefont
  {Beenakker}}]{Baxevanis2015a}%
  \BibitemOpen
  \bibfield  {author} {\bibinfo {author} {\bibfnamefont {B.}~\bibnamefont
  {Baxevanis}}, \bibinfo {author} {\bibfnamefont {V.~P.}\ \bibnamefont
  {Ostroukh}}, \ and\ \bibinfo {author} {\bibfnamefont {C.~W.~J.}\ \bibnamefont
  {Beenakker}},\ }\href {\doibase 10.1103/PhysRevB.91.041409} {\bibfield
  {journal} {\bibinfo  {journal} {Phys. Rev. B}\ }\textbf {\bibinfo {volume}
  {91}},\ \bibinfo {pages} {041409} (\bibinfo {year} {2015})}\BibitemShut
  {NoStop}%
\bibitem [{\citenamefont {Adroguer}\ \emph {et~al.}(2010)\citenamefont
  {Adroguer}, \citenamefont {Grenier}, \citenamefont {Carpentier},
  \citenamefont {Cayssol}, \citenamefont {Degiovanni},\ and\ \citenamefont
  {Orignac}}]{Adroguer2010a}%
  \BibitemOpen
  \bibfield  {author} {\bibinfo {author} {\bibfnamefont {P.}~\bibnamefont
  {Adroguer}}, \bibinfo {author} {\bibfnamefont {C.}~\bibnamefont {Grenier}},
  \bibinfo {author} {\bibfnamefont {D.}~\bibnamefont {Carpentier}}, \bibinfo
  {author} {\bibfnamefont {J.}~\bibnamefont {Cayssol}}, \bibinfo {author}
  {\bibfnamefont {P.}~\bibnamefont {Degiovanni}}, \ and\ \bibinfo {author}
  {\bibfnamefont {E.}~\bibnamefont {Orignac}},\ }\href {\doibase
  10.1103/PhysRevB.82.081303} {\bibfield  {journal} {\bibinfo  {journal} {Phys.
  Rev. B}\ }\textbf {\bibinfo {volume} {82}},\ \bibinfo {pages} {081303}
  (\bibinfo {year} {2010})}\BibitemShut {NoStop}%
\bibitem [{\citenamefont {Reinthaler}\ \emph {et~al.}(2013)\citenamefont
  {Reinthaler}, \citenamefont {Recher},\ and\ \citenamefont
  {Hankiewicz}}]{Reinthaler2013a}%
  \BibitemOpen
  \bibfield  {author} {\bibinfo {author} {\bibfnamefont {R.~W.}\ \bibnamefont
  {Reinthaler}}, \bibinfo {author} {\bibfnamefont {P.}~\bibnamefont {Recher}},
  \ and\ \bibinfo {author} {\bibfnamefont {E.~M.}\ \bibnamefont {Hankiewicz}},\
  }\href {\doibase 10.1103/PhysRevLett.110.226802} {\bibfield  {journal}
  {\bibinfo  {journal} {Phys. Rev. Lett.}\ }\textbf {\bibinfo {volume} {110}},\
  \bibinfo {pages} {226802} (\bibinfo {year} {2013})}\BibitemShut {NoStop}%
\bibitem [{Note1()}]{Note1}%
  \BibitemOpen
  \bibinfo {note} {Note that deviations of integer or semi-integer values of
  $f$ can be attributed to finite size effects.}\BibitemShut {Stop}%
\bibitem [{\citenamefont {Beenakker}\ and\ \citenamefont {van
  Houten}(1991)}]{Beenakker1991a}%
  \BibitemOpen
  \bibfield  {author} {\bibinfo {author} {\bibfnamefont {C.~W.~J.}\
  \bibnamefont {Beenakker}}\ and\ \bibinfo {author} {\bibfnamefont
  {H.}~\bibnamefont {van Houten}},\ }\href {\doibase
  10.1103/PhysRevLett.66.3056} {\bibfield  {journal} {\bibinfo  {journal}
  {Phys. Rev. Lett.}\ }\textbf {\bibinfo {volume} {66}},\ \bibinfo {pages}
  {3056} (\bibinfo {year} {1991})}\BibitemShut {NoStop}%
\bibitem [{\citenamefont {Brouwer}\ and\ \citenamefont
  {Beenakker}(1997)}]{Brouwer1997}%
  \BibitemOpen
  \bibfield  {author} {\bibinfo {author} {\bibfnamefont {P.}~\bibnamefont
  {Brouwer}}\ and\ \bibinfo {author} {\bibfnamefont {C.}~\bibnamefont
  {Beenakker}},\ }\href {\doibase
  https://doi.org/10.1016/S0960-0779(97)00018-0} {\bibfield  {journal}
  {\bibinfo  {journal} {Chaos, Solitons and Fractals}\ }\textbf {\bibinfo
  {volume} {8}},\ \bibinfo {pages} {1249} (\bibinfo {year} {1997})},\ \bibinfo
  {note} {chaos and Quantum Transport in Mesoscopic Cosmos}\BibitemShut
  {NoStop}%
\bibitem [{\citenamefont {He}\ \emph {et~al.}(2014)\citenamefont {He},
  \citenamefont {Ng}, \citenamefont {Lee},\ and\ \citenamefont
  {Law}}]{He2014a}%
  \BibitemOpen
  \bibfield  {author} {\bibinfo {author} {\bibfnamefont {J.~J.}\ \bibnamefont
  {He}}, \bibinfo {author} {\bibfnamefont {T.~K.}\ \bibnamefont {Ng}}, \bibinfo
  {author} {\bibfnamefont {P.~A.}\ \bibnamefont {Lee}}, \ and\ \bibinfo
  {author} {\bibfnamefont {K.~T.}\ \bibnamefont {Law}},\ }\href {\doibase
  10.1103/PhysRevLett.112.037001} {\bibfield  {journal} {\bibinfo  {journal}
  {Phys. Rev. Lett.}\ }\textbf {\bibinfo {volume} {112}},\ \bibinfo {pages}
  {037001} (\bibinfo {year} {2014})}\BibitemShut {NoStop}%
\bibitem [{\citenamefont {Fisher}\ and\ \citenamefont
  {Lee}(1981)}]{Fisher1981a}%
  \BibitemOpen
  \bibfield  {author} {\bibinfo {author} {\bibfnamefont {D.~S.}\ \bibnamefont
  {Fisher}}\ and\ \bibinfo {author} {\bibfnamefont {P.~A.}\ \bibnamefont
  {Lee}},\ }\href {\doibase 10.1103/PhysRevB.23.6851} {\bibfield  {journal}
  {\bibinfo  {journal} {Phys. Rev. B}\ }\textbf {\bibinfo {volume} {23}},\
  \bibinfo {pages} {6851} (\bibinfo {year} {1981})}\BibitemShut {NoStop}%
\bibitem [{\citenamefont {Iida}\ \emph {et~al.}(1990)\citenamefont {Iida},
  \citenamefont {Weidenm\"uller},\ and\ \citenamefont {Zuk}}]{Iida1990a}%
  \BibitemOpen
  \bibfield  {author} {\bibinfo {author} {\bibfnamefont {S.}~\bibnamefont
  {Iida}}, \bibinfo {author} {\bibfnamefont {H.}~\bibnamefont
  {Weidenm\"uller}}, \ and\ \bibinfo {author} {\bibfnamefont {J.}~\bibnamefont
  {Zuk}},\ }\href {\doibase https://doi.org/10.1016/0003-4916(90)90275-S}
  {\bibfield  {journal} {\bibinfo  {journal} {Annals of Physics}\ }\textbf
  {\bibinfo {volume} {200}},\ \bibinfo {pages} {219 } (\bibinfo {year}
  {1990})}\BibitemShut {NoStop}%
\bibitem [{\citenamefont {Sticlet}\ \emph {et~al.}(2012)\citenamefont
  {Sticlet}, \citenamefont {Bena},\ and\ \citenamefont {Simon}}]{Sticlet2012a}%
  \BibitemOpen
  \bibfield  {author} {\bibinfo {author} {\bibfnamefont {D.}~\bibnamefont
  {Sticlet}}, \bibinfo {author} {\bibfnamefont {C.}~\bibnamefont {Bena}}, \
  and\ \bibinfo {author} {\bibfnamefont {P.}~\bibnamefont {Simon}},\ }\href
  {\doibase 10.1103/PhysRevLett.108.096802} {\bibfield  {journal} {\bibinfo
  {journal} {Phys. Rev. Lett.}\ }\textbf {\bibinfo {volume} {108}},\ \bibinfo
  {pages} {096802} (\bibinfo {year} {2012})}\BibitemShut {NoStop}%
\bibitem [{\citenamefont {Prada}\ \emph {et~al.}(2017)\citenamefont {Prada},
  \citenamefont {Aguado},\ and\ \citenamefont {San-Jose}}]{Prada2017a}%
  \BibitemOpen
  \bibfield  {author} {\bibinfo {author} {\bibfnamefont {E.}~\bibnamefont
  {Prada}}, \bibinfo {author} {\bibfnamefont {R.}~\bibnamefont {Aguado}}, \
  and\ \bibinfo {author} {\bibfnamefont {P.}~\bibnamefont {San-Jose}},\ }\href
  {\doibase 10.1103/PhysRevB.96.085418} {\bibfield  {journal} {\bibinfo
  {journal} {Phys. Rev. B}\ }\textbf {\bibinfo {volume} {96}},\ \bibinfo
  {pages} {085418} (\bibinfo {year} {2017})}\BibitemShut {NoStop}%
\bibitem [{Note2()}]{Note2}%
  \BibitemOpen
  \bibinfo {note} {Naively, one could argue that these trivial states are
  present at zero energy only for some specific values of the Zeeman field,
  whereas the topological states should stick at zero energy, provided that the
  system is long enough. However, in an interacting scenario, the trivial
  states can be pinned at zero energy for a finite range of Zeeman fields
  because, in contrast with Majorana modes, these trivial states are charged
  and can interact with the electrostatic environment, changing the
  compressibility of the system once they are close to the Fermi
  level.}\BibitemShut {Stop}%
\bibitem [{\citenamefont {Dom\'inguez}\ \emph
  {et~al.}(2017{\natexlab{b}})\citenamefont {Dom\'inguez}, \citenamefont
  {Cayao}, \citenamefont {San-Jose}, \citenamefont {Aguado}, \citenamefont
  {Yeyati},\ and\ \citenamefont {Prada}}]{Dominguez2017a}%
  \BibitemOpen
  \bibfield  {author} {\bibinfo {author} {\bibfnamefont {F.}~\bibnamefont
  {Dom\'inguez}}, \bibinfo {author} {\bibfnamefont {J.}~\bibnamefont {Cayao}},
  \bibinfo {author} {\bibfnamefont {P.}~\bibnamefont {San-Jose}}, \bibinfo
  {author} {\bibfnamefont {R.}~\bibnamefont {Aguado}}, \bibinfo {author}
  {\bibfnamefont {A.~L.}\ \bibnamefont {Yeyati}}, \ and\ \bibinfo {author}
  {\bibfnamefont {E.}~\bibnamefont {Prada}},\ }\href {\doibase
  10.1038/s41535-017-0012-0} {\bibfield  {journal} {\bibinfo  {journal} {npj
  Quantum Materials}\ }\textbf {\bibinfo {volume} {2}},\ \bibinfo {pages} {13}
  (\bibinfo {year} {2017}{\natexlab{b}})}\BibitemShut {NoStop}%
\bibitem [{\citenamefont {Escribano}\ \emph {et~al.}(2018)\citenamefont
  {Escribano}, \citenamefont {Yeyati},\ and\ \citenamefont
  {Prada}}]{Escribano2018a}%
  \BibitemOpen
  \bibfield  {author} {\bibinfo {author} {\bibfnamefont {S.~D.}\ \bibnamefont
  {Escribano}}, \bibinfo {author} {\bibfnamefont {A.~L.}\ \bibnamefont
  {Yeyati}}, \ and\ \bibinfo {author} {\bibfnamefont {E.}~\bibnamefont
  {Prada}},\ }\href {\doibase 10.3762/bjnano.9.203} {\bibfield  {journal}
  {\bibinfo  {journal} {Beilstein J. Nanotechnol.}\ }\textbf {\bibinfo {volume}
  {9}} (\bibinfo {year} {2018}),\ 10.3762/bjnano.9.203}\BibitemShut {NoStop}%
\bibitem [{\citenamefont {Martin-Rodero}\ \emph {et~al.}(1994)\citenamefont
  {Martin-Rodero}, \citenamefont {Garcia-Vidal},\ and\ \citenamefont
  {Levy~Yeyati}}]{Martin1994a}%
  \BibitemOpen
  \bibfield  {author} {\bibinfo {author} {\bibfnamefont {A.}~\bibnamefont
  {Martin-Rodero}}, \bibinfo {author} {\bibfnamefont {F.~J.}\ \bibnamefont
  {Garcia-Vidal}}, \ and\ \bibinfo {author} {\bibfnamefont {A.}~\bibnamefont
  {Levy~Yeyati}},\ }\href {\doibase 10.1103/PhysRevLett.72.554} {\bibfield
  {journal} {\bibinfo  {journal} {Phys. Rev. Lett.}\ }\textbf {\bibinfo
  {volume} {72}},\ \bibinfo {pages} {554} (\bibinfo {year} {1994})}\BibitemShut
  {NoStop}%
\bibitem [{\citenamefont {Levy~Yeyati}\ \emph {et~al.}(1995)\citenamefont
  {Levy~Yeyati}, \citenamefont {Mart\'{\i}n-Rodero},\ and\ \citenamefont
  {Garc\'{\i}a-Vidal}}]{Yeyati1995a}%
  \BibitemOpen
  \bibfield  {author} {\bibinfo {author} {\bibfnamefont {A.}~\bibnamefont
  {Levy~Yeyati}}, \bibinfo {author} {\bibfnamefont {A.}~\bibnamefont
  {Mart\'{\i}n-Rodero}}, \ and\ \bibinfo {author} {\bibfnamefont {F.~J.}\
  \bibnamefont {Garc\'{\i}a-Vidal}},\ }\href {\doibase
  10.1103/PhysRevB.51.3743} {\bibfield  {journal} {\bibinfo  {journal} {Phys.
  Rev. B}\ }\textbf {\bibinfo {volume} {51}},\ \bibinfo {pages} {3743}
  (\bibinfo {year} {1995})}\BibitemShut {NoStop}%
\bibitem [{\citenamefont {Keldysh}(1964)}]{Keldysh1964a}%
  \BibitemOpen
  \bibfield  {author} {\bibinfo {author} {\bibfnamefont {L.~V.}\ \bibnamefont
  {Keldysh}},\ }\href@noop {} {\bibfield  {journal} {\bibinfo  {journal} {Zh.
  Eksp. Teor. Fiz.}\ }\textbf {\bibinfo {volume} {47}},\ \bibinfo {pages}
  {1515} (\bibinfo {year} {1964})}\BibitemShut {NoStop}%
\bibitem [{\citenamefont {Nambu}(1960)}]{Nambu1960a}%
  \BibitemOpen
  \bibfield  {author} {\bibinfo {author} {\bibfnamefont {Y.}~\bibnamefont
  {Nambu}},\ }\href {\doibase 10.1103/PhysRev.117.648} {\bibfield  {journal}
  {\bibinfo  {journal} {Phys. Rev.}\ }\textbf {\bibinfo {volume} {117}},\
  \bibinfo {pages} {648} (\bibinfo {year} {1960})}\BibitemShut {NoStop}%
\bibitem [{\citenamefont {Sancho}\ \emph {et~al.}(1985)\citenamefont {Sancho},
  \citenamefont {Sancho},\ and\ \citenamefont {Rubio}}]{Lopez1984a}%
  \BibitemOpen
  \bibfield  {author} {\bibinfo {author} {\bibfnamefont {M.~P.~L.}\
  \bibnamefont {Sancho}}, \bibinfo {author} {\bibfnamefont {J.~M.~L.}\
  \bibnamefont {Sancho}}, \ and\ \bibinfo {author} {\bibfnamefont
  {J.}~\bibnamefont {Rubio}},\ }\href {\doibase 10.1088/0305-4608/15/4/009}
  {\bibfield  {journal} {\bibinfo  {journal} {Journal of Physics F: Metal
  Physics}\ }\textbf {\bibinfo {volume} {15}},\ \bibinfo {pages} {851}
  (\bibinfo {year} {1985})}\BibitemShut {NoStop}%
\bibitem [{\citenamefont {Asano}(2001)}]{Asano2001a}%
  \BibitemOpen
  \bibfield  {author} {\bibinfo {author} {\bibfnamefont {Y.}~\bibnamefont
  {Asano}},\ }\href {\doibase 10.1103/PhysRevB.63.052512} {\bibfield  {journal}
  {\bibinfo  {journal} {Phys. Rev. B}\ }\textbf {\bibinfo {volume} {63}},\
  \bibinfo {pages} {052512} (\bibinfo {year} {2001})}\BibitemShut {NoStop}%
\bibitem [{\citenamefont {Hart}\ \emph {et~al.}(2014)\citenamefont {Hart},
  \citenamefont {Ren}, \citenamefont {Wagner}, \citenamefont {Leubner},
  \citenamefont {Mühlbauer}, \citenamefont {Br\"une}, \citenamefont {Buhmann},
  \citenamefont {Molenkamp},\ and\ \citenamefont {Yacoby}}]{Hart2014a}%
  \BibitemOpen
  \bibfield  {author} {\bibinfo {author} {\bibfnamefont {S.}~\bibnamefont
  {Hart}}, \bibinfo {author} {\bibfnamefont {H.}~\bibnamefont {Ren}}, \bibinfo
  {author} {\bibfnamefont {T.}~\bibnamefont {Wagner}}, \bibinfo {author}
  {\bibfnamefont {P.}~\bibnamefont {Leubner}}, \bibinfo {author} {\bibfnamefont
  {M.}~\bibnamefont {Mühlbauer}}, \bibinfo {author} {\bibfnamefont
  {C.}~\bibnamefont {Br\"une}}, \bibinfo {author} {\bibfnamefont
  {H.}~\bibnamefont {Buhmann}}, \bibinfo {author} {\bibfnamefont {L.~W.}\
  \bibnamefont {Molenkamp}}, \ and\ \bibinfo {author} {\bibfnamefont
  {A.}~\bibnamefont {Yacoby}},\ }\href {\doibase 10.1038/nphys3036} {\bibfield
  {journal} {\bibinfo  {journal} {Nat. Phys.}\ }\textbf {\bibinfo {volume}
  {10}},\ \bibinfo {pages} {638} (\bibinfo {year} {2014})}\BibitemShut
  {NoStop}%
\bibitem [{\citenamefont {Pribiag}\ \emph {et~al.}(2015)\citenamefont
  {Pribiag}, \citenamefont {Beukman}, \citenamefont {Qu}, \citenamefont
  {Cassidy}, \citenamefont {Charpentier}, \citenamefont {Wegscheider},\ and\
  \citenamefont {Kouwenhoven}}]{Pribiag2015a}%
  \BibitemOpen
  \bibfield  {author} {\bibinfo {author} {\bibfnamefont {V.~S.}\ \bibnamefont
  {Pribiag}}, \bibinfo {author} {\bibfnamefont {A.~J.~A.}\ \bibnamefont
  {Beukman}}, \bibinfo {author} {\bibfnamefont {F.}~\bibnamefont {Qu}},
  \bibinfo {author} {\bibfnamefont {M.~C.}\ \bibnamefont {Cassidy}}, \bibinfo
  {author} {\bibfnamefont {C.}~\bibnamefont {Charpentier}}, \bibinfo {author}
  {\bibfnamefont {W.}~\bibnamefont {Wegscheider}}, \ and\ \bibinfo {author}
  {\bibfnamefont {L.~P.}\ \bibnamefont {Kouwenhoven}},\ }\href {\doibase
  10.1038/nnano.2015.86} {\bibfield  {journal} {\bibinfo  {journal} {Nat.
  Nanotech.}\ }\textbf {\bibinfo {volume} {10}},\ \bibinfo {pages} {593}
  (\bibinfo {year} {2015})}\BibitemShut {NoStop}%
\bibitem [{\citenamefont {Bai}\ \emph {et~al.}(2020)\citenamefont {Bai},
  \citenamefont {Yang}, \citenamefont {Luysberg}, \citenamefont {Feng},
  \citenamefont {Bliesener}, \citenamefont {Lippertz}, \citenamefont {Taskin},
  \citenamefont {Mayer},\ and\ \citenamefont {Ando}}]{Bai2020a}%
  \BibitemOpen
  \bibfield  {author} {\bibinfo {author} {\bibfnamefont {M.}~\bibnamefont
  {Bai}}, \bibinfo {author} {\bibfnamefont {F.}~\bibnamefont {Yang}}, \bibinfo
  {author} {\bibfnamefont {M.}~\bibnamefont {Luysberg}}, \bibinfo {author}
  {\bibfnamefont {J.}~\bibnamefont {Feng}}, \bibinfo {author} {\bibfnamefont
  {A.}~\bibnamefont {Bliesener}}, \bibinfo {author} {\bibfnamefont
  {G.}~\bibnamefont {Lippertz}}, \bibinfo {author} {\bibfnamefont {A.~A.}\
  \bibnamefont {Taskin}}, \bibinfo {author} {\bibfnamefont {J.}~\bibnamefont
  {Mayer}}, \ and\ \bibinfo {author} {\bibfnamefont {Y.}~\bibnamefont {Ando}},\
  }\href {\doibase 10.1103/PhysRevMaterials.4.094801} {\bibfield  {journal}
  {\bibinfo  {journal} {Phys. Rev. Materials}\ }\textbf {\bibinfo {volume}
  {4}},\ \bibinfo {pages} {094801} (\bibinfo {year} {2020})}\BibitemShut
  {NoStop}%
\bibitem [{\citenamefont {Alicea}(2010)}]{Alicea2010a}%
  \BibitemOpen
  \bibfield  {author} {\bibinfo {author} {\bibfnamefont {J.}~\bibnamefont
  {Alicea}},\ }\href {\doibase 10.1103/PhysRevB.81.125318} {\bibfield
  {journal} {\bibinfo  {journal} {Phys. Rev. B}\ }\textbf {\bibinfo {volume}
  {81}},\ \bibinfo {pages} {125318} (\bibinfo {year} {2010})}\BibitemShut
  {NoStop}%
\bibitem [{\citenamefont {Mortensen}\ \emph {et~al.}(1999)\citenamefont
  {Mortensen}, \citenamefont {Flensberg},\ and\ \citenamefont
  {Jauho}}]{Mortensen1999a}%
  \BibitemOpen
  \bibfield  {author} {\bibinfo {author} {\bibfnamefont {N.~A.}\ \bibnamefont
  {Mortensen}}, \bibinfo {author} {\bibfnamefont {K.}~\bibnamefont
  {Flensberg}}, \ and\ \bibinfo {author} {\bibfnamefont {A.-P.}\ \bibnamefont
  {Jauho}},\ }\href {\doibase 10.1103/PhysRevB.59.10176} {\bibfield  {journal}
  {\bibinfo  {journal} {Phys. Rev. B}\ }\textbf {\bibinfo {volume} {59}},\
  \bibinfo {pages} {10176} (\bibinfo {year} {1999})}\BibitemShut {NoStop}%
\bibitem [{\citenamefont {Reuther}\ \emph {et~al.}(2013)\citenamefont
  {Reuther}, \citenamefont {Alicea},\ and\ \citenamefont
  {Yacoby}}]{Reuther2013a}%
  \BibitemOpen
  \bibfield  {author} {\bibinfo {author} {\bibfnamefont {J.}~\bibnamefont
  {Reuther}}, \bibinfo {author} {\bibfnamefont {J.}~\bibnamefont {Alicea}}, \
  and\ \bibinfo {author} {\bibfnamefont {A.}~\bibnamefont {Yacoby}},\ }\href
  {\doibase 10.1103/PhysRevX.3.031011} {\bibfield  {journal} {\bibinfo
  {journal} {Phys. Rev. X}\ }\textbf {\bibinfo {volume} {3}},\ \bibinfo {pages}
  {031011} (\bibinfo {year} {2013})}\BibitemShut {NoStop}%
\bibitem [{\citenamefont {Chen}\ \emph {et~al.}(2011)\citenamefont {Chen},
  \citenamefont {Shen}, \citenamefont {Sheng}, \citenamefont {Wang},\ and\
  \citenamefont {Xing}}]{Chen2011a}%
  \BibitemOpen
  \bibfield  {author} {\bibinfo {author} {\bibfnamefont {W.}~\bibnamefont
  {Chen}}, \bibinfo {author} {\bibfnamefont {R.}~\bibnamefont {Shen}}, \bibinfo
  {author} {\bibfnamefont {L.}~\bibnamefont {Sheng}}, \bibinfo {author}
  {\bibfnamefont {B.~G.}\ \bibnamefont {Wang}}, \ and\ \bibinfo {author}
  {\bibfnamefont {D.~Y.}\ \bibnamefont {Xing}},\ }\href {\doibase
  10.1103/PhysRevB.84.115420} {\bibfield  {journal} {\bibinfo  {journal} {Phys.
  Rev. B}\ }\textbf {\bibinfo {volume} {84}},\ \bibinfo {pages} {115420}
  (\bibinfo {year} {2011})}\BibitemShut {NoStop}%
\bibitem [{\citenamefont {Novik}\ \emph {et~al.}(2014)\citenamefont {Novik},
  \citenamefont {Guigou},\ and\ \citenamefont {Recher}}]{Novik2014a}%
  \BibitemOpen
  \bibfield  {author} {\bibinfo {author} {\bibfnamefont {E.~G.}\ \bibnamefont
  {Novik}}, \bibinfo {author} {\bibfnamefont {M.}~\bibnamefont {Guigou}}, \
  and\ \bibinfo {author} {\bibfnamefont {P.}~\bibnamefont {Recher}},\ }\href
  {\doibase 10.1103/PhysRevB.90.165442} {\bibfield  {journal} {\bibinfo
  {journal} {Phys. Rev. B}\ }\textbf {\bibinfo {volume} {90}},\ \bibinfo
  {pages} {165442} (\bibinfo {year} {2014})}\BibitemShut {NoStop}%
\bibitem [{\citenamefont {Pientka}\ \emph {et~al.}(2017)\citenamefont
  {Pientka}, \citenamefont {Keselman}, \citenamefont {Berg}, \citenamefont
  {Yacoby}, \citenamefont {Stern},\ and\ \citenamefont
  {Halperin}}]{Pientka2017a}%
  \BibitemOpen
  \bibfield  {author} {\bibinfo {author} {\bibfnamefont {F.}~\bibnamefont
  {Pientka}}, \bibinfo {author} {\bibfnamefont {A.}~\bibnamefont {Keselman}},
  \bibinfo {author} {\bibfnamefont {E.}~\bibnamefont {Berg}}, \bibinfo {author}
  {\bibfnamefont {A.}~\bibnamefont {Yacoby}}, \bibinfo {author} {\bibfnamefont
  {A.}~\bibnamefont {Stern}}, \ and\ \bibinfo {author} {\bibfnamefont {B.~I.}\
  \bibnamefont {Halperin}},\ }\href {\doibase 10.1103/PhysRevX.7.021032}
  {\bibfield  {journal} {\bibinfo  {journal} {Phys. Rev. X}\ }\textbf {\bibinfo
  {volume} {7}},\ \bibinfo {pages} {021032} (\bibinfo {year}
  {2017})}\BibitemShut {NoStop}%
\bibitem [{\citenamefont {Pikulin}\ \emph {et~al.}(2016)\citenamefont
  {Pikulin}, \citenamefont {Komijani},\ and\ \citenamefont
  {Affleck}}]{Pikulin2016a}%
  \BibitemOpen
  \bibfield  {author} {\bibinfo {author} {\bibfnamefont {D.~I.}\ \bibnamefont
  {Pikulin}}, \bibinfo {author} {\bibfnamefont {Y.}~\bibnamefont {Komijani}}, \
  and\ \bibinfo {author} {\bibfnamefont {I.}~\bibnamefont {Affleck}},\ }\href
  {\doibase 10.1103/PhysRevB.93.205430} {\bibfield  {journal} {\bibinfo
  {journal} {Phys. Rev. B}\ }\textbf {\bibinfo {volume} {93}},\ \bibinfo
  {pages} {205430} (\bibinfo {year} {2016})}\BibitemShut {NoStop}%
\end{thebibliography}

\end{document}